\documentclass[prd,nofootinbib,twocolumn,superscriptaddress,preprintnumbers,balancelastpage]{revtex4-1}
\pdfoutput=1
\usepackage{amsmath,amssymb}
\usepackage{amsfonts}
\usepackage{bm,bbm}
\usepackage{graphicx}
\usepackage{makecell}
\usepackage{mathrsfs}
\usepackage{slashed}
\usepackage{booktabs}
\usepackage[dvipsnames]{xcolor}
\usepackage[normalem]{ulem}
\usepackage{tikz}
\usepackage{soul}
\usepackage{orcidlink}
\usepackage{hyperref}
\usepackage[capitalise]{cleveref}
\hypersetup{colorlinks,citecolor= blue,linkcolor= blue, urlcolor=blue}
\usepackage{xcolor}
\usepackage{ulem}
\usepackage{array}
\usepackage{verbatim}
\usepackage{epsfig}
\usepackage{multirow}
\usepackage{pifont}
\usepackage{ulem,fancyvrb}
\usepackage{xcolor} 
\allowdisplaybreaks[4]

\newcommand{\xmark}{\ding{55}}%
\newcommand{\calO}{\mathcal{O} }

\begin{document}

\title{Systematic study of lepton-flavor-violating dark matter interactions via indirect detection in effective field theories}
\author{Sahabub Jahedi\,\orcidlink{0000-0003-1016-8264}}
\affiliation{State Key Laboratory of Nuclear Physics and
Technology, Institute of Quantum Matter, South China Normal
University, Guangzhou 510006, China}
\affiliation{Guangdong Basic Research Center of Excellence for
Structure and Fundamental Interactions of Matter, Guangdong
Provincial Key Laboratory of Nuclear Science, Guangzhou
510006, China}
\author{Jin-Han Liang\,\orcidlink{0000-0002-6141-216X}}
\email{jinhanliang@m.scnu.edu.cn}
\affiliation{State Key Laboratory of Nuclear Physics and
Technology, Institute of Quantum Matter, South China Normal
University, Guangzhou 510006, China}
\affiliation{Guangdong Basic Research Center of Excellence for
Structure and Fundamental Interactions of Matter, Guangdong
Provincial Key Laboratory of Nuclear Science, Guangzhou
510006, China}
\author{Yi Liao\,\orcidlink{0000-0002-1009-5483}}
\email{liaoy@m.scnu.edu.cn}
\affiliation{State Key Laboratory of Nuclear Physics and
Technology, Institute of Quantum Matter, South China Normal
University, Guangzhou 510006, China}
\affiliation{Guangdong Basic Research Center of Excellence for
Structure and Fundamental Interactions of Matter, Guangdong
Provincial Key Laboratory of Nuclear Science, Guangzhou
510006, China}
\author{Xiao-Dong Ma\,\orcidlink{0000-0001-7207-7793}}
\email{maxid@scnu.edu.cn}
\affiliation{State Key Laboratory of Nuclear Physics and
Technology, Institute of Quantum Matter, South China Normal
University, Guangzhou 510006, China}
\affiliation{Guangdong Basic Research Center of Excellence for
Structure and Fundamental Interactions of Matter, Guangdong
Provincial Key Laboratory of Nuclear Science, Guangzhou
510006, China}
\author{Yoshiki Uchida\,\orcidlink{0000-0003-4540-7595}}
\affiliation{State Key Laboratory of Nuclear Physics and
Technology, Institute of Quantum Matter, South China Normal
University, Guangzhou 510006, China}
\affiliation{Guangdong Basic Research Center of Excellence for
Structure and Fundamental Interactions of Matter, Guangdong
Provincial Key Laboratory of Nuclear Science, Guangzhou
510006, China}

\begin{abstract}
Lepton-flavor-violating (LFV) interactions involving dark matter (DM) particles remain a largely unexplored area. 
In this study, we systematically investigate LFV DM interactions within the framework of effective field theories by analyzing astrophysical photons and positrons produced from DM annihilation. 
Employing the astrophysical photon and positron data collected by Fermi-LAT, INTEGRAL, XMM-Newton, and AMS-02, we place meaningful constraints on all leading-order effective operators involving a DM pair and a flavor-violating charged lepton pair. 
Our analysis covers the three well-known DM candidates: a scalar, a fermion, and a vector particle. 
For the photon flux, we consider contributions from final-state radiation, radiative decay, and inverse Compton scattering and examine their respective sensitivity regions across different DM masses and photon energies. 
We find that, for DM masses below $\calO(20\,\rm GeV)$, INTEGRAL provides the most stringent constraints on annihilation cross sections and effective operators in all three LFV channels, whereas AMS-02 offers the strongest constraints above $\calO(20~\rm GeV)$.
\end{abstract}

\maketitle

\section{Introduction}

The existence of dark matter (DM) is strongly supported by astrophysical observations. However, despite decades of theoretical and experimental studies, its microscopic nature remains elusive. 
Among various DM scenarios, the 
$\mathbb{Z}_2$-symmetry-protected DM candidate represents a compelling possibility that naturally ensures its longevity on cosmological timescales.
Over the years, this type of DM candidate has been extensively studied experimentally through direct and indirect detection methods as well as via direct production at colliders. 
Although no positive signals have been observed so far, stringent upper limits have been established on the relevant observables and couplings~\cite{Bertone:2004pz,Feng:2010gw,Roszkowski:2017nbc,Arcadi:2017kky,Boveia:2018yeb,Schumann:2019eaa,Arbey:2021gdg,Cirelli:2024ssz}.  
A key assumption underlying these searches is that the DM pairs couple to pairs of standard model (SM) particles in a flavor-conserving manner.

Given the stringent limits on flavor-conserving DM interactions
and our limited understanding of its true nature,
extending the investigation of DM-SM interactions into the flavor-violating sector may offer valuable new insights. 
Flavor-changing neutral current DM-quark interactions serve as an illuminating example, as they could potentially connect DM to flavor anomalies such as the $B\to K\nu\bar\nu$ excess reported in Belle II measurements~\cite{Belle-II:2023esi,He:2023bnk,He:2024iju,Berezhnoy:2023rxx,Calibbi:2025rpx,Ding:2025eqq,Liu:2025lbw}. 
In contrast, lepton-flavor-violating (LFV) DM interactions remain largely unexplored.
Such interactions are common in various new physics (NP) models,
including scotogenic seesaw scenarios that simultaneously address neutrino masses and DM candidates~\cite{Tao:1996vb,Ma:2006km},
lepton-flavored DM models~\cite{Acaroglu:2022hrm}, 
the axion- or axionlike-particle-mediated frameworks~\cite{Davoudiasl:2025gxn},
and others.
Given the variety of model frameworks, it is worthwhile to systematically study such interactions from the perspective of effective field theory (EFT), making minimal assumptions about the underlying dynamics. 

EFT is a tailored tool for studying low-energy processes with small momentum transfer, such as those relevant to DM direct and indirect detections.
In the low-energy EFT extended with an additional DM particle, known as dark sector EFT (DSEFT)~\cite{He:2022ljo,Liang:2023yta}, the LFV DM-charged lepton interactions take a factorized form, ${\tt DM}^2 \overline{\ell_i}\Gamma \ell_{j\neq i}$,
where the lepton flavor indices $i,j=e,\mu,\tau$. 
Recently, we initiated a study of these interactions by examining the LFV decays of the $\mu$ and $\tau$ leptons involving a pair of light DM in the final state~\cite{Jahedi:2025hnu}, focusing on the DM mass range $m_{\tt DM}\lesssim |m_i-m_j|/2$.
Beyond this mass threshold, indirect detection of DM becomes a valuable approach to probe the relevant parameter space~\cite{Liang:2025mfk}. 
In the present work, we aim to extend this investigation to probe the heavier DM mass range by analyzing indirect detection signals of DM annihilation, including photons and positrons, 
as observed by Fermi-LAT~\cite{Fermi-LAT:2012edv}, INTEGRAL~\cite{Bouchet:2011fn}, XMM-Newton~\cite{Foster:2021ngm}, and AMS-02~\cite{AMS:2019rhg}. 

In this analysis, we consider three well-known DM candidates with spin up to 1, including a spin-0 scalar DM $\phi$ (either complex or real), a spin-1/2 fermion DM $\chi$ (either Dirac or Majorana), and a spin-1 vector DM $X$ (either complex or real). For each case, we identify the leading-order effective operators and study constraints on their Wilson coefficients (WCs) from indirect detection signatures. 
We calculate the photon flux induced by various mechanisms, including prompt photons from final-state radiation ({\tt FSR}) and radiative decays ({\tt Rad}) of charged leptons 
and nonprompt photons from inverse Compton scattering of electrons or positrons off the background photons.
In particular, we will demonstrate that the normalized {\tt FSR} photon spectra from all s-wave operators in each specific annihilation channel exhibit a nearly degenerate pattern. 

For the two-to-two annihilation processes ${\tt DM+DM}\to \ell_i^\mp\ell_j^{\pm}$,
these operators exhibit rich partial-wave contributions, including s, p, and d waves. 
We find that 20 operators induce DM annihilation via the s wave at leading order, while 12 operators lead to the p-wave annihilation. In addition, in the fermion DM case,  
two operators $\calO_{\ell\chi 2}^{\tt V(A)}$ involve both s- and p-wave contributions at leading order, whereas in the vector DM case, two operators $\calO_{\ell X3}^{\tt V(A)}$ involve both p- and d-wave contributions.
Unlike the common assumption of s-wave dominance, the p- and d-wave contributions are highly sensitive to the DM velocity distribution and generally lead to weaker constraints on the relevant annihilation cross sections and EFT WCs.   
Our results indicate that the constraints on effective scales associated with p-wave operators are weaker by 1--2 orders of magnitude compared to those on s-wave operators of the same dimension. 
For the dimension-6 (dim-6) operators in the fermion DM scenario, the effective scale can be probed from a few tens of GeV to TeV range for the s-wave operators, with the exact constraints varying according to the DM masses.
These results complement existing constraints from charged LFV decays for DM masses below the annihilation thresholds. 

The remainder of this paper is organized as follows. 
In \cref{sec:DSEFT}, we present all leading-order operators involving a DM pair and a LFV charged lepton current, and derive DM annihilation cross sections and their velocity expansions. 
The calculation of photon and positron fluxes arising from various mechanisms is detailed in \cref{sec:spectrum}. The datasets used in our analysis,
along with our final constraints on 
thermally averaged cross sections and EFT operators are presented in \cref{sec:constraint}. 
Finally, our summary is offered in \cref{sec:summary}. 
\cref{app:thermal} provides a concise derivation of the thermally averaged cross section for a general velocity-dependent annihilation scenario. 

\section{LFV DM annihilation in DSEFT}
\label{sec:DSEFT}

\subsection{Effective lepton-DM operators in DSEFT}

\begin{table}[h]
\center
\resizebox{\linewidth}{!}{
\renewcommand\arraystretch{1.2}
\begin{tabular}{  l  l }
\hline
\multicolumn{2}{c}{\bf Scalar DM case}
\\
\hline
$\calO_{\ell\phi}^{{\tt S},ij}=
(\overline{\ell_i}\ell_j)(\phi^\dagger \phi)$ & 
$\calO_{\ell\phi}^{{\tt P},ij}=
(\overline{\ell_i} i \gamma_5\ell_j)(\phi^\dagger \phi)$
\\
$\calO_{\ell\phi}^{{\tt V},ij}=
(\overline{\ell_i}\gamma^\mu\ell_j) 
(\phi^\dagger i \overleftrightarrow{\partial_\mu} \phi)$(\xmark) & 
$ \calO_{\ell\phi}^{{\tt A},ij}=
(\overline{\ell_i}\gamma^\mu\gamma_5\ell_j) 
(\phi^\dagger i \overleftrightarrow{\partial_\mu} \phi)$(\xmark)  
\\
\hline
\multicolumn{2}{c}{\bf Fermion DM case}
\\
\hline
$\calO_{\ell\chi1}^{{\tt S},ij} =
(\overline{\ell_i}\ell_j)(\overline{\chi}\chi)$ &
$\calO_{\ell\chi1}^{{\tt P},ij} =
(\overline{\ell_i} i \gamma_5\ell_j)(\overline{\chi}\chi)$
\\
$\calO_{\ell\chi2}^{{\tt S},ij} =
(\overline{\ell_i}\ell_j)(\overline{\chi}i \gamma_5\chi)$ &
$\calO_{\ell\chi2}^{{\tt P},ij} =
(\overline{\ell_i} \gamma_5\ell_j)(\overline{\chi} \gamma_5\chi)$
\\
$\calO_{\ell\chi1}^{{\tt V},ij} =
(\overline{\ell_i}\gamma^\mu \ell_j)(\overline{\chi}\gamma_\mu\chi)$ (\xmark) &
$\calO_{\ell\chi1}^{{\tt A},ij} =
(\overline{\ell_i}\gamma^\mu\gamma_5 \ell_j)(\overline{\chi}\gamma_\mu \chi)$(\xmark)
\\
$\calO_{\ell\chi2}^{{\tt V},ij} =
(\overline{\ell_i}\gamma^\mu\ell_j)(\overline{\chi}\gamma_\mu  \gamma_5\chi)$  &
$\calO_{\ell\chi2}^{{\tt A},ij} =
(\overline{\ell_i}\gamma^\mu\gamma_5\ell_j)(\overline{\chi}\gamma_\mu \gamma_5\chi)$
\\
$\calO_{\ell\chi1}^{{\tt T},ij} =
(\overline{\ell_i}\sigma^{\mu\nu} \ell_j)(\overline{\chi}\sigma_{\mu\nu}\chi)$(\xmark) &
$\calO_{\ell\chi2}^{{\tt T},ij} =
(\overline{\ell_i}\sigma^{\mu\nu}\ell_j)(\overline{\chi}\sigma_{\mu\nu} \gamma_5\chi)$\,(\xmark)
\\\hline 
\multicolumn{2}{c}{\bf Vector DM case A}
\\
\hline
$\calO_{\ell X}^{{\tt S},ij} =
(\overline{\ell_i}\ell_j)(X_\mu^\dagger X^\mu)$ &
$ \calO_{\ell X}^{{\tt P},ij} =
(\overline{\ell_i}i \gamma_5\ell_j)(X_\mu^\dagger X^\mu)$
\\
$ \calO_{\ell X1}^{{\tt T},ij} =
\frac{i}{2} (\overline{\ell_i} \sigma^{\mu\nu}\ell_j)(X_\mu^\dagger X_\nu - X_\nu^\dagger X_\mu)$(\xmark)  &
$ \calO_{\ell X2}^{{\tt T},ij} =
\frac{1}{2} (\overline{\ell_i}\sigma^{\mu\nu}\gamma_5\ell_j)(X_\mu^\dagger X_\nu - X_\nu^\dagger X_\mu)$(\xmark)
\\
$ \calO_{\ell X1}^{{\tt V},ij}=
\frac{1}{2} [ \overline{\ell_i}\gamma_{(\mu} i \overleftrightarrow{D_{\nu)} } \ell_j] (X^{\mu \dagger} X^\nu + X^{\nu \dagger} X^\mu)$ & 
$\calO_{\ell X1}^{{\tt A},ij} =
\frac{1}{2} [\overline{\ell_i}\gamma_{(\mu} \gamma_5 i \overleftrightarrow{D_{\nu)} } \ell_j ](X^{\mu \dagger} X^\nu + X^{\nu \dagger} X^\mu  )$
\\ 
$ \calO_{\ell X2}^{{\tt V},ij} =
(\overline{\ell_i}\gamma_\mu\ell_j)\partial_\nu (X^{\mu \dagger} X^\nu + X^{\nu \dagger} X^\mu)$ & 
$\calO_{\ell X2}^{{\tt A},ij} =
(\overline{\ell_i}\gamma_\mu \gamma_5\ell_j)\partial_\nu (X^{\mu \dagger} X^\nu + X^{\nu \dagger} X^\mu)$
\\
$ \calO_{\ell X3}^{{\tt V},ij} =
(\overline{\ell_i}\gamma_\mu\ell_j)( X_\rho^\dagger \overleftrightarrow{\partial_\nu} X_\sigma )\epsilon^{\mu\nu\rho\sigma}$ & 
$\calO_{\ell X3}^{{\tt A},ij} =
(\overline{\ell_i}\gamma_\mu\gamma_5\ell_j) (X_\rho^\dagger \overleftrightarrow{ \partial_\nu} X_\sigma )\epsilon^{\mu\nu\rho\sigma}$
\\
$ \calO_{\ell X4}^{{\tt V},ij} = (\overline{\ell_i}\gamma^\mu\ell_j)(X_\nu^\dagger  i \overleftrightarrow{\partial_\mu} X^\nu)$(\xmark) & 
$\calO_{\ell X4}^{{\tt A},ij}=
(\overline{\ell_i}\gamma^\mu\gamma_5\ell_j)(X_\nu^\dagger  i \overleftrightarrow{\partial_\mu} X^\nu)$(\xmark)
\\
$ \calO_{\ell X5}^{{\tt V},ij} =
(\overline{\ell_i}\gamma_\mu\ell_j)i\partial_\nu (X^{\mu \dagger} X^\nu - X^{\nu \dagger} X^\mu)$(\xmark) &
$\calO_{\ell X5}^{{\tt A},ij}=
(\overline{\ell_i}\gamma_\mu \gamma_5\ell_j)i \partial_\nu (X^{\mu \dagger} X^\nu - X^{\nu \dagger} X^\mu)$(\xmark)
\\
$\calO_{\ell X6}^{{\tt V},ij} =
(\overline{\ell_i}\gamma_\mu\ell_j) i \partial_\nu ( X^\dagger_\rho X_\sigma )\epsilon^{\mu\nu\rho\sigma}$(\xmark) & 
$ \calO_{\ell X 6}^{{\tt A},ij} =
(\overline{\ell_i}\gamma_\mu\gamma_5\ell_j)i \partial_\nu (  X^\dagger_\rho X_\sigma)\epsilon^{\mu\nu\rho\sigma}$(\xmark)
\\\hline
\multicolumn{2}{c}{\bf Vector DM case B}
\\\hline
$\tilde \calO_{\ell X1}^{{\tt S},ij}=
(\overline{\ell_i}\ell_j)X_{\mu\nu}^\dagger  X^{\mu\nu}$ &
$\tilde \calO_{\ell X2}^{{\tt S},ij}=
(\overline{\ell_i}\ell_j)X_{\mu\nu}^\dagger \tilde X^{ \mu\nu}$
\\
$\tilde \calO_{\ell X1}^{{\tt P},ij}=
(\overline{\ell_i}i\gamma_5\ell_j)X_{\mu\nu}^\dagger X^{ \mu\nu}$ &
$\tilde \calO_{\ell X2}^{{\tt P},ij}=
(\overline{\ell_i}i \gamma_5\ell_j)X_{\mu\nu}^\dagger \tilde X^{ \mu\nu}$
\\
$\tilde \calO_{\ell X1}^{{\tt T},ij}=
\frac{i}{2}  ( \overline{\ell_i} \sigma^{\mu\nu}\ell_j)
(X^{\dagger}_{ \mu\rho} X^{\rho}_{\,\nu}-X^{\dagger}_{ \nu\rho} X^{\rho}_{\,\mu})$ (\xmark)  & 
$ \tilde \calO_{\ell X2}^{{\tt T},ij}=
\frac{1}{2}  (\overline{\ell_i} \sigma^{\mu\nu}\gamma_5 \ell_j)
(X^{\dagger}_{ \mu\rho} X^{\rho}_{\,\nu}-X^{\dagger}_{ \nu\rho} X^{\rho}_{\,\mu})$(\xmark)
\\
\hline
\end{tabular} } 
\caption{The leading-order effective operators containing a lepton bilinear and a pair of DM particles in the DSEFT framework~\cite{He:2022ljo}.
Here, the flavor indices $ij=e\mu(\mu e), e\tau(\tau e),\mu\tau(\tau\mu)$. } 
\label{tab:operators}
\end{table}

We work in the DSEFT framework and focus on the leading-order effective operators involving a pair of DM particles and a pair of charged leptons with distinct flavors~\cite{He:2022ljo,Liang:2023yta}. 
The relevant operators respecting the unbroken SM gauge symmetry $\rm SU(3)_c\otimes U(1)_{em}$ are summarized in \cref{tab:operators}, 
where the SM charged lepton fields are denoted by $\ell_{i,j}=e,\mu,\tau$. 
For the vector DM, we follow Ref.\,\cite{He:2022ljo} and consider two parametrizations depending on whether the vector field is represented by a four-vector potential $X_\mu$ (case A) or a field strength tensor $X_{\mu\nu}\equiv\partial_\mu X_\nu - \partial_\nu X_\mu$ (case B). 
In vector DM case A, the covariant derivative for the operator $\calO_{\ell X1}^{\tt V}$ is defined as 
$\overline{\ell_i}\gamma_{(\mu} i \overleftrightarrow{D_{\nu)} }\ell_j\equiv  i (\overline{\ell_i}\gamma_{\mu}  D_{\nu}\ell_j-\overline{ D_{\nu} \ell_i}\gamma_{\mu} \ell_j)+ \mu \leftrightarrow \nu $,  
and similarly for the operator $\calO_{\ell X1}^{\tt A}$ by replacing $\gamma_\mu$ with $\gamma_\mu\gamma_5$. 
In vector DM case B, the dual field strength tensor is given by $\tilde X^{\mu\nu} = (1/2)\epsilon^{\mu\nu\rho\sigma} X_{\rho\sigma}$.
In the table, the ``\xmark'' symbol indicates that the relevant operator vanishes when the DM particle is its own antiparticle.

When the full SM gauge symmetry $\rm SU(3)_c\times SU(2)_L\times U(1)_Y$ is imposed, the DSEFT framework can be naturally extended to a framework similar to the standard model effective field theory, i.e., the extension of SMEFT. In this case, the local effective operators are constructed from the DM and SM fields that respect the full gauge symmetry. 
Since the DM field may carry nontrivial SM charges, these SMEFT-like operators must properly incorporate this possibility. As examples, the operator bases for various charge assignments and DM scenarios may be found in~\cite{Brod:2017bsw,Criado:2021trs,Aebischer:2022wnl}.
For instance, if the DM field is an electroweak scalar singlet, the SM gauge-invariant counterparts of the four DSEFT operators in \cref{tab:operators} take the form  
$(\overline{L_i}e_jH)(\phi^\dagger \phi)$,
$(H^\dagger\overline{e_i}L_j)(\phi^\dagger \phi)$, 
$(\overline{L_i}\gamma^\mu L_j)(\phi^\dagger i\overleftrightarrow{\partial_\mu} \phi)$,
and $(\overline{e_i}\gamma^\mu e_j)(\phi^\dagger i\overleftrightarrow{\partial_\mu} \phi)$,
where $H$, $L$, and $e$ represent the SM Higgs, left-handed lepton doublet, and right-handed lepton singlet fields, respectively.

The ultraviolet-complete models that simultaneously account for neutrino mass and provide a DM candidate, such as the scotogenic seesaw mechanism and its extensions~\cite{Tao:1996vb,Ma:2006km}, can yield the effective DM-lepton interactions considered here.
Another example is provided in~\cite{Liang:2025mfk}, where a leptophilic Higgs doublet and a scalar DM candidate $\phi$ are introduced that can generate the LFV dim-6 operator $(\overline{L_i}e_jH)(\phi^\dagger \phi)$  after integrating out the new heavy Higgs doublet.
In this work, we focus exclusively on the indirect detection analysis within the EFT framework, leaving a comprehensive investigation of ultraviolet scenarios that generate these LFV DM interactions for future study.

According to the effective operators listed in \cref{tab:operators}, it is straightforward to calculate the DM annihilation cross section in each case and its corresponding partial-wave expansion in terms of the DM relative velocity $v_{\text{rel}}$. For our analysis, the relevant annihilation channels are two-to-two processes $\texttt{DM}+\texttt{DM} \to \ell_i^- \ell_j^+$. Their thermally averaged annihilation cross sections $\langle \sigma v_{\text{rel}} \rangle_{ij}$ directly determine the predicted photon and positron fluxes, which will be detailed in the next section. Consequently, to derive constraints on the WCs from existing astrophysical observations, it is necessary to calculate the annihilation cross sections, as we discuss in the following subsection.

\subsection{Calculation of annihilation cross sections}

The total annihilation cross section of a specific process $\texttt{DM}(k_1)+\texttt{DM}(k_2) \to \ell_i^-(p_i) \ell_j^+(p_j)$ is expressed as
\begin{align}
\sigma=\frac{1}{16\pi \lambda(s,m_{\tt DM}^2,m_{\tt DM}^2)}\int^{\frac{1}{2}(t_{+}+t_{-})}_{\frac{1}{2}(t_{+}-t_{-})} \overline{|\mathcal{M}|^2} dt,
\label{eq:sigmatot}
\end{align}
where $\overline{|\mathcal{M}|^2}$ denotes the spin-averaged and -summed matrix element squared.
$s\equiv(k_1+k_2)^2$ is the invariant mass squared of the DM pair, and
$t\equiv(k_1-p_i)^2$ with
\begin{subequations}
\begin{align}
t_{-}&=\frac{1}{s}
\sqrt{\lambda(s,m_{\tt DM}^2,m_{\tt DM}^2)\lambda(s,m_i^2,m_j^2)},
\\
t_{+}&=2m_{\tt DM}^2+m_i^2+m_j^2-s.
\end{align}
\end{subequations}
Here, $\lambda(x,y,z)=x^2+y^2+z^2-2(xy+yz+zx)$
denotes the triangle function.
Defining $\rho_{\pm}\equiv s-(m_i\mp m_j)^2$, it follows that  $\lambda({s,m_i^2,m_j^2})=\rho_+ \rho_-$.
The total cross section induced by each DSEFT operator can be calculated directly from 
\cref{eq:sigmatot}, and the final results are summarized in the second column in \cref{tab:cross-section}. 
In the table, $\kappa_f \equiv 1-4m_{\tt DM}^2/s
=\lambda(s,m_{\tt DM}^2, m_{\tt DM}^2)/s^2$.
Since DM particles annihilate nonrelativistically, it is a good approximation to expand the  velocity-weighted cross section in terms of the DM relative velocity $v_{\rm rel}$ to simplify subsequent calculations.  
In the center-of-mass frame, $s =4 m_{\tt DM}^2 [1+v_{\text{rel}}^2/(4-v_{\text{rel}}^2)] = 4 m_{\tt DM}^2[1+v_{\text{rel}}^2/4+v_{\text{rel}}^4/16+\calO(v_{\text{rel}}^6)]$ 
in the nonrelativistic limit. 
Consequently, $\sigma v_{\text{rel}}$ can be Taylor-expanded in powers of $v_{\text{rel}}^2$ as 
$\sigma v_{\text{rel}}=\hat a + \hat b\,v_{\text{rel}}^2 + \hat d\,v_{\text{rel}}^4 +\calO(v_{\text{rel}}^6)$, where $\hat a$, $\hat b$, and $\hat d$ denote the s-, p-, and d-wave contributions, respectively.
The leading-order expansion results of $\sigma v_{\rm rel}$ are tabulated in the third column in \cref{tab:cross-section}.

\begin{figure*}[t]
\centering
\includegraphics[width=0.329\linewidth]{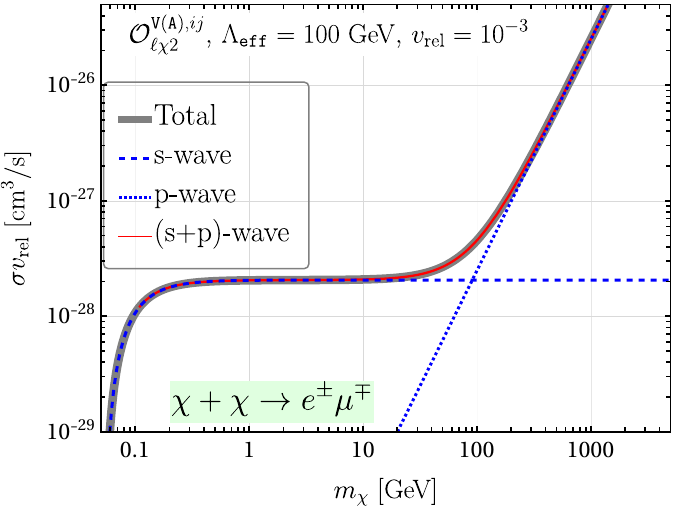} 
\includegraphics[width=0.329\linewidth]{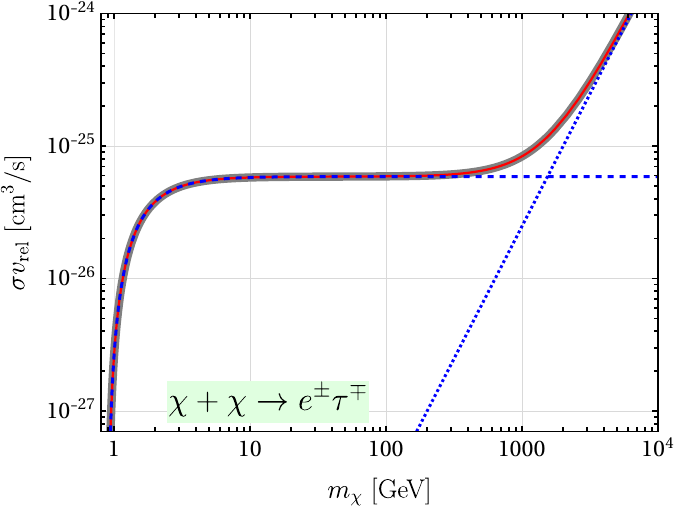}
\includegraphics[width=0.329\linewidth]{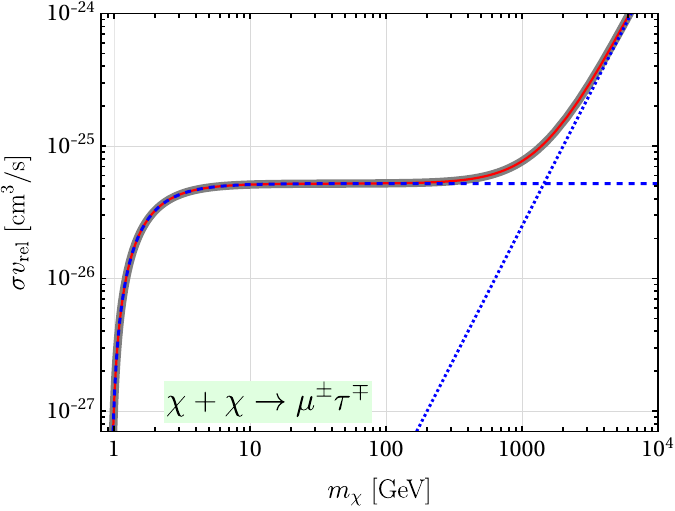}
\caption{ 
Comparison of $\sigma v_{\text{rel}}$ as a function of $m_{\tt DM}$ between the full calculation (thick gray line) and the partial-wave components for the operators $\calO_{\ell\chi2}^{{\tt V (A)},ij}$.  
In the plots, we use typical values of $v_{\rm rel}=10^{-3}$ and $\Lambda_{\text{eff}}=100$\,GeV.
Similar behavior applies to the vector DM operators $\calO_{\ell X3}^{{\tt V (A)},ij}$, where the labels   ``s-wave'' and ``p-wave'' are replaced by ``p-wave'' and ``d-wave'', respectively. 
}
\label{fig:sigmav}
\end{figure*}

We make a few comments on the results. 
First, the calculation of the squared matrix elements for the process $\texttt{DM}+\texttt{DM} \to \ell_i^- \ell_j^+$,
induced by the DSEFT operators, is performed using 
{\tt FeynCalc} package~\cite{Shtabovenko:2016sxi}, followed by the phase space integration. 
Second, in the Galactic DM halo, the typical DM velocity is approximately $10^{-3}c$ where $c$ is the speed of light, implying that DM is highly nonrelativistic. It is, therefore, reasonable to expand $\sigma v_{\text{rel}}$ in terms of $v_{\text{rel}}^2$ and keep the leading-order terms. 
From the last column in the table, one observes that 
the scalar DM operators $\calO_{\ell \phi}^{{\tt S(P)}}$ ($\calO_{\ell \phi}^{{\tt V(A)}}$) lead to annihilation processes that are dominated by the s (p) wave, due to the DM current structure. 
In the case of fermion DM, every operator contributes primarily through s-wave annihilation except $\calO^{\tt S(P)}_{\ell \chi 1}$, which instead produce a p-wave-dominated contribution. For the vector DM case A (B), contributions to $\sigma v_{\rm rel}$ from operators $\calO^{\tt S(P)}_{\ell X}$, $\calO^{\tt T}_{\ell X1,2}$, and  $\calO^{\tt V(A)}_{\ell X 6}$ ($\mathcal{\tilde{O}}^{\tt S(P)}_{\ell X 1}$ and $\mathcal{\tilde{O}}^{\tt T}_{\ell X 1,2}$) are s-wave dominated, whereas the remaining operators $\calO^{\tt V(A)}_{\ell X1,2,3,4}$ ($\mathcal{\tilde{O}}^{\tt S(P)}_{\ell X 2}$) contribute predominantly through p-wave annihilation. 
Lastly, for the operators $\calO_{\ell \chi 2}^{\tt V(A)}$ ($\calO_{\ell X 3}^{\tt V(A)}$) in the fermion (vector) case, 
it is necessary to include both the leading- and next-to-leading-order expansion to accurately capture the behavior of the full calculation. This is because 
their leading-order term is proportional to the lepton mass squared via $\eta_{ij}\equiv {\rm max}[m_i^2,m_j^2]/(4m_{\tt DM}^2)$, and the next-to-leading-order term becomes dominant when $m_{\tt DM} \gg {\rm max}[m_i,m_j]/v_{\rm rel}$. 
For illustration, 
\cref{fig:sigmav} compares the contributions to $\sigma v_{\rm rel}$ from different partial waves with the full calculation for the fermion DM operators $\calO_{\ell \chi 2}^{\tt V(A)}$ at a fixed velocity $v_{\rm rel}=10^{-3}$.
The left, middle, and right panels display the results for the three annihilation modes corresponding to $e\mu$, $e\tau$, and $\mu\tau$ flavor combinations, respectively.
As shown in the left (middle or right) panel, the s-wave contribution is valid only when the DM mass is below $m_{\mu(\tau)}/v_{\rm rel}$, whereas the p-wave contribution becomes significant when $m_{\tt DM}\gtrsim m_{\mu(\tau)}/v_{\rm rel}$.

\subsection{Thermally averaged cross sections $\langle \sigma v_{\rm rel}\rangle$}

Since the DM phase space distribution in the galaxy is a function of both position and velocity, and the observed cosmic-ray signal represents a statistical average over all  velocity states,
it is, therefore, convenient to define a position-dependent thermally averaged annihilation cross section $\langle\sigma v_{\rm rel}\rangle(\pmb{r})$ as follows: 
\begin{align}
\langle \sigma v_{\rm rel} \rangle (\pmb{r}) \equiv
\int d^3 \pmb{v}_1 f(\pmb{r}, \pmb{v}_1) \int d^3 \pmb{v}_2 f(\pmb{r}, \pmb{v}_2) \sigma v_{\rm rel},    
\label{eq:sigmav1}
\end{align}
where $\pmb{r}$ denotes the DM annihilation position in the galactic frame. 
The DM velocity distribution $f(\pmb{r}, \pmb{v})$ at position $\pmb{r}$ is approximated by a normalized, isotropic
Maxwell-Boltzmann distribution with a velocity dispersion $v_0(r)$ \cite{Robertson:2009bh,Zhao:2016xie,Arguelles:2019ouk}, which is $r$ dependent and can be obtained by solving the Jeans equation~\cite{binney2011galactic,Ferrer:2013cla}.
According to the nonrelativistic expansion,
$\sigma v_{\rm rel}  = \hat a + \hat b\, v_{\rm rel}^2 + \hat d\, v_{\rm rel}^4+\cdots $, 
the thermal average becomes
\begin{align}
\langle \sigma v_{\rm rel} \rangle(r)
=\hat{a} + 6\hat{b}\, v_0^2(r) + 60\hat{d} \, v_0^4(r)+\cdots,
\label{eq:sigmavexp}
\end{align}
which is also isotropic and $v_0(r)$ dependent for p- and d-wave contributions.
More details for the derivation are given in \cref{app:thermal}.

\begin{table*}[t]
\center
\resizebox{\linewidth}{!}{
\renewcommand\arraystretch{1.1}
\begin{tabular}{ c l l}
\hline
Operator & \multicolumn{1}{l}{Total cross section $\sigma\,\big[|C_i^j|^2\big]$} 
& \multicolumn{1}{l}{Leading-order $\sigma v_{\text{rel}}\,\big[|C_i^j|^2\big]$}
\\
\hline
\multicolumn{3}{c}{\bf Scalar DM case}
\\
\hline
$\calO^{{\tt S},ij}_{\ell\phi}(\calO^{{\tt P},ij}_{\ell\phi})$
& $\frac{\sqrt{\rho_{+}\rho_{-}}}{8\pi s^2 \sqrt{\kappa_f}} \rho_{-},$
& $\frac{1}{4\pi}(1-\eta_{ij})^2.$
\\
$\calO^{{\tt V},ij}_{\ell\phi}(\calO^{{\tt A},ij}_{\ell\phi})$
& $\frac{\sqrt{\kappa_f \rho_{+} \rho_{-}}}{24\pi s^2} \rho_{+} (3s-\rho_{-}),$
& $\frac{m_\phi^2}{12\pi}(1-\eta_{ij})^2(2+\eta_{ij}) v_{\rm rel}^2.$
\\
\hline
\multicolumn{3}{c}{\bf Fermion DM case}
\\
\hline
$\calO_{\ell \chi 1}^{{\tt S},ij}(\calO_{\ell \chi 1}^{{\tt P},ij})$
& $\frac{\sqrt{\kappa_f \rho_{+} \rho_{-}}}{16\pi s}\rho_{-} ,$
& $\frac{m_\chi^2}{8 \pi} (1-\eta_{ij})^2  v_{\rm rel}^2.$
\\
$\calO_{\ell \chi 2}^{{\tt S},ij}(\calO_{\ell \chi 2}^{{\tt P},ij})$
& $\frac{\sqrt{\rho_{+} \rho_{-}}}{16\pi s \sqrt{\kappa_f}}\rho_{-},$
& $\frac{m_\chi^2}{2\pi}(1-\eta_{ij})^2.$
\\
$\calO_{\ell \chi 1}^{{\tt V},ij}(\calO_{\ell \chi 1}^{{\tt A},ij})$
& $\frac{\sqrt{\rho_{+} \rho_{-}}}{48\pi s^2\sqrt{\kappa_f}} (3-\kappa_f) \rho_{+} (3s-\rho_{-}),$
& $\frac{m_\chi^2}{2\pi}(1-\eta_{ij})^2(2+\eta_{ij}).$
\\
$\calO_{\ell \chi 2}^{{\tt V},ij}(\calO_{\ell \chi 2}^{{\tt A},ij})$
& $\frac{\sqrt{\rho_{+} \rho_{-} }}{48\pi s^2\sqrt{\kappa_f}} \big\{ 3s[ 
 2\kappa_f\rho_{+} + (1-\kappa_f)\rho_{-} ]-(3-\kappa_f)\rho_{+}  \rho_{-} \big\},$
& $\makecell[l]{\frac{m_\chi^2}{2\pi}\eta_{ij}(1-\eta_{ij})^2 \\ 
+\frac{m_\chi^2}{24\pi}(1-\eta_{ij})(4-5 \eta_{ij}+ 7 \eta_{ij}^2) v_{\rm rel}^2.}$
\\
$\calO_{\ell \chi 1}^{{\tt T},ij}(\calO_{\ell \chi 2}^{{\tt T},ij})$
& $\frac{\sqrt{\rho_{+} \rho_{-} }}{12\pi s^2 \sqrt{\kappa_f}} \big\{3 s[ (3-2\kappa_f)\rho_{+}
+ \kappa_f \rho_{-}] - 2(3-\kappa_f)\rho_{+} \rho_{-} \big\},$
& $\frac{2m_\chi^2}{\pi}(1-\eta_{ij})^2(1+2\eta_{ij}).$
\\
\hline
\multicolumn{3}{c}{\bf Vector DM case A}
\\
\hline
$\calO_{\ell X}^{{\tt S},ij}(\calO_{\ell X}^{{\tt P},ij})$
& $\frac{\sqrt{\rho_{+} \rho_{-} }}{1512\pi m_X^4 \sqrt{\kappa_f}} (3-2 \kappa_f+3 \kappa_f^2)\rho_{-},$
& $\frac{1}{12\pi}(1-\eta_{ij})^2.$
\\
$\calO_{\ell X1}^{{\tt T},ij}(\calO_{\ell X2}^{{\tt T},ij})$
& $\makecell[l]{\frac{\sqrt{\rho_{+} \rho_{-} }}{1728\pi m_X^4 s \sqrt{\kappa_f}}\big\{
3s[ 2(2\kappa_f-\kappa_f^2)\rho_{+} +(3-4\kappa_f + \kappa_f^2)\rho_{-} ]
- 2 (3-\kappa_f^2) \rho_{+} \rho_{-}
\big\},}$
& $\frac{1}{18\pi}(1-\eta_{ij})^2(1+2\eta_{ij}).$
\\
$\calO_{\ell X1}^{{\tt V},ij}(\calO_{\ell X1}^{{\tt A},ij})$
&$\makecell[l]{\frac{\sqrt{\rho_{+} \rho_{-}}}{4320 \pi m_X^4 s^2\sqrt{\kappa_f}}\big\{
15s^3[ (\kappa_f-\kappa_f^2)\rho_{+} +\kappa_f^2 \rho_{-}] \\
- 5s^2[ 3(\kappa_f-\kappa_f^2)\rho_{+}^2 
+ 3 \kappa_f^2 \rho_{-}^2 + 4 \kappa_f\rho_{+} \rho_{-} ] \\
+ 10 s[(3-2\kappa_f)\rho_{+} + (\kappa_f+\kappa_f^2)\rho_{-}]\rho_{+}\rho_{-}
-(15-10\kappa_f+3\kappa_f^2)\rho_{+}^2\rho_{-}^2\big\},}$
& $\frac{4m_X^2}{9\pi}(1-\eta_{ij})^4(1+\eta_{ij}).$
\\
$\calO_{\ell X2}^{{\tt V},ij}(\calO_{\ell X2}^{{\tt A},ij})$
&$\frac{\sqrt{\kappa_f\rho_{+} \rho_{-}}}{864 \pi m_X^4}\big\{3s[(1-\kappa_f)\rho_{+} + \kappa_f\rho_{-}]
-(1+2\kappa_f)\rho_{+}\rho_{-}\big\},$
& $\frac{m_X^2}{27\pi}(1-\eta_{ij})^2(2+\eta_{ij}) v_{\rm rel}^2.$
\\
$\calO_{\ell X3}^{{\tt V},ij}(\calO_{\ell X3}^{{\tt A},ij})$
&
$\frac{\sqrt{\kappa_f \rho_{+} \rho_{-} }}{432\pi m_X^2s}\big\{
 3s[2\kappa_f \rho_{+} + (1-\kappa_f)\rho_{-}]
 -(3-\kappa_f)\rho_{+} \rho_{-} \big\},$
& $\makecell[l]{\frac{m_X^2}{18\pi}\eta_{ij}(1-\eta_{ij})^2 v_{\rm rel}^2 \\
+\frac{m_X^2}{108 \pi}(1-\eta_{ij})(2-\eta_{ij}+2\eta_{ij}^2) v^4_{\text{rel}}. } $
\\
$\calO_{\ell X4}^{{\tt V},ij}(\calO_{\ell X4}^{{\tt A},ij})$
&
$\frac{\sqrt{\kappa_f \rho_{+}\rho_{-}}}{3456 \pi m_X^4}
(3-2\kappa_f+3\kappa_f^2)\rho_{+}(3s-\rho_{-}),$
&$\frac{m_X^2}{36\pi}(1-\eta_{ij})^2(2+\eta_{ij}) v_{\rm rel}^2.$
\\
$\calO_{\ell X5}^{{\tt V},ij}(\calO_{\ell X5}^{{\tt A},ij})$
&
$\frac{\sqrt{\kappa_f \rho_{+} \rho_{-} }}{864 \pi m_X^4}
(2-\kappa_f) \rho_{+}(3s-\rho_{-}),$
&$\frac{2m_X^2}{27\pi}(1-\eta_{ij})^2(2+\eta_{ij}) v_{\rm rel}^2.$
\\
$\calO_{\ell X6}^{{\tt V},ij}(\calO_{\ell X6}^{{\tt A},ij})$
&
$\frac{\sqrt{\rho_{+} \rho_{-} }}{432 \pi m_X^2 s\sqrt{\kappa_f}}(3-\kappa_f) \rho_{+}(3s-\rho_{-}),$
&$\frac{2m_X^2}{9\pi}(1-\eta_{ij})^2(2+\eta_{ij}).$
\\
\hline
\multicolumn{3}{c}{\bf Vector DM case B}
\\
\hline
$\mathcal{\tilde{O}}_{\ell X1}^{{\tt S},ij}(\mathcal{\tilde{O}}_{\ell X1}^{{\tt P},ij})$
& $\frac{\sqrt{\rho_{+} \rho_{-} }}{288\pi \sqrt{\kappa_f}}(3+2 \kappa_f+3 \kappa_f^2)\rho_{-},$
& $\frac{m_X^4}{3\pi}(1-\eta_{ij})^2.$
\\
$\mathcal{\tilde{O}}_{\ell X2}^{{\tt S},ij}(\mathcal{\tilde{O}}_{\ell X2}^{{\tt P},ij})$
& $\frac{\sqrt{\kappa_f\rho_{+} \rho_{-}}}{36\pi}\rho_{-},$
& $\frac{2m_X^4}{9\pi}(1-\eta_{ij})^2 v_{\rm rel}^2.$
\\
$\mathcal{\tilde{O}}_{\ell X1}^{{\tt T},ij}(\mathcal{\tilde{O}}_{\ell X2}^{{\tt T},ij})$
& $\frac{\sqrt{\rho_{+} \rho_{-}}}{1728\pi s \sqrt{\kappa_f}}\big\{
3 s [ 2 (3\kappa_f -\kappa_f^2)\rho_{+} + (3+\kappa_f^2)\rho_{-} ]
- 2(3 + 6 \kappa_f - \kappa_f^2)\rho_{+} \rho_{-} \big\},$
& $\frac{m_X^4}{18\pi}(1-\eta_{ij})^2(1+2\eta_{ij}).$
\\
\hline
\end{tabular} } 
\caption{
The annihilation cross section for the process
${\tt DM+ DM} \to \ell^{-}_i \ell^{+}_{j\neq i}$ induced by each DSEFT operator with its WC $C_i^j$, for the case of a complex scalar or vector and Dirac fermion DM field.
For the $\sigma$ in the second column, the expression for the operators in round brackets is obtained by $\rho_{-}\leftrightarrow \rho_{+}$. 
For $\sigma v_{\rm rel}$, we have neglected the lighter lepton's mass and kept the leading-order results in the expansion of $v_{\rm rel}^2$. Note that $\eta_{ij}\equiv {\rm max}[m_i^2,m_j^2]/(4m_{\tt DM}^2)$.
In the real scalar or vector and Majorana fermion DM cases, the above results for $\sigma$ and $\sigma v_{\rm rel}$ from nonvanishing operators need to be multiplied by a factor of 4.  
} 
\label{tab:cross-section}
\end{table*}

\section{Calculation of the photon and positron spectra}
\label{sec:spectrum}

In our analysis, the relevant observational signals include diffuse photons and positrons. Below, we detail their primary production mechanisms from DM origins and the calculation of the spectra at Earth after accounting for the astrophysical propagation effects.

\subsection{Photon flux}

We begin with the calculation of the photon flux. 
The diffuse x- and gamma-ray background from DM annihilation in the Universe has two main contributions: the localized galactic component and the extragalactic component due to the smooth DM distribution throughout the Universe.  
For telescopes relevant to our discussion, the Milky Way's DM halo provides the dominant contribution; therefore, we neglect the subdominant extragalactic contribution. 
For LFV DM-lepton operators in \cref{tab:operators}, the induced photon flux is mainly generated through three mechanisms: 
{\tt FSR} via the three-body annihilation ${\tt DM+DM}\to \ell_i^\pm\ell_j^\mp \gamma$, 
{\tt Rad} of final-state leptons in the two-body annihilation ${\tt DM+DM}\to \ell_i^\pm\ell_j^\mp$, and inverse Compton scattering ({\tt ICS}) of background soft photons upscattered by 
energetic $e^\pm$ generated in the two-body annihilation and subsequent decays of $\mu$ and/or $\tau$ leptons.
The first two correspond to the so-called prompt photons, while the {\tt ICS} photons are nonprompt. 
The total differential photon flux is a sum of them: 
\begin{align}
{d^2\Phi_\gamma \over d\Omega dE_\gamma}  
= {d^2\Phi_\gamma^{\tt FSR} \over d\Omega dE_\gamma} 
+ {d^2\Phi_\gamma^{\tt Rad} \over d\Omega dE_\gamma} 
+ {d^2\Phi_\gamma^{\tt ICS} \over d\Omega dE_\gamma}.
\end{align}

For the prompt photons from {\tt FSR} and {\tt Rad}, the flux is related to the thermally averaged cross section in \cref{eq:sigmav1} via the following form:
\begin{align}
{d^2\Phi_\gamma^I \over d\Omega dE_\gamma} =
\frac{1}{4\pi}
\frac{d N_\gamma^I}{dE_\gamma} 
\int_{\rm LOS} \frac{\rho^2(\pmb{r})}{{\tt S}} 
\frac{ \langle \sigma v_{\rm rel} \rangle_{ij+ji} (\pmb{r})}{(1+\delta_{ij})m_{\tt DM}^2} d s^\prime.
\label{eq:dPhidEprompt}
\end{align}
Here, $\tt S = 4\,(2)$ for complex (real) scalar , vector, or Dirac (Majorana) DM particles, reflecting the number of independent combinations of DM annihilation. 
$d N_\gamma^I/dE_\gamma$ denotes the normalized photon spectrum in each annihilation process relative to the corresponding two-to-two annihilation, with $I$ representing either {\tt FSR} or {\tt Rad}. 
The calculation will be detailed below.
The integration of $s^\prime$  is along the line of sight (LOS) that accounts for all contributions along a specific direction. 
For the DM distribution $\rho(\pmb{r})$, we adopt the standard Navarro-Frenk-White ({\tt NFW}) profile, 
$ \rho_{\tt NFW}(r)\equiv \rho_s (r/r_s)^{-1}(1+r/r_s)^{-2}$ \cite{Navarro:1995iw,Navarro:1996gj},
where $r$ denotes the distance from the Galactic Center, 
and $r_s$ is the scale radius defined by $(d \ln \rho / d\ln r)_{r=r_s}=-2$. Numerically, $r_s = 24.42~\rm kpc$ and $\rho_s = 0.184~\rm GeV/cm^3$~\cite{Cirelli:2010xx}, leading to a local DM density $\rho \approx 0.3~\rm GeV/cm^3$.
In the galactic coordinate, $r$ is related to the LOS distance $s'$ via
$r(s^\prime) =\sqrt{r_\odot^2 + {s^\prime}^2 -2 r_\odot s^\prime \cos\psi}$. Here, $\psi$ is the angle between the LOS and the direction to the Galactic Center, 
$r_\odot = 8.3~\rm kpc$ is the distance of Earth to
the galactic center, and $\cos \psi \equiv \cos b\cos l$, where $b$ and $l$ are the galactic latitude and longitude, respectively. 
The integration range for $s^\prime$ is from 0 to a
maximal value determined by the virial radius
$r_{\rm vir} = 300~\rm kpc$ of the DM halo \cite{Lin:2019uvt}, 
\begin{align}
s^\prime_{\tt max} = r_\odot \cos\psi 
+ \sqrt{r_{\rm vir}^2 - r_\odot^2 \sin^2\psi }. 
\label{eq:smax}
\end{align}

\begin{figure*}[t]
\centering
\includegraphics[width=0.329\linewidth]{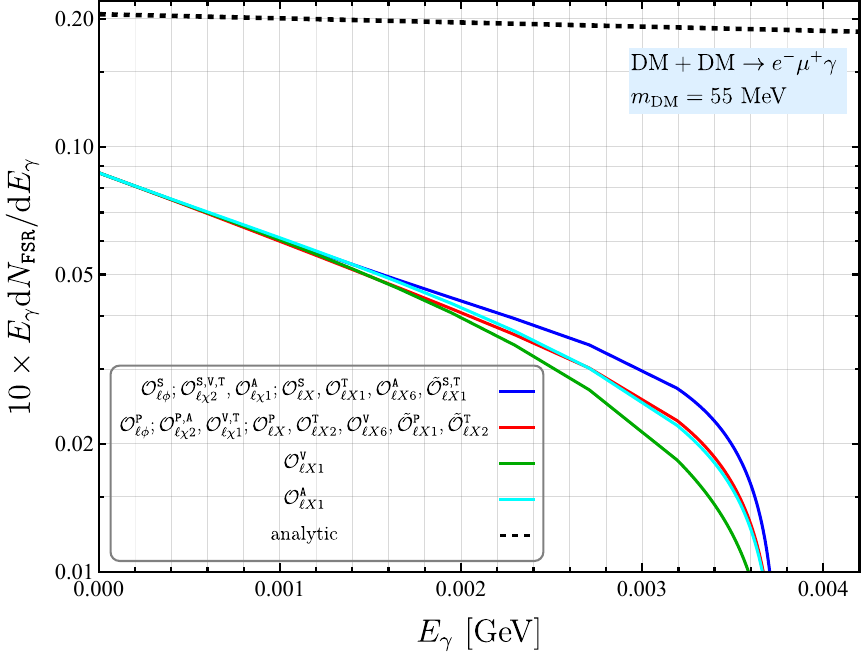} 
\includegraphics[width=0.329\linewidth]{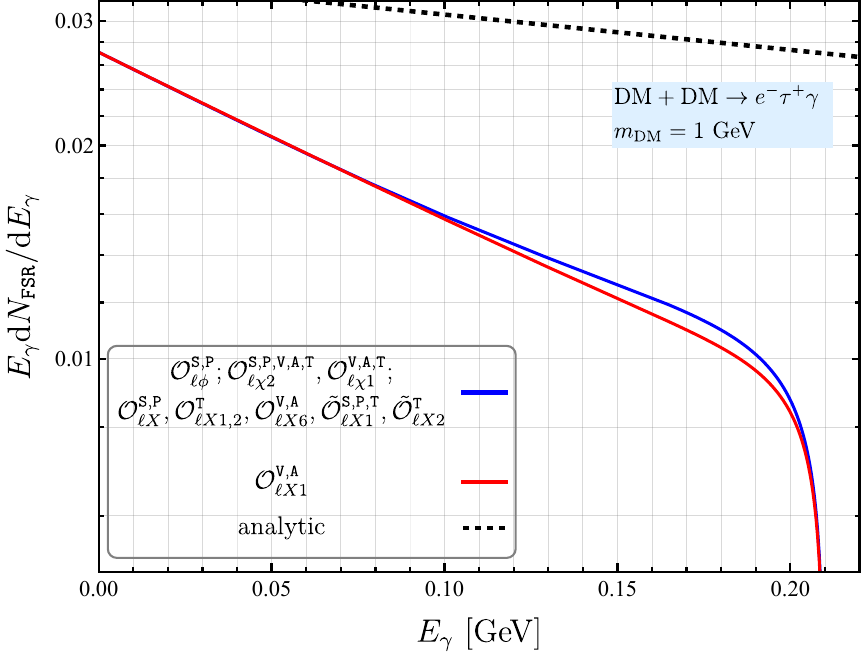} 
\includegraphics[width=0.329\linewidth]{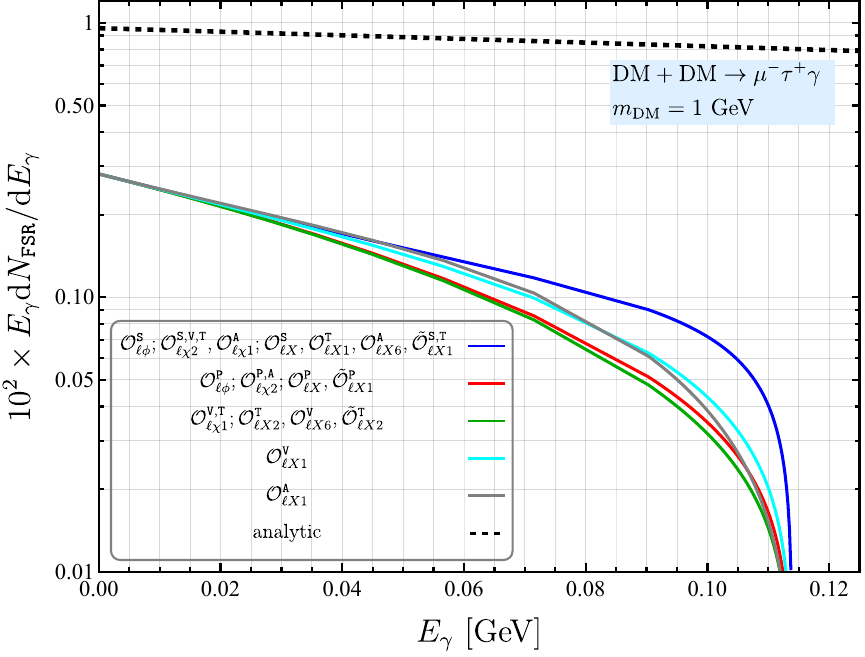}
\\
\includegraphics[width=0.329\linewidth]{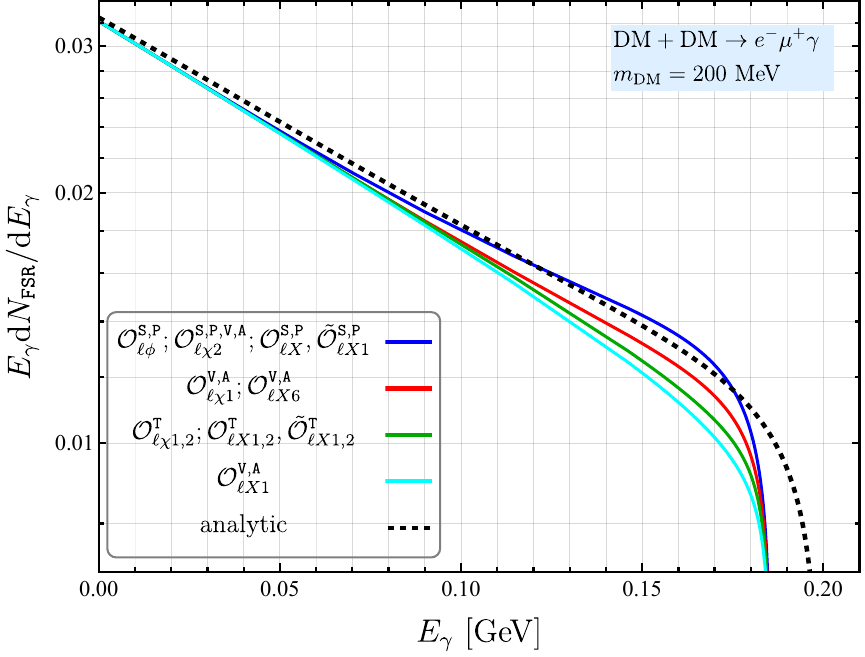} 
\includegraphics[width=0.329\linewidth]{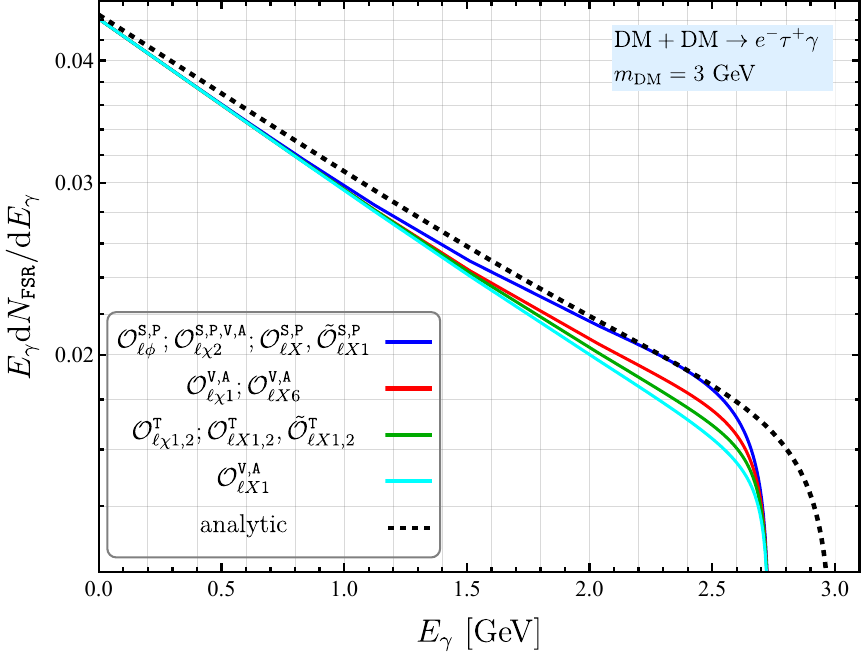} 
\includegraphics[width=0.329\linewidth]{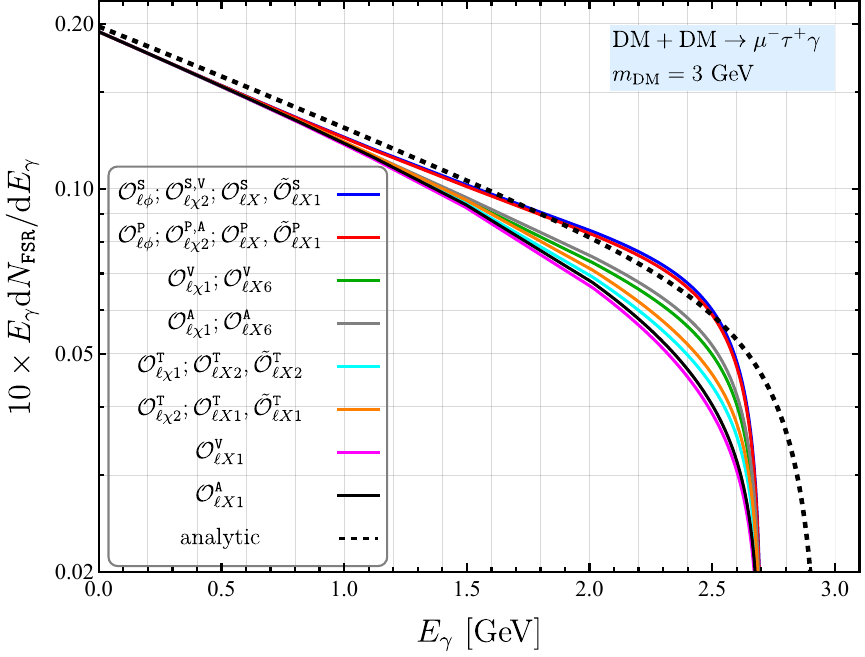}
\\
\includegraphics[width=0.329\linewidth]{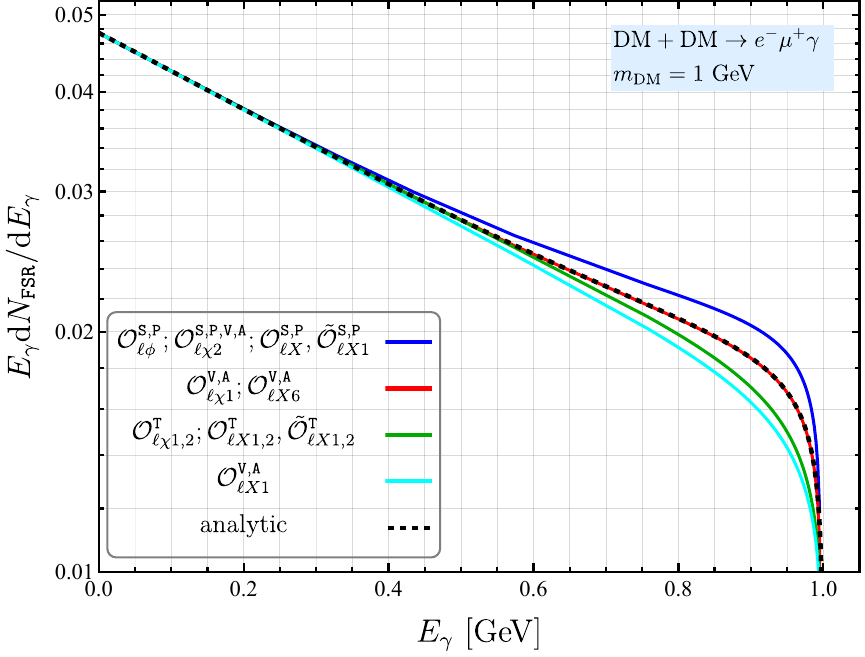} 
\includegraphics[width=0.329\linewidth]{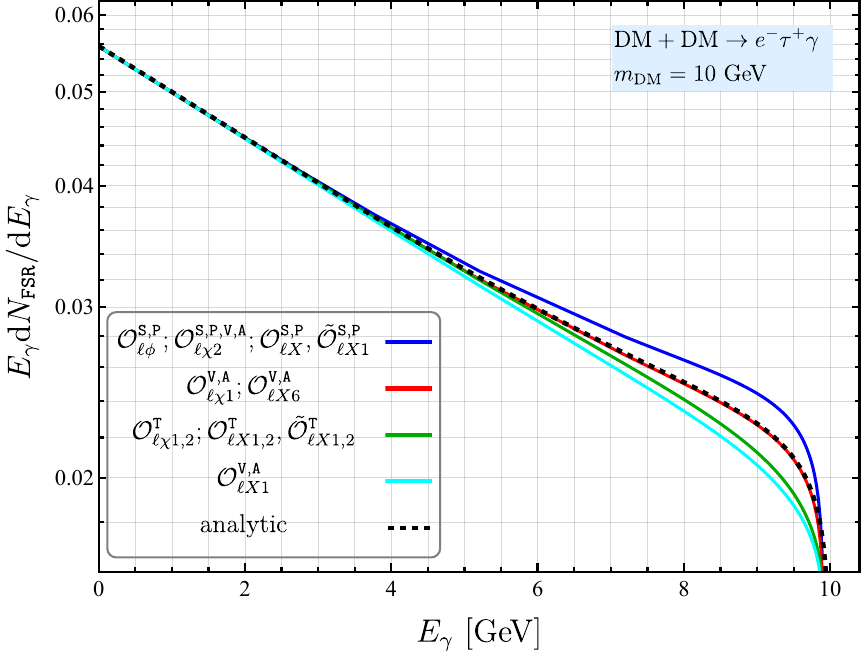} 
\includegraphics[width=0.329\linewidth]{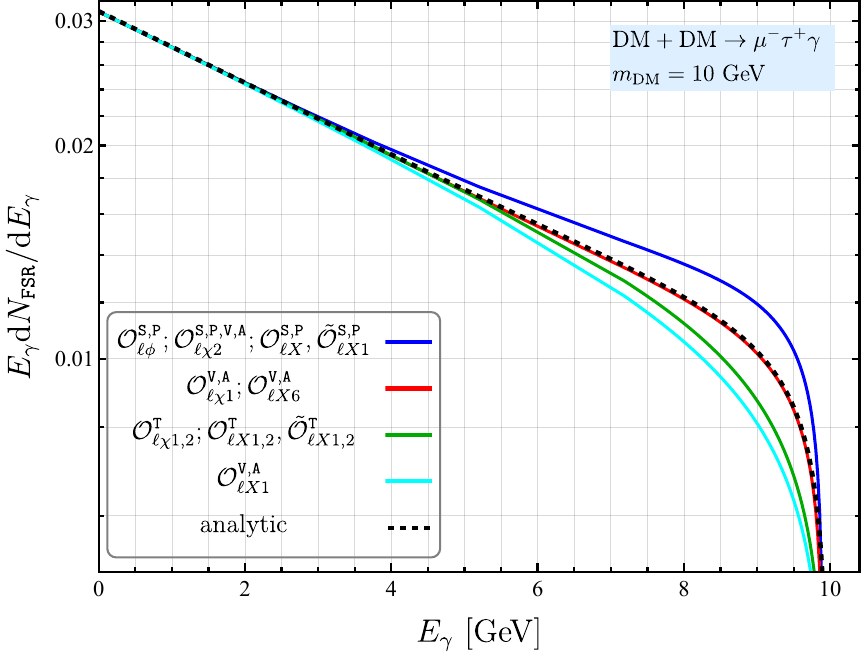} 
\caption{Comparison of normalized {\tt FSR} spectra
across various s-wave operators and different benchmark DM masses for the three LFV annihilation channels. 
The dashed curves represent the analytical calculation based on \cref{eq:ana}.}
\label{fig:E_vs_EdNdE}
\end{figure*}

\subsubsection{Final-state radiation}
\label{sec:fsr}

We now proceed to the determination of the {\tt FSR} photon spectra from operators that yield an s-wave annihilation, which can be readily identified from \cref{tab:cross-section} as those whose leading-order expansion for $\sigma v_{\rm rel}$ exhibits no velocity suppression. 
In our calculation, the normalized {\tt FSR} spectra are obtained from a direct numerical approach.
For each s-wave operator, we input the corresponding matrix element for the radiative process ${\tt DM+DM}\to \ell^-_i \ell_j^+\gamma$ into the {\tt FeynCalc} package~\cite{Shtabovenko:2016sxi} and compute the squared amplitude. This is followed by numerical integration over the final-state lepton phase space for a given DM mass, and division by the flux factor to obtain the differential cross section as a function of the photon's energy in the center-of-mass frame.
Finally, the normalized photon spectrum is obtained by dividing the differential cross section $d\sigma({\tt DM+DM}\to \ell_i^- \ell_j^+\gamma)/d E_\gamma$ by the total cross section of the corresponding two-to-two annihilation process ${\tt DM+DM}\to \ell_i^- \ell_j^+$ mediated by the same operator. The {\it Mathematica} notebook used to generate the spectra in our analysis is available as auxiliary files from the arXiv source file.

Our numerical results indicate that when the DM mass is significantly greater than the mass of the heavier lepton, all s-wave EFT operators produce nearly identical normalized {\tt FSR} spectra for a given lepton flavor combination. 
The differences in their spectra mainly occur as the 
photon energy approaches its maximum,
where the leptonic-current structures of the operators and the masses of the final-state leptons play a significant role. 
For illustration, \cref{fig:E_vs_EdNdE} presents a comparison across different operators for three benchmark DM masses in each channel. 
As observed in the plots,  for the typical photon energy below half of its maximum value $E_\gamma^{\tt max} \equiv
m_{\tt DM}-(m_i+m_j)^2/(4m_{\tt DM})$, the spectra from different operators become degenerate.

This degenerate behavior for the spectra can be understood analytically as follows. 
Based on the lepton structure function
method~\cite{Nicrosini:1987sw},
the normalized spectra can be approximated by the following analytic expression [see Eq.\,(14) in that reference]:
\begin{align}
\frac{dN_{ij}^\gamma(x_\gamma,s)}{d x_\gamma} & = H_{i,j}(x_\gamma,\hat{s}),~
\hat{s}\equiv (1-x_\gamma)s,
\label{eq:Nphoton_spectrum}
\end{align}
where $s$ ($\hat{s}$) denotes the invariant mass squared of the DM pair (lepton pair) and $x_\gamma = E_\gamma/m_{\tt DM}$.
$H_{i,j}(x_\gamma,\hat{s})$ is a final-state radiator, which accounts for the radiative corrections to the two-to-two scattering cross section caused by {\tt FSR}. By extending the lepton-flavor-conserving (LFC) result in~\cite{Nicrosini:1987sw} to our LFV cases, we have 
\begin{align}
H_{i,j}(x_\gamma,\hat{s}) = \int_{1-x_\gamma}^1 \frac{dz}{z} D_i(z,\hat{s}) D_j\Big(\frac{1-x_\gamma}{z},\hat{s}\Big).
\end{align}
Here, $D_i(x,\hat{s})$ denotes the structure function of the lepton $i$ at the scale $\hat{s}$ with a longitudinal momentum fraction of $x$. Its exact expression up to $\calO(\alpha^2)$ is given in Eq.\,(3) in Ref.\,\cite{Nicrosini:1987sw}.
Expanding the $D_{i,j}$ function to the linear order in the fine structure constant $\alpha$, we obtain
\begin{align}
H_{i,j}(x_\gamma,\hat{s}) &= d(x_\gamma,\hat{s},m_{i}) 
+ d(x_\gamma,\hat{s},m_{j}),
\label{eq:ana}
\end{align}
where
\begin{align}
d(x_\gamma,\hat{s},m_\ell) &= \frac{\alpha}{2\pi} (\log[\hat{s}/m_\ell^2]-1) \frac{1+(1-x_\gamma)^2}{x_\gamma}.
\label{eq:d}
\end{align}
As shown by the black dashed curves in \cref{fig:E_vs_EdNdE}, the spectra obtained from \cref{eq:ana}, which is essentially operator independent, agree very well with our direct numerical calculation when $m_{\tt DM}\gg (m_i+m_j)/2$ and $E_\gamma \lesssim E_{\tt DM}/2$. Even in the high-energy tail where $E_\gamma \gtrsim E_{\tt DM}/2$, the differences remain below 20\,\%.

\begin{figure*}[t]
\centering
\includegraphics[width=0.329\textwidth]{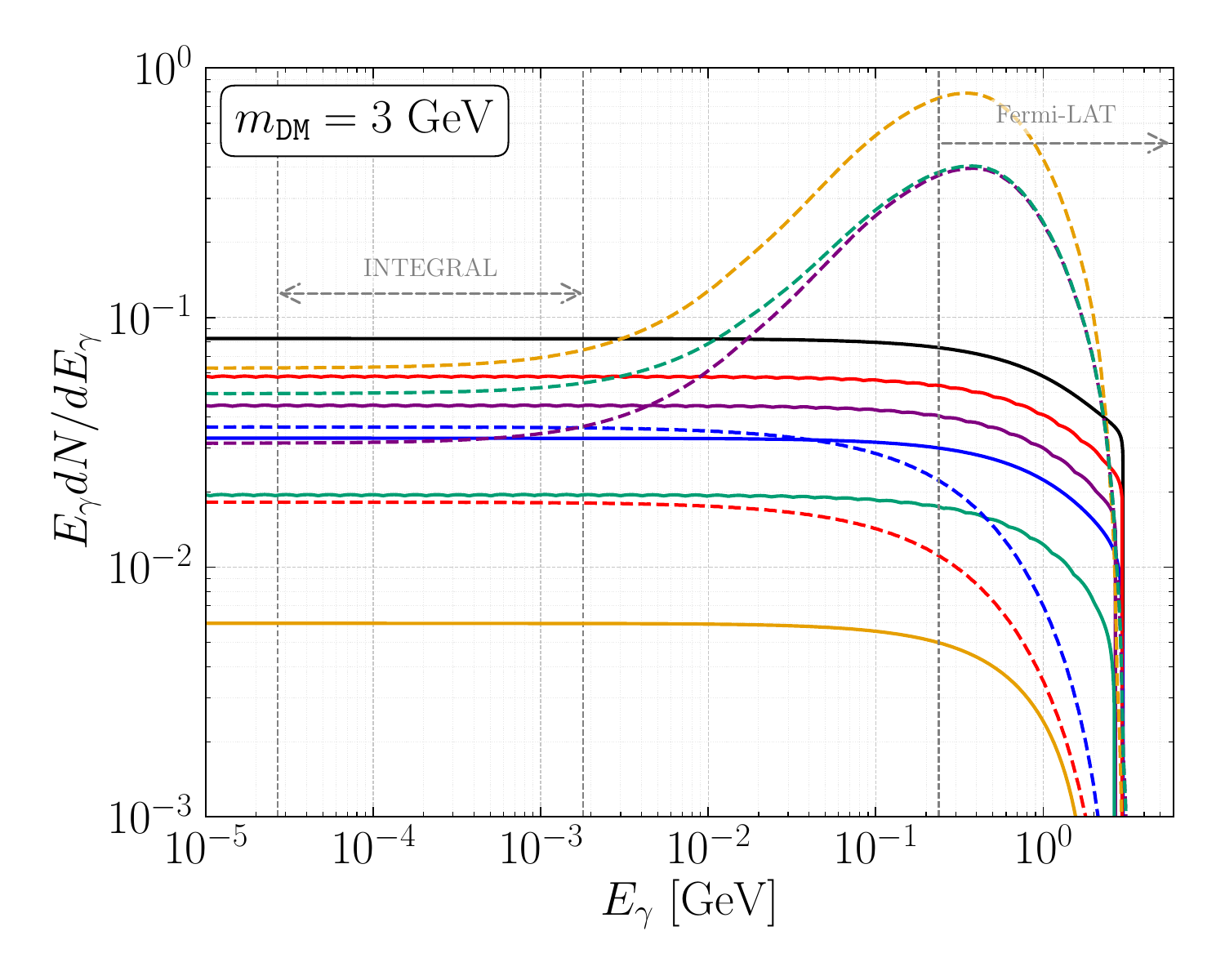}
\includegraphics[width=0.329\textwidth]{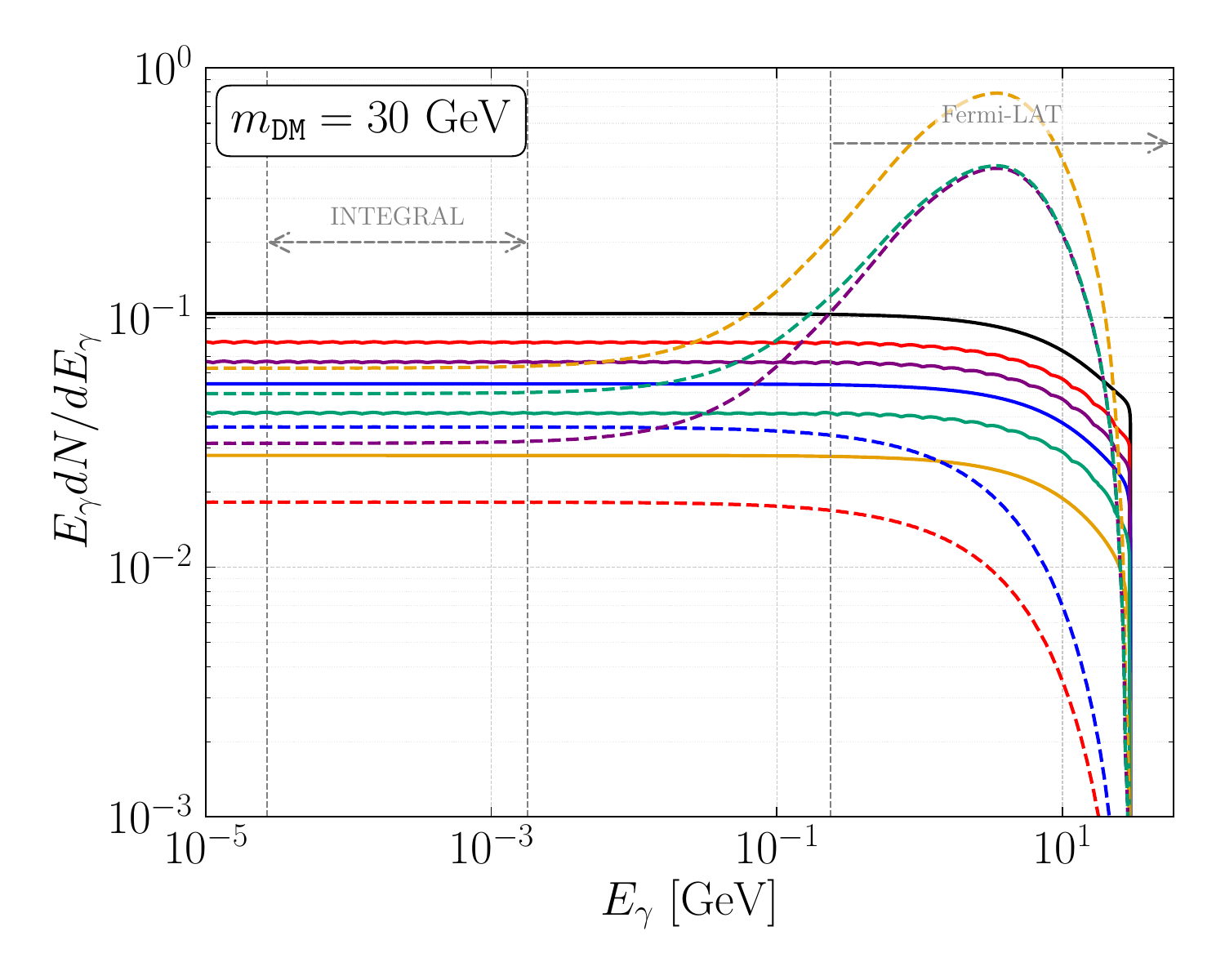}
\includegraphics[width=0.329\textwidth]{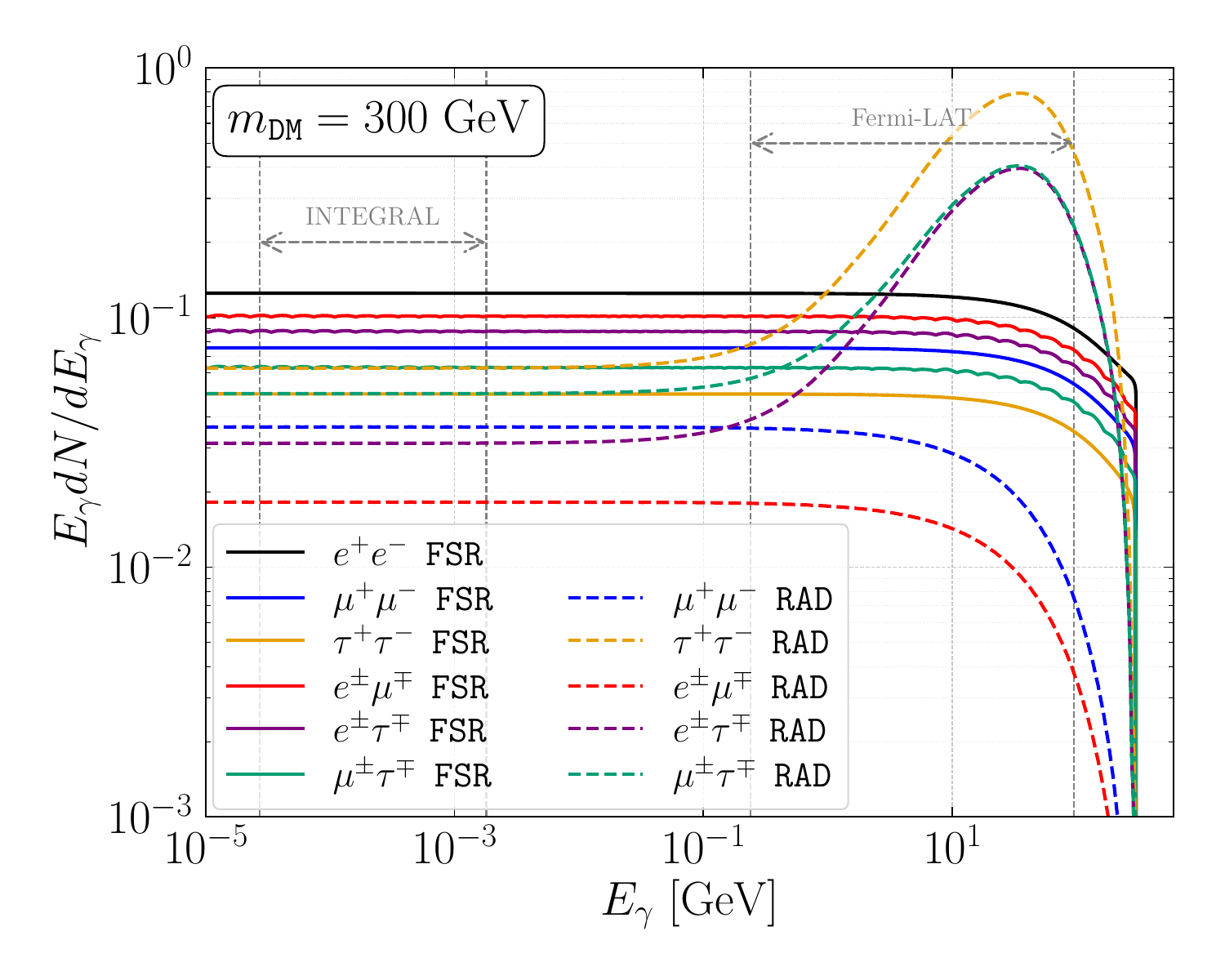}
\caption{Comparison of photon spectra for different annihilation channels, including both LFV and LFC cases. 
The {\tt FSR} spectra are shown as solid curves, while the {\tt Rad} spectra are represented by dashed curves. The left, middle, and right panels correspond to DM masses of 3~GeV, 30~GeV, and 300~GeV, respectively. }
\label{fig:all_mode_comparison}
\end{figure*}

Before closing this subsection, we make two additional remarks. 
First, for operators that share the same Lorentz structure but involve different flavor combinations, the normalized {\tt FSR} spectrum for each LFV annihilation mode with $\ell^-_i\ell^+_{j}\gamma$ final states always lies between those of the two LFC modes involving $\ell^-_i\ell^+_i\gamma$ and $\ell^-_j\ell^+_j\gamma$ ($j\neq i$), 
where the mode containing the lighter lepton pair dominates. 
The solid curves in \cref{fig:all_mode_comparison} 
show the {\tt FSR} spectra for the six channels of lepton flavor, which are calculated based on the operator $\bar{\ell_i} \gamma_\mu \gamma_5 \ell_j \bar{\chi} \gamma^\mu\gamma_5 \chi$ with three benchmark DM masses: 3 (left panel), 30 (middle panel), and 300~GeV (right panel). As clearly illustrated by the plots, both LFV and LFC annihilation modes produce {\tt FSR} spectra with similar shapes.
Second, we do not show the corresponding {\tt FSR} spectra for the p- and d-wave operators in Fig.\,3, as their calculation depends nontrivially on the DM velocity distribution.
To assess this effect, we generate the spectra using MadGraph and find that, after normalization, they exhibit only weak dependence on the initial DM velocity. 
In the photon energy region of interest, these spectra closely resemble those of the s-wave operators and are well described by the analytic formula when $m_{\rm DM} \gg m_\ell$.
However, the MadGraph results suffer from large fluctuations in the low-energy (keV) region.
Therefore, in the following calculations, we adopt the {\tt FSR} spectra of the s-wave operators as a proxy for those of the p- and d-wave operators.

\subsubsection{Radiative decays}

The {\tt Rad} photons arise from the radiative decays of the primary charged leptons, $\mu^\pm$ and $\tau^\pm$, produced in DM annihilation, and are independent of EFT operators. 
For the muon, the dominant radiative decay process is  $\mu^- \to e^{-} \bar\nu_{e} \nu_\mu \gamma$,
whereas for the $\tau$ lepton, both the 
leptonic decay modes $\tau^- \to e^{-}(\mu^-) \bar\nu_{e(\mu)} \nu_\tau \gamma$ and hadronic decay channels contribute significantly. 
For the four-body leptonic decay process $\ell^- \to \ell'^{-} \bar\nu_{\ell'} \nu_\ell \gamma$ (and its charge-conjugation counterpart),
the {\tt Rad} photon spectrum can be calculated analytically and is given by~\cite{Kuno:1999jp},
\begin{align}
{d N_\gamma^{{\tt Rad},\ell} \over d E_\gamma^* } 
& = 
{\alpha (1-x_\ell)\over 36\pi E_\gamma^*}
\Big\{
x_\ell(1-x_\ell)(46-55x_\ell)-102
\nonumber
\\
& + 12\big[3-2x_\ell (1-x_\ell)^2\big]\ln{{1-x_\ell\over r_{\ell\ell'}}}
\Big\},
\label{eq:dNdEga}
\end{align}
where $E_\gamma^*$ is the photon energy in the lepton rest frame, $x_\ell = 2E_\gamma^*/ m_\ell$, and $r_{\ell\ell'}= (m_{\ell'}/m_\ell)^2$ with $\ell\ell'=(\mu e,\tau e,\tau \mu)$ for the three LFV annihilation modes.

For the hadronic radiative decays of the $\tau$ lepton, 
no analytical formulas exist for the resulting photon spectrum. Instead, we obtain it through a detailed \textsc{pythia} simulation~\cite{Bierlich:2022pfr}.\footnote{
Since the hadronic matrix elements in  hadronic $\tau$ decays involve nonperturbative QCD effects, their calculations suffer from uncertainties in the choice of hadronic current parametrization, resonance modeling, and form factors \cite{Kuhn:1990ad,Finkemeier:1996dh,CLEO:1999rzk}.  
For the dominant mode $\tau^- \to \nu_\tau \pi^- \pi^0$, the uncertainty in the branching ratio is about 4\% \cite{Kuhn:1990ad}, 
while for other modes such as $\tau^- \to \nu_\tau \pi^0 \pi^0 \pi^-$, the $\pi^0\pi^-$ invariant mass distributions from \textsc{pythia} and Herwig++ can differ by up to 15\% \cite{Ilten:2013yed}.
These effects introduce uncertainties into the photon spectrum from $\tau$ radiative decays at most at a similar level.
} 
We generate $10^5$ $\tau^-$ leptons at rest using \textsc{pythia}, collect the final-state photons from their decay products, and bin them according to the photon energy to obtain the normalized spectrum. 
We find that the leptonic radiative decay modes $\tau^- \to e^- (\mu^-) \bar{\nu}_{e(\mu)} \nu_\tau \gamma$ are not included in the \textsc{pythia} simulation. Therefore, we combine the simulated photon spectrum with the analytic expressions in \cref{eq:dNdEga} for these leptonic modes to derive the total radiative decay spectrum for $\tau$ decay at rest, $dN_\gamma^{{\tt Rad},\tau}/dE_\gamma^*$.

For the produced charged lepton $\ell$ moving in motion, 
the photon spectrum discussed above in its rest frame needs to be boosted to the DM rest frame, which is given by 
\begin{align}
\frac{dN_\gamma^{{\tt Rad},\ell}}{d y}
= \int_y^1 \frac{d \omega}{\omega}\,\frac{dN_\gamma^{{\tt Rad},\ell}}{d \omega} \, ,
\label{eq:photon-boost}
\end{align}
where $y \equiv E_\gamma / \sqrt{E_\ell^2 - m_\ell^2}$ and $\omega \equiv E_\gamma^*/(2 m_\ell)$.
In the DM annihilation channel ${\tt DM+DM}\to \ell_i^-\ell_j^+$, $\ell=\ell_i^-,\ell_j^+$, and the energies of the charged leptons are
\begin{align}
E_{i,j} = m_{\tt DM} \pm \frac{m_i^2 - m_j^2}{4 m_{\tt DM}}.
\label{eq:Eij}
\end{align}

The dashed lines in \cref{fig:all_mode_comparison} show the {\tt Rad} spectra for the three benchmark DM masses used for the {\tt FSR} calculation. 
For a given LFV channel $\ell_i^- \ell_j^+$, the normalized spectrum from {\tt Rad} also falls between those of the corresponding LFC channels $\ell_i^-\ell_i^+$ and $\ell_j^- \ell_j^+$, with the {\tt Rad} contribution vanishing for the $e^+e^-$ pair. Unlike in the case of {\tt FSR}, the dominant contribution now comes from the heavier lepton pair.
Since the {\tt FSR} mainly originates from the collinear and soft photon emission of energetic final-state lighter leptons, its contribution is enhanced in the lower photon energy tail.
Consequently, increasing the DM mass, which makes the final-state leptons more energetic, enhances collinear and soft radiation, resulting in a more significant {\tt FSR} contribution.
In contrast, photons from $\tau$ leptons are predominantly produced through hadronic $\tau$ decays and carry substantial energies.
As a result, for LFV channels involving $\tau$ leptons ($e \tau$ and $\mu \tau$), the {\tt Rad} contribution dominates in the high-energy region. 
In the low-energy region, the {\tt FSR} contribution dominates for all three benchmark DM masses in the $e \tau$ channel, while for the $\mu \tau$ channel it dominates only when $m_{\tt DM} \gtrsim 300~\rm GeV$.
Finally, it should be noted that the {\tt FSR} always dominates over the {\tt Rad} in the photon spectrum for the $e\mu$ case. 

\subsubsection{The inverse Compton scattering}

In addition to the {\tt FSR} and {\tt Rad} photons, the {\tt ICS} photons---produced primarily by the scattering of high-energy $e^\pm$'s from DM annihilation off the galactic background low-energy photons---dominate the photon flux at lower photon energies, as will be shown below. 
The background photons mainly consist of the cosmic microwave background (CMB), optical starlight emitted by stars in the galactic disk, and  infrared light resulting from the absorption and reemission of starlight by galactic dust. 
The {\tt ICS} flux can be written as an integral of the photon emissivity $j(E_\gamma, \pmb{r})$ at position $\bm{r}$ in the galactic coordinate, along the LOS direction,
\begin{align}
{d^2\Phi_\gamma^{\tt ICS} \over d\Omega dE_\gamma} = {1\over 4\pi}\int_{\rm LOS} ds^\prime\,     
\frac{j(E_\gamma, \pmb{r})}{E_\gamma}.
\end{align}
The emissivity characterizes the photon emission capability of a cell at $\pmb{r}$ and is obtained by convolving the $e^\pm$ spectral number density ($dn_{e^\pm}/dE_e$) with the corresponding differential radiation power into photons
(${\cal P}_{\tt ICS}(E_\gamma, E_e, \bm{r})$),
\begin{align}
j(E_\gamma, \pmb{r}) = \int_{m_e}^{E_e^{\tt max}}
\hspace{-1.5mm} d E_e   
~{\cal P}_{\tt ICS}(E_\gamma, E_e, \bm{r}) 
\frac{dn_{e^\pm}}{dE_e}(E_e, \pmb{r}),
\end{align}
where the maximum energy of the electron or positron is determined by the kinematics of DM annihilation, with $E_e^{\tt max} = {\rm max}[ E_i, E_j]$, where $E_{i,j}$ are given in \cref{eq:Eij}.

The $e^\pm$ spectral distribution $dn_{e^\pm}/dE_e$ includes contributions from both electrons and positrons produced through the annihilation channel ${\tt DM+DM}\to \ell_i^- \ell_j^+$ and the subsequent decays of charged muons and taus. 
It relates to the $e^\pm$ phase-space distribution 
function $f_{e}(p,\pmb{r})$ via the relation $dn_{e^\pm}/dE_e=4\pi E_e p f_{e}(p,\pmb{r})$, where $p$ and $E_e$ denote the magnitude of the momentum and the energy of $e^\pm$, respectively. 
The function $f_{e}(p,\pmb{r})$ is governed by a transport equation, whose main components include a diffusion term, an energy loss term, and an $e^\pm$-source term~\cite{Regis:2008ij}. 
In addition, two other terms---advection and convection---describe the flow of $e^\pm$'s near the Galactic Center;
however, their effect can be safely neglected in the region of interest for DM annihilation and observation~\cite{Cirelli:2009vg}. 

Furthermore, following Ref.\,\cite{Cirelli:2020bpc}, we adopt the ``on the spot'' approximation in our analysis. 
This implies neglecting the diffusion term and assuming that the scattering occurs exactly at the location where these $e^\pm$'s are produced from DM annihilation. 
It is anticipated that diffusion would partially redistribute the $e^\pm$ population from regions of high DM density to areas of lower density, effectively shifting the photon flux generated by ICS from the Galactic Center region toward larger radii. 
References.\,\cite{Meade:2009iu,Cirelli:2009vg} demonstrate that the flux predicted by fully numerical \textsc{galprop} calculations \cite{Cholis:2008wq,Borriello:2009fa} agrees with the on the spot approximation within a factor of 2, 
making this approximation reasonably accurate.
By neglecting the diffusion term, the $e^\pm$ spectral distribution can then be solved semianalytically: 
\begin{align}
\frac{dn_{e^\pm}}{dE_e}(E_e, \pmb{r}) = 
\frac{Y_{e^\pm}(E_e)}{b_{\tt tot}(E_e,\bm{r})} \frac{ \langle \sigma v_{\rm rel} \rangle_{ij+ji} (\pmb{r}) }{{\tt S}
(1+\delta_{ij}) m_{\tt DM}^2}\rho^2(\pmb{r}),
\label{eq:dPhidEprompt}
\end{align}
where $b_{\rm tot}(E_e, \pmb{r})$ represents the total energy loss function of the $e^\pm$s~\cite{Blumenthal:1970gc}, including contributions from synchrotron radiation, bremsstrahlung, ionization, and {\tt ICS}.
We adopt the numerical interpolation function provided by \textsc{PPPC4} \cite{Cirelli:2010xx} in our calculation.
$Y_{e^\pm}(E_e)$ is defined by
\begin{align}
Y_{e^\pm}(E_e) \equiv \int_{E_e}^{E_e^{\tt max}} d \tilde{E}_e \frac{dN_{e^\pm}}{d\tilde{E}_e}(\tilde{E}_e).
\end{align}
Here, $dN_{e^\pm} / d\tilde{E}_e$ is the energy spectrum of $e^\pm$s produced per DM annihilation event in the DM rest frame, including contributions from the primary $e^\pm$ (when $\ell_i^-=e^-$ or $\ell_j^+=e^+$) and the secondary ones due to decays of $\mu$ and $\tau$.
The spectrum of primary $e^\pm$ is simply represented by a Dirac delta function.
For muon decay at rest, the $e^\pm$ spectrum is determined by the process $\mu^- \to e^- \nu_\mu \bar{\nu}_e$ and its conjugate.
For $\tau$ decay at rest, we obtain the $e^\pm$ spectra  using \textsc{pythia8} simulations.
Similarly to the $\tt Rad$ component of the photon flux, we need to boost the $e^\pm$ spectrum in the lepton rest frame to the DM rest frame via
\begin{equation}
    \frac{dN_{e^\pm}}{d\tilde E_e} = \frac{1}{2 \beta_\ell \gamma_\ell} 
\int_{E^*_{\rm min}}^{E^*_{\rm max}}  \frac{dN_{e^\pm}}{dE^*_e} \, \frac{dE^*_e}{p^*_e},
\label{eq:boost}
\end{equation}
where $\gamma_\ell =E_\ell /m_\ell$ is the Lorentz boost factor of the parent lepton $\ell$, $\beta_\ell=(1-1/\gamma_\ell^2)^{1/2}$, and  $p_e^*=\sqrt{(E_e^*)^2-m_e^2}$. The integration limits are given by $E^*_{\rm max(min)} = \gamma_\ell (\tilde{E}_e \pm \beta_\ell \tilde{p}_e)$ with $\tilde{p}_e = \sqrt{\tilde{E}_e^2 - m_e^2}$.

\begin{figure*}[t]
\centering
\includegraphics[width=0.329\linewidth]{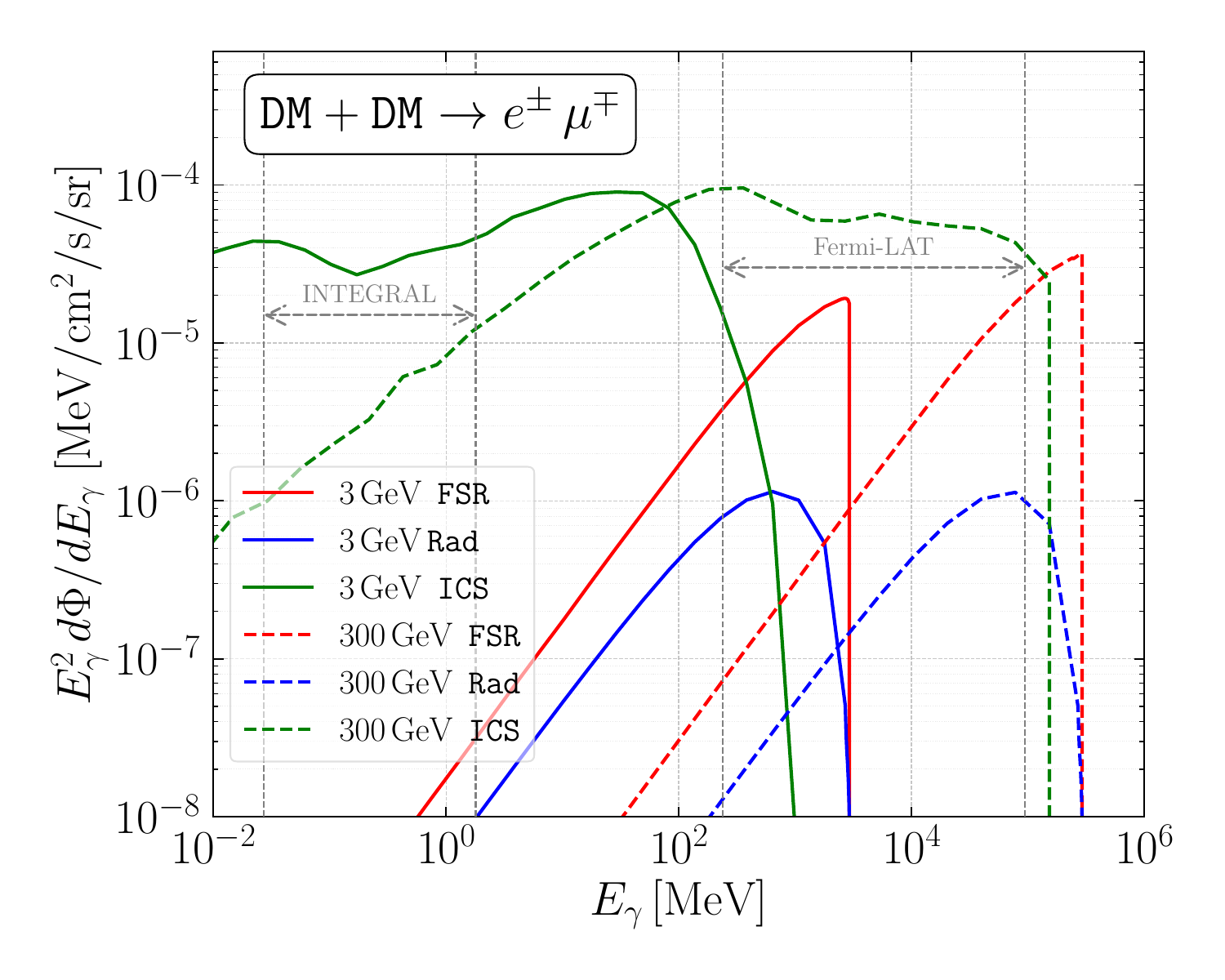}
\includegraphics[width=0.329\linewidth]{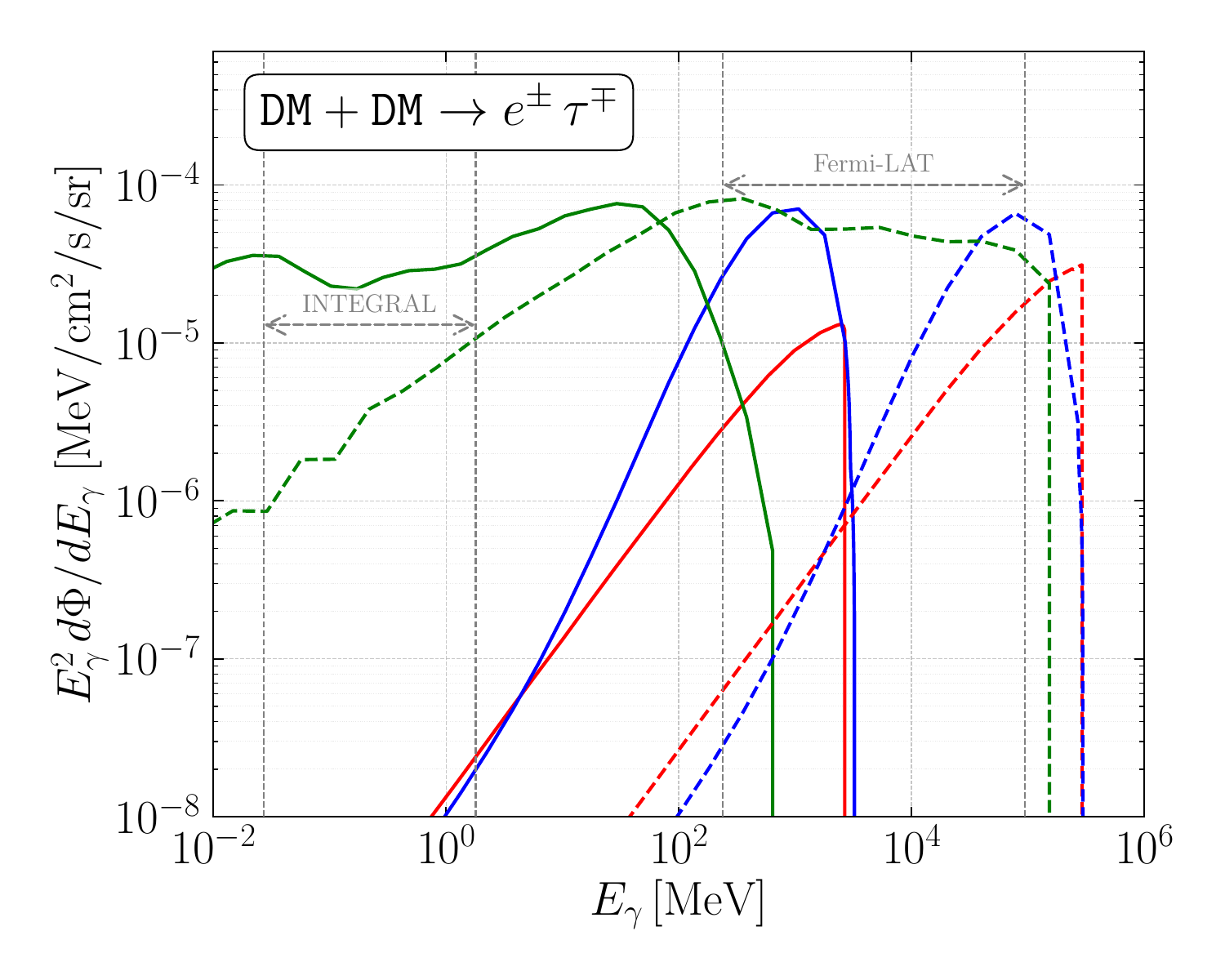}
\includegraphics[width=0.329\linewidth]{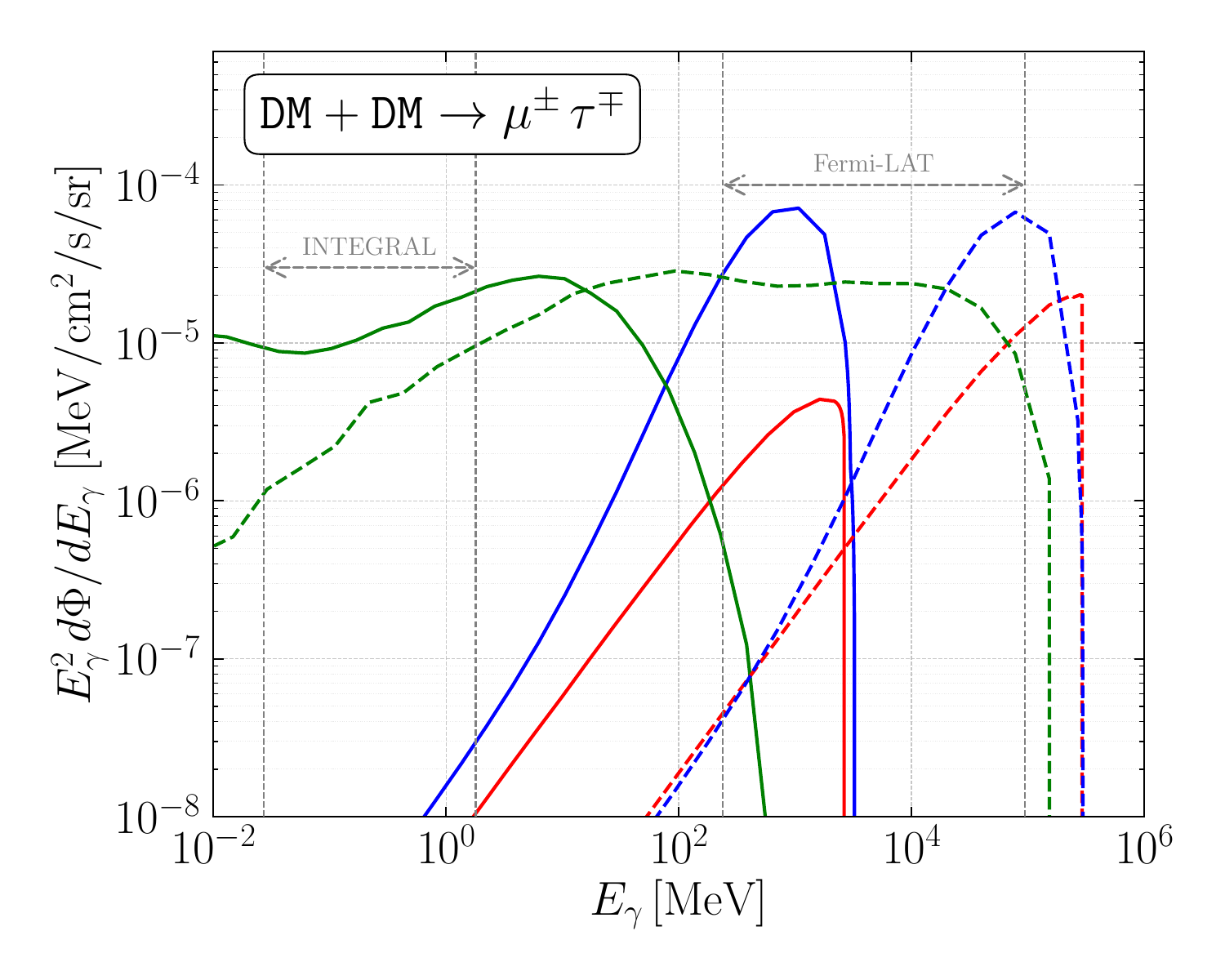}
\caption{ Photon fluxes from {\tt FSR} (red curves), {\tt Rad} (blue curves), and {\tt ICS} (green curves) are shown for the three LFV DM annihilation channels.
The solid and dashed curves correspond to $m_{\tt DM}=3~\rm{GeV}$ and $300~\rm{GeV}$, respectively. 
The gray vertical dashed lines denote the energy ranges covered by INTEGRAL and Fermi-LAT.
We fix $\langle \sigma v_{\rm rel} \rangle_{ij+ji}$ to $3\times10^{-26}$ and $3\times10^{-24}~{\rm cm}^3/{\rm s}$ for $m_{\tt DM}=3$ and 300~GeV cases, respectively, and adopt the sky region for the Fermi-LAT dataset ($8^\circ < |b| < 90^\circ$, $0 < l < 360^\circ$).
}
\label{fig:flux}
\end{figure*}

Next, we turn to the function ${\cal P}_{\tt ICS}$, which physically represents the scattering rate of high-energy $e^\pm$ interacting with background photons. It can be expressed as 
\begin{align}
{\cal P}_{\tt ICS} = 
\int\hspace{-1.5mm} d \epsilon_\gamma  
(E_\gamma-\epsilon_\gamma)\frac{dn_\gamma}{d\epsilon_\gamma}(\epsilon_\gamma, \bm{r}) \frac{d\sigma_{\rm eff}}{d E_\gamma}(\epsilon_\gamma,E_\gamma, E_e).    
\end{align}
Here, $\epsilon_\gamma$ denotes
the energy of background photons with the number density spectrum $dn_\gamma/d\epsilon_\gamma$ which comprises the three components described above. 
The isotropic CMB component follows a thermal Bose-Einstein distribution at a temperature $T=2.73~\rm K$. 
For the infrared and starlight components, we employ the data provided in \textsc{galprop} \cite{Vladimirov:2010aq}.
$d\sigma_{\rm eff}/d E_\gamma$ denotes an effective ICS cross section that incorporates the effects of Lorentz boosts and frame transformations. 
Its expression can be derived from Eq.\,(2.48) in Ref.\,\cite{Blumenthal:1970gc}, with appropriate adjustments to account for differences in notation:   
\begin{align}
\frac{d\sigma_{\rm eff}}{d E_\gamma} 
 &= \frac{3\sigma_\mathrm{T} }{4 \gamma^2 \epsilon_\gamma}
\Big[2 q \ln q+q+1-2 q^2
\nonumber\\
&+\frac{1}{2} \frac{\kappa_\gamma^2}{1-\kappa_\gamma}(1-q)\Big],
\end{align}
where $\kappa_\gamma \equiv E_\gamma/(\gamma m_e)$, and $q\equiv \kappa_\gamma/[\Gamma_\epsilon(1-\kappa_\gamma)]$ with $\Gamma_\epsilon\equiv 4\epsilon_\gamma \gamma/m_e$.
$\gamma=E_e/m_e$ is the Lorentz boost factor of the electron or positron in the reference frame of the photon gas. 
$\sigma_\mathrm{T} = 0.6652~\rm b$ is the total Thomson cross section.
Trading the integral variable $\epsilon_\gamma$ by the dimensionless $q$ variable,  
the final differential power 
$\mathcal{P}_{\tt ICS}$ takes the following form:
\begin{align}
\mathcal{P}_{\tt ICS} 
= 
\sigma_{\mathrm{T}} E_\gamma \int_{\frac{1}{4 \gamma^2} }^1 d q \,\frac{d n_\gamma}{d\epsilon_\gamma}(\epsilon_\gamma(q), r)\,f_{\tt ICS} (q),
\label{eq:power}
\end{align}
where 
\begin{align}
f_{\tt ICS} &\equiv \frac{3 }{4 \gamma^2} \left[1-\frac{1}{4 q \gamma^2\left(1-\kappa_\gamma\right)}\right]
\times\frac{1}{q}
\nonumber\\
&\times \Big[2 q \ln q+q+1-2 q^2+\frac{1}{2} \frac{\kappa_\gamma^2}{1-\kappa_\gamma}(1-q)\Big].
\end{align}

\subsubsection{Comparison of photon flux}

Having detailed the three main photon sources, it is enlightening to compare their relative contributions to the produced photon flux in various scenarios.
In \cref{fig:flux}, we show the predicted photon fluxes for the three LFV channels from the three photon components: {\tt FSR} (red curves), {\tt Rad} (blue curves), and {\tt ICS} (green curves). 
The fluxes are calculated based on a Dirac DM scenario for 
two benchmark DM masses: $m_{\tt DM} = 3$ (solid curves) and 300~GeV (dashed curves). The values of $\langle \sigma v_{\rm rel} \rangle=\langle\sigma v_{\rm rel}\rangle_{ij+ji}$
are fixed at $3\times10^{-26}$ and $3\times10^{-24}~{\rm cm}^3/{\rm s}$ for the two mass points, respectively.
In the calculation, we choose a sky region with $8^\circ < |b| < 90^\circ$ and $0 < l < 360^\circ$, which corresponds to the observation window of Fermi-LAT.

\begin{table*}[t]
\centering
\begin{tabular}{lccc}
\hline
Experiment & Data energy range & Sky coverage & Data source \\
\hline
INTEGRAL
& 27--1800 keV
& $|l|<23.1^\circ$ ($E_\gamma < 600$ keV); $|l|<60^\circ$ (600 keV  $< E_\gamma < 1800$ keV)
& Ref.~\cite{Bouchet:2011fn} \\

XMM-Newton
& 2.5--8 keV
& All sky excluding $|b|\leq 2^\circ$
& Refs.~\cite{Foster:2021ngm,XMM_BSO_DATA} \\

Fermi-LAT
& 200 MeV--10 GeV
& $0<l<360^\circ$, $8^\circ<|b|<90^\circ$
& Ref.~\cite{Fermi-LAT:2012edv} \\

AMS-02
& 20 GeV--1 TeV
& ...
& Ref.~\cite{AMS:2019rhg} \\
\hline
\end{tabular}
\caption{Summary of the datasets used in this work, including the energy range, sky coverage, and data sources for each experiment.
}
\label{tab:datasets}
\end{table*}

As shown in the plots, the photon spectra exhibit distinct behavior for the different radiation components, reflecting their underlying production mechanisms. In the low-energy range (keV--100~MeV), {\tt ICS} dominates for all three LFV channels. 
This is because electrons and positrons produced in GeV-scale DM annihilation already carry sufficient energy to upscatter low-energy background photons into the x-ray range, resulting in the {\tt ICS} flux that greatly exceeds those from {\tt Rad} and {\tt FSR}.
In the high-energy range (100 MeV-10 GeV), the dominant contribution depends on both annihilation channel and DM mass. 
For $\tau$-related channels ($e\tau$ and $\mu\tau$), {\tt Rad} dominates in both the light and heavy DM scenarios, as hadronic $\tau$ decays produce energetic photons. 
In the $e\mu$ channel, the dominant component varies with DM mass: {\tt FSR} prevails in the light DM region, while {\tt ICS} becomes dominant in the heavy DM region. 
These features highlight the complementary roles played by the three photon components across different energy ranges.
More importantly, they will help us to understand the final constraints presented in the next section. 

\subsection{Positron flux}

Taking into account both energy losses and spatial diffusion, the solution of the transport equation leads to the positron flux from DM annihilation given by \cite{Cirelli:2010xx}
\begin{align}
\frac{d \Phi_{e^+}}{dE_e } (E_e, \pmb{r}) = 
\int_{E_e}^{E_e^{\tt max}}
\hspace{-1.5mm} d \tilde{E}_e 
\frac{ Q_{e} (\tilde{E}_e, \pmb{r}) I(E_e, \tilde{E}_e, \pmb{r})}{4\pi\, b_{\tt tot} (E_e, \pmb{r})},
\label{eq:Phie}
\end{align}
where the positron source term $Q_{e}$ is related to the DM annihilation cross section through the relation
\begin{align}
Q_e(\tilde{E}_e, \pmb{r}) = \frac{dN_{e^+}}{d \tilde{E}_e}\frac{\langle \sigma v_{\rm rel} \rangle_{ij+ji} (\pmb{r}) }{{\tt S}(1+\delta_{ij}) m_{\texttt{DM}}^2}\rho^2(\pmb{r}).
\end{align}
Here, $dN_{e^+}/d\tilde{E}_e$ is  half of the spectrum in \cref{eq:boost}, since only positrons contribute to the signal from the $ij$ and $ji$ modes with the assumption  $\langle \sigma v_{\rm rel} \rangle_{ij} = \langle \sigma v_{\rm rel} \rangle_{ji}$. 
$I(E_e, \tilde{E}_e, \pmb{r})$ in \cref{eq:Phie}  is a generalized dimensionless halo function that  
describes the probability of observing a positron with energy $E_e$ at position $\pmb{r}$, after it has been  injected with energy $\tilde{E}_e$.
We evaluate the expression at $\pmb{r} = \pmb{r}_\odot$ with $\pmb{r}_\odot$ being the location of the Sun and use the corresponding interpolation halo function $I(E_e, \tilde{E}_e, \pmb{r}_\odot)$ offered in \cite{Cirelli:2010xx}.

\section{Constraints}
\label{sec:constraint}

\begin{figure*}[t]
\centering
\includegraphics[width=0.329\linewidth]{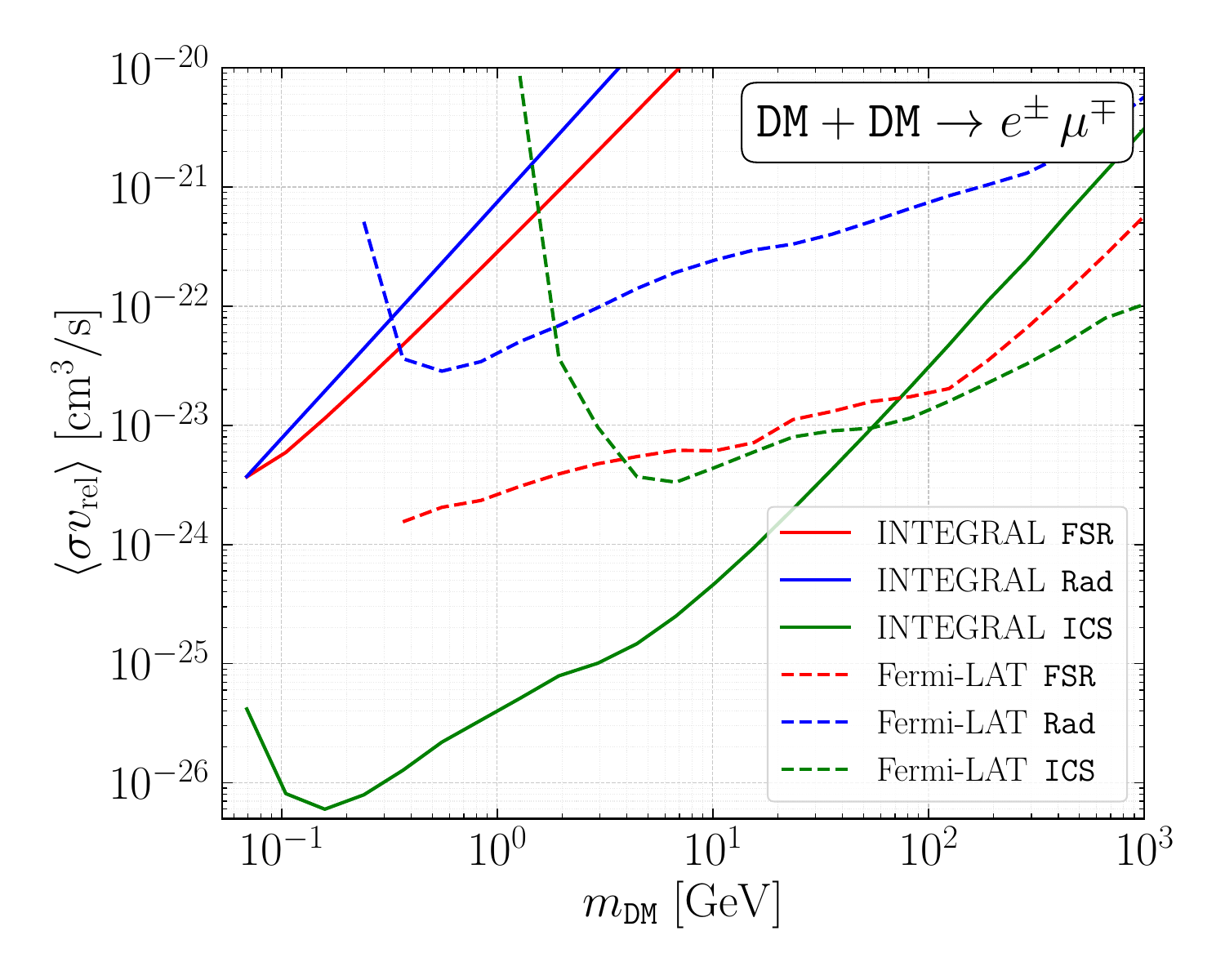}
\includegraphics[width=0.329\linewidth]{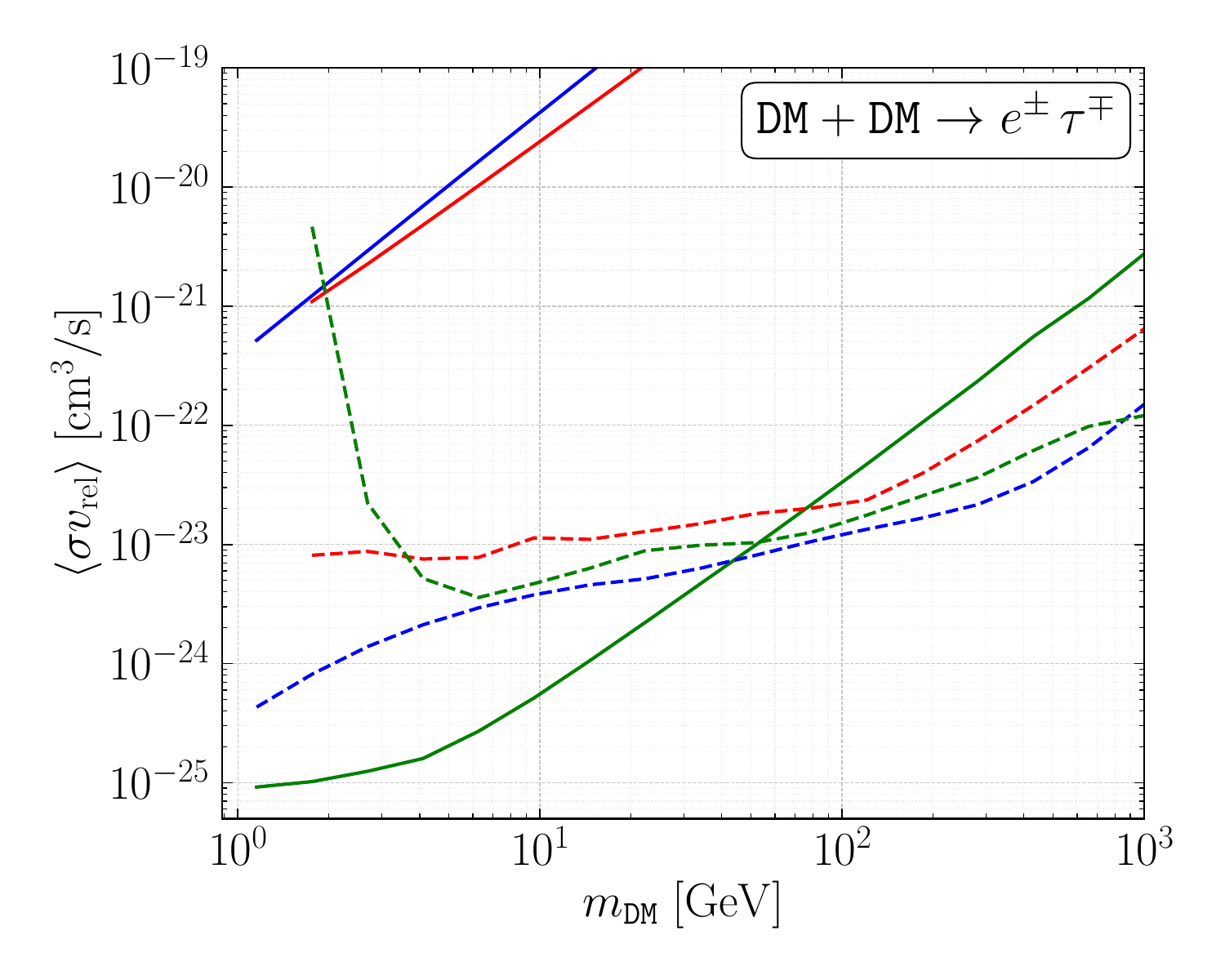}
\includegraphics[width=0.329\linewidth]{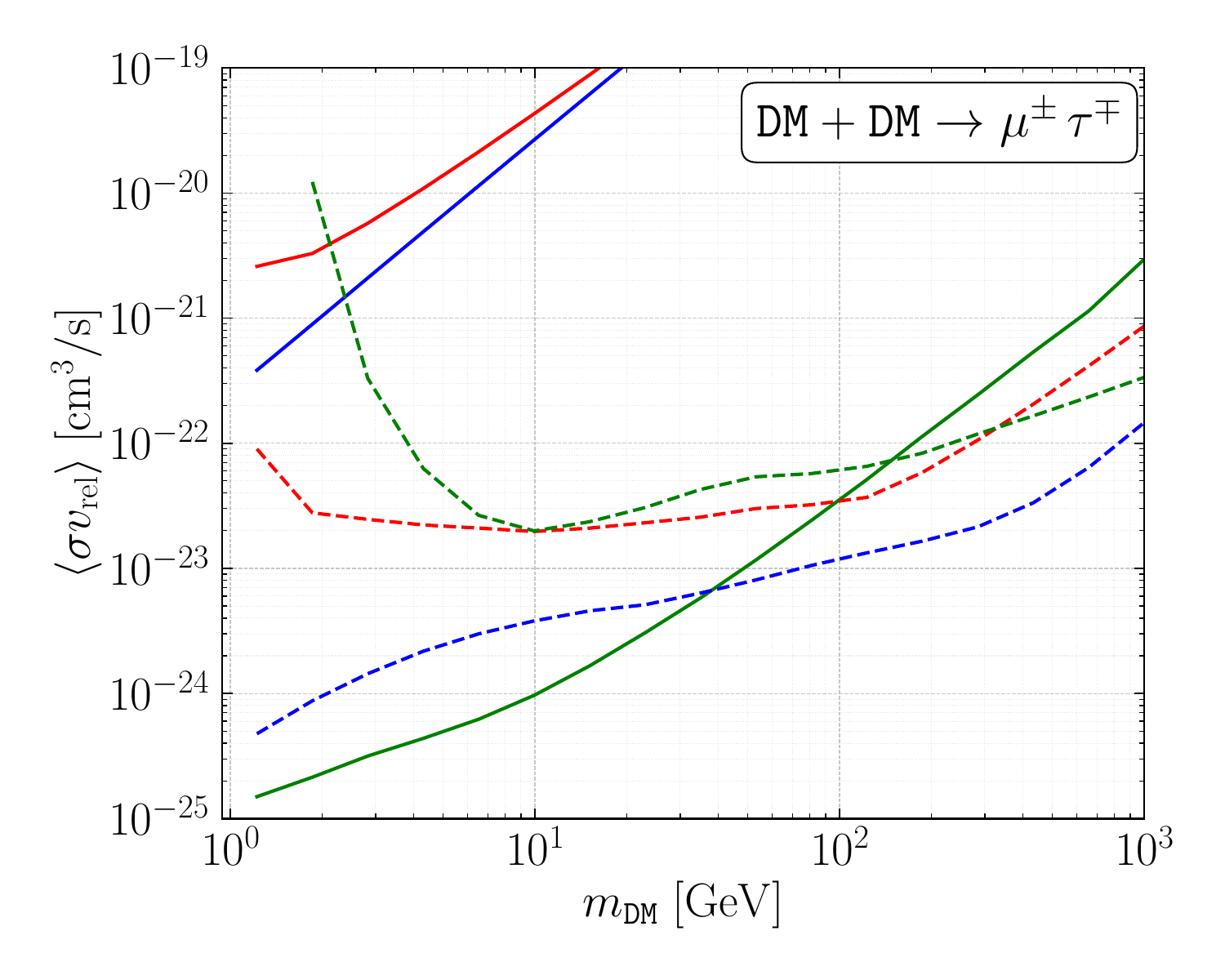}
\caption{
Comparison of the constraints on $\langle \sigma v_ {\rm rel} \rangle_{ij+ji} $ from the three photon components: 
{\tt FSR} (red curves), {\tt Rad} (blue curves), and {\tt ICS} (green curves).
The constraints are derived from the INTEGRAL (solid lines) and Fermi-LAT (dashed lines) datasets.
}
\label{fig:con-component}
\end{figure*}

We employ four complementary datasets to derive our constraints, including three x- or gamma-ray telescopes---INTEGRAL, XMM-Newton, and Fermi-LAT, together with a cosmic-ray positron dataset measured by AMS-02. 
Their sensitive energy ranges and sky coverages are summarized in \cref{tab:datasets}. 
INTEGRAL data are taken from Ref.\,\cite{Bouchet:2011fn}, covering photon energies from 27 to 1800 keV. 
The fluxes are provided in latitude bins and integrated over $|l|<23.1^\circ$ for energies below 600 keV and over $|l|<60^\circ$ for energies between 600 and 1800 keV.
For XMM-Newton, we use the publicly available all-sky dataset from the MOS and PN cameras \cite{Foster:2021ngm,XMM_BSO_DATA}, excluding the latitude region $|b| \leq 2^\circ$. 
To suppress dominant instrumental backgrounds, the energy range used in the analysis is restricted to 2.5--8~keV for the MOS camera and 2.5--7~keV for the PN camera \cite{Foster:2021ngm}.
We adopt the Fermi-LAT 2012 dataset \cite{Fermi-LAT:2012edv} in the energy range 200 MeV--10 GeV, covering the sky region $0<l<360^\circ$ and $8^\circ<|b|<90^\circ$.
Finally, we use the AMS-02 positron spectrum measured up to 1 TeV from~\cite{AMS:2019rhg}, restricting the analysis to energies above 20 GeV to suppress solar modulation effects.
Further details for these datasets can be found in~\cite{Liang:2025mfk}.
Please note that the XMM-Newton dataset is publicly accessible online at~\cite{XMM_BSO_DATA}, while the datasets we extracted from INTEGRAL, Fermi-LAT,  and AMS-02 are available in the ancillary files of the arXiv version of our previous work~\cite{Liang:2025mfk}. 

In our analysis,  we neglect astrophysical background components due to their large uncertainties and assume that all observed fluxes originate from DM annihilation in order to derive a conservative limit. It is expected that a dedicated analysis including background contributions will lead to more stringent bounds.
Following Refs.\,\cite{Cirelli:2020bpc,Cirelli:2023tnx}, we use a conservative chi-square test to derive our constraints:
\begin{align}
\chi^2 \equiv \sum_i \Big( \frac{{\tt max}\left[ S_i (w, m_{\tt DM}) - O_i, 0 \right]}{\sigma_i} \Big),
\label{eq:sta}
\end{align}
where $S_i$ and $O_i$ are the DM-induced and observed photon fluxes or event counts in the $i$th bin with uncertainty $\sigma_i$.
Here, $w=\hat a$, $\hat b$, and $\hat d$ for deriving constraints on the general thermally averaged cross section for s-, p-, and d-wave cases, respectively, while $w=\Lambda_{\rm eff}$ applies in the case of EFT operators.
For each DM mass point, we derive the $2\sigma$ bound on the parameter $w$ by requiring $\chi^2 = 4$.

\subsection{General constraints on $\langle\sigma v_{\rm rel}\rangle$} 

\begin{figure*}[t]
\centering
\includegraphics[width=0.329\textwidth]{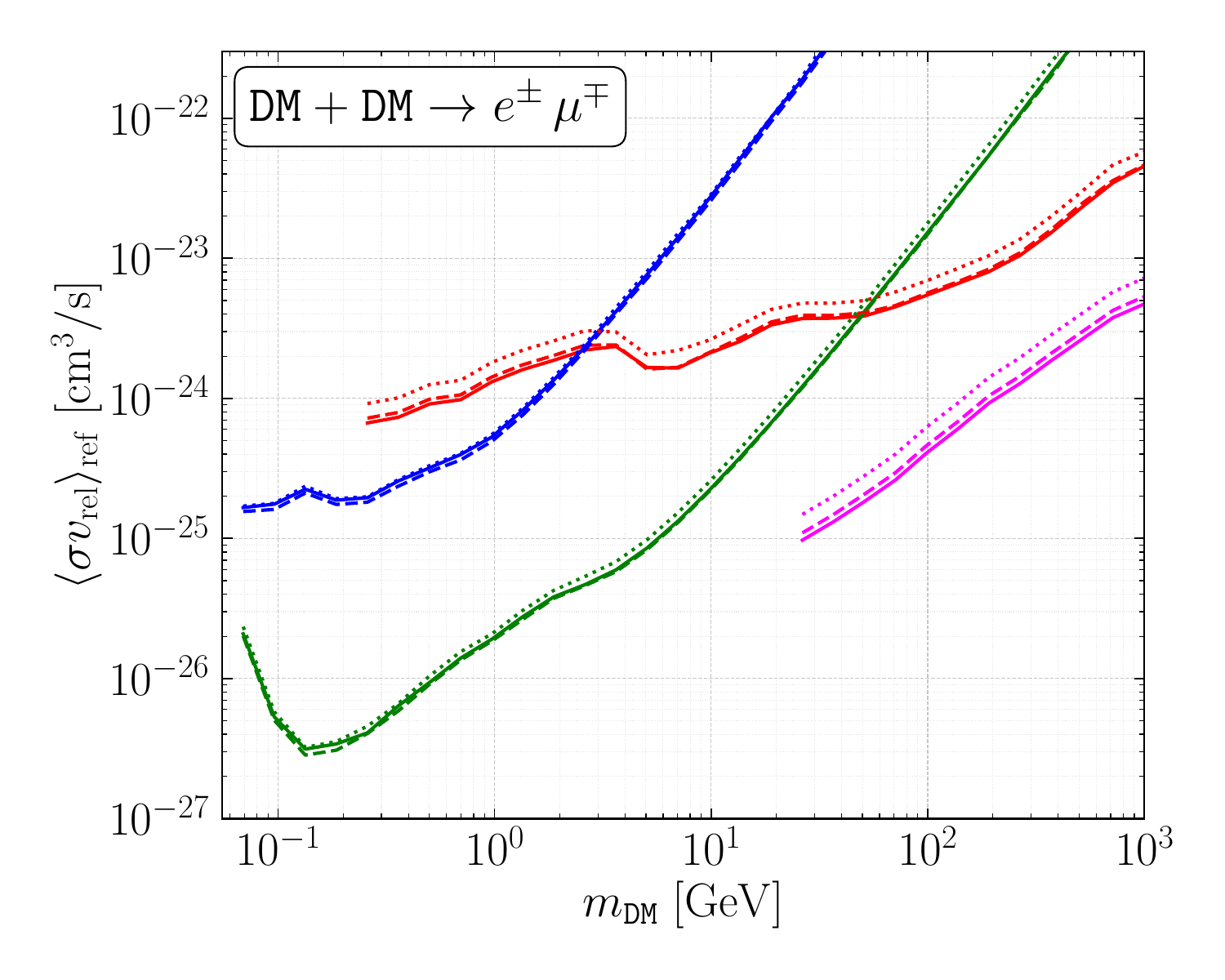} 
\includegraphics[width=0.329\textwidth]{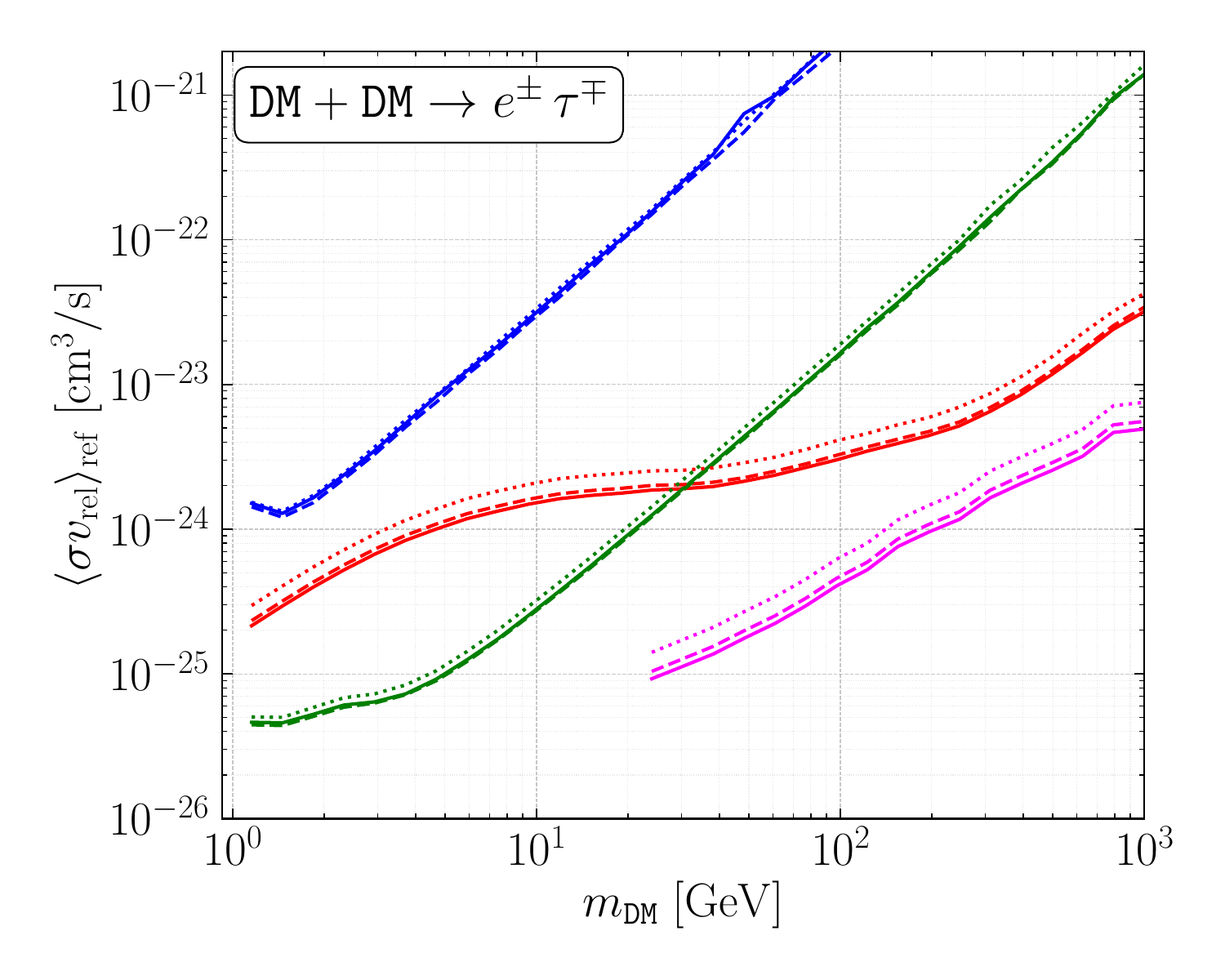}
\includegraphics[width=0.329\textwidth]{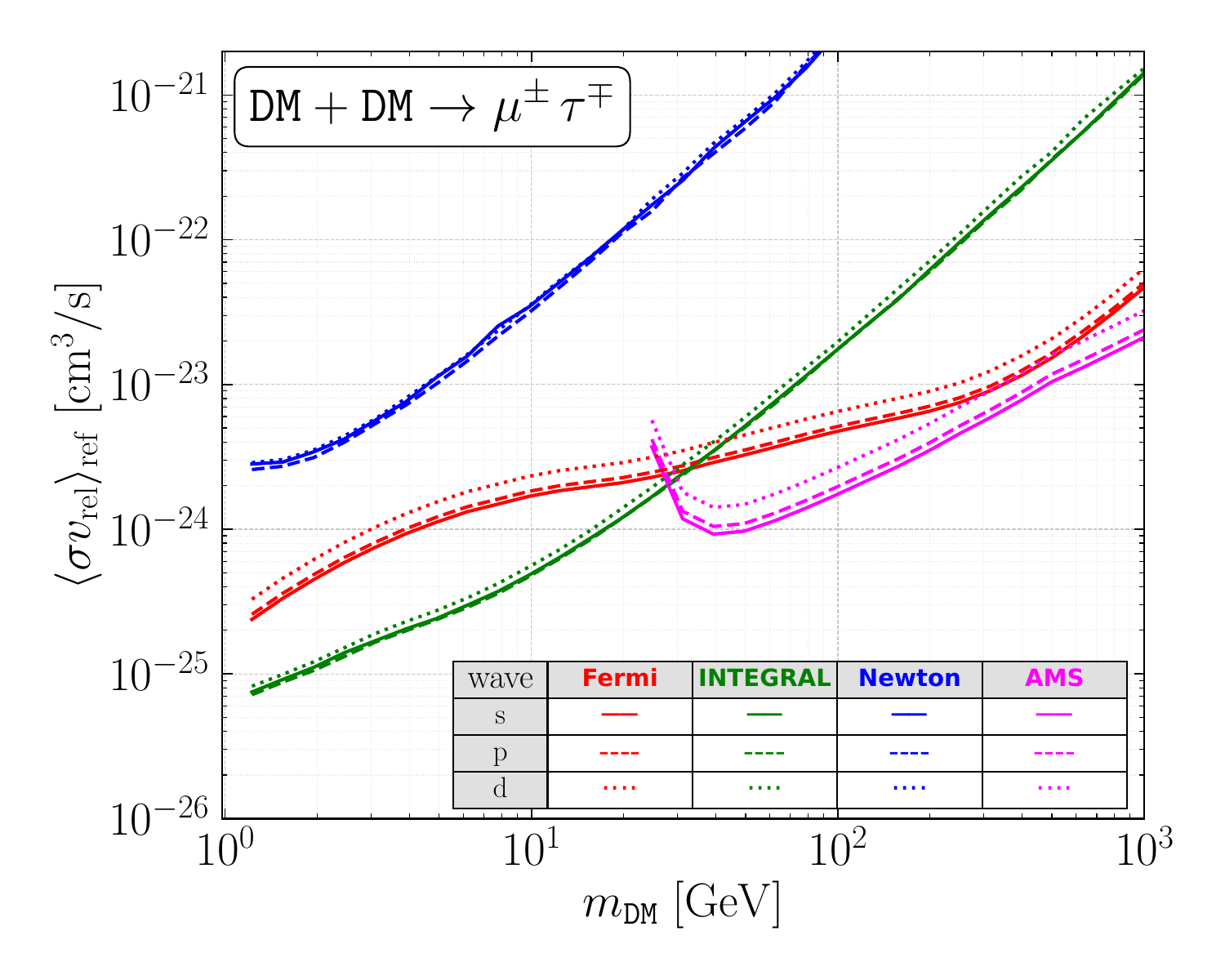}
\caption{
The $2\sigma$ constraints on the annihilation cross sections associated with the s-wave (solid curves), p-wave (dashed curves), and d-wave (dotted curves) contributions from Fermi-LAT (red curves), INTEGRAL (green curves), XMM-Newton (blue curves), and AMS-02(magenta curves).
To facilitate comparison among different partial waves, a reference annihilation cross section is defined as $\langle \sigma v_{\rm rel} \rangle_{\rm ref} \equiv \hat a_{2\sigma}$, $\hat b_{2\sigma}(v_{\rm ref}^{\rm p})^2$, and $\hat d_{2\sigma}(v_{\rm ref}^{\rm d})^4$ for the s, p, and d waves, respectively, with $v_{\rm ref}^{\rm p(d)}=1.4(1.7)\times10^{-3}$. 
}
\label{fig:sigma_spd}
\end{figure*}

As discussed in the previous section and illustrated in \cref{fig:flux}, the three components of the photon signal exhibit varying relative magnitudes depending on the DM mass, the annihilation channel, and the photon energy range.
The telescopes
INTEGRAL and Fermi-LAT are sensitive to low-energy and high-energy photons, respectively; making them complementary to each other.
Motivated by this complementarity, in \cref{fig:con-component}, we present a comparison of the constraints on the s wave $\langle \sigma v_{\rm rel}\rangle$ from the three photon components---{\tt FSR} (red curves), {\tt Rad} (blue curves), and {\tt ICS} (green curves)---through the INTEGRAL (solid curves) and Fermi-LAT (dashed curves) datasets.
The behavior of the constraints in \cref{fig:con-component} can be directly understood in relation to the photon flux comparison presented in \cref{fig:flux}. 
In particular, owing to the significant x-ray flux induced by {\tt ICS}, the INTEGRAL constraints derived from the {\tt ICS} component provide the strongest limits for all three LFV channels in the mass range $m_{\tt DM} \lesssim 50~{\rm GeV}$.
In the higher mass region $m_{\tt DM} \gtrsim 50~{\rm GeV}$, the constraints are instead dominated by Fermi-LAT. 
In this regime, {\tt ICS} remains the most important component for the photon flux in the $e\mu$ channel, while the dominant contribution shifts to the {\tt Rad} component for the $\tau$-related channels ($e\tau$ and $\mu\tau$).

In \cref{fig:sigma_spd}, we present the final constraints on $\langle \sigma v_{\rm rel}\rangle$ related to the s-, p-, and d-wave contributions, indicated by the solid, dashed, and dotted curves, respectively. 
The constraints derived from INTEGRAL, XMM-Newton, Fermi-LAT, and AMS-02 are represented by red, green, blue, and magenta curves, respectively. 
For the photon telescope experiments (INTEGRAL, XMM-Newton, and Fermi-LAT), the constraints are derived by combining all three photon components.
The results for the three partial-wave components are evaluated according to the nonrelativistic expansion $\sigma v_{\rm rel}  = \hat a + \hat b\, v_{\rm rel}^2 + \hat d\, v_{\rm rel}^4+\cdots$.  We derive the $2\sigma$ constraints on $\hat a$, $\hat b$, and $\hat d$, which are denoted as $\hat a_{2\sigma}$, $\hat b_{2\sigma}$, and $\hat d_{2\sigma}$ for the s-, p-, and d-wave contributions, respectively. 
We find that the ratios $\hat a_{2\sigma}/\hat b_{2\sigma}$ and $\hat a_{2\sigma}/\hat d_{2\sigma}$ across the entire mass range can be largely accounted for by two constant velocity-suppression factors: 
$\left(v_{\rm ref}^{\rm p}\right)^2$ and $\left(v_{\rm ref}^{\rm d}\right)^4$.
This suggests, with good approximation, that 
\begin{equation}
\hat a_{2\sigma} \simeq \hat b_{2\sigma} \left(v_{\rm ref}^{\rm p}\right)^2
\simeq \hat d_{2\sigma}\left(v_{\rm ref}^{\rm d}\right)^4. 
\label{eq:scale}
\end{equation}
For INTEGRAL and XMM-Newton, we find that $v_{\rm ref}^{\rm p} = 1.4 \cdot 10^{-3}$ and $v_{\rm ref}^{\rm d} = 1.7 \cdot 10^{-3}$ for the p- and d-wave cases, respectively. 
We present the p- and d-wave results in \cref{fig:sigma_spd} by rescaling $\hat b_{2\sigma}$ and $\hat d_{2\sigma}$ by $\left(v_{\rm ref}^{\rm p}\right)^2$ and $\left(v_{\rm ref}^{\rm d}\right)^4$, respectively.
From the plots, it can be observed that the required reference velocities for AMS-02 and Fermi-LAT will be slightly smaller.
The constraints are enhanced by a factor of 2 in the real scalar or vector and Majorana fermion DM cases. 
Notice that the constraints on the s-wave annihilation cross section coincide with the results in our previous work~\cite{Liang:2025mfk}.

\begin{figure*}[t]
\centering
\includegraphics[width=0.329\textwidth]{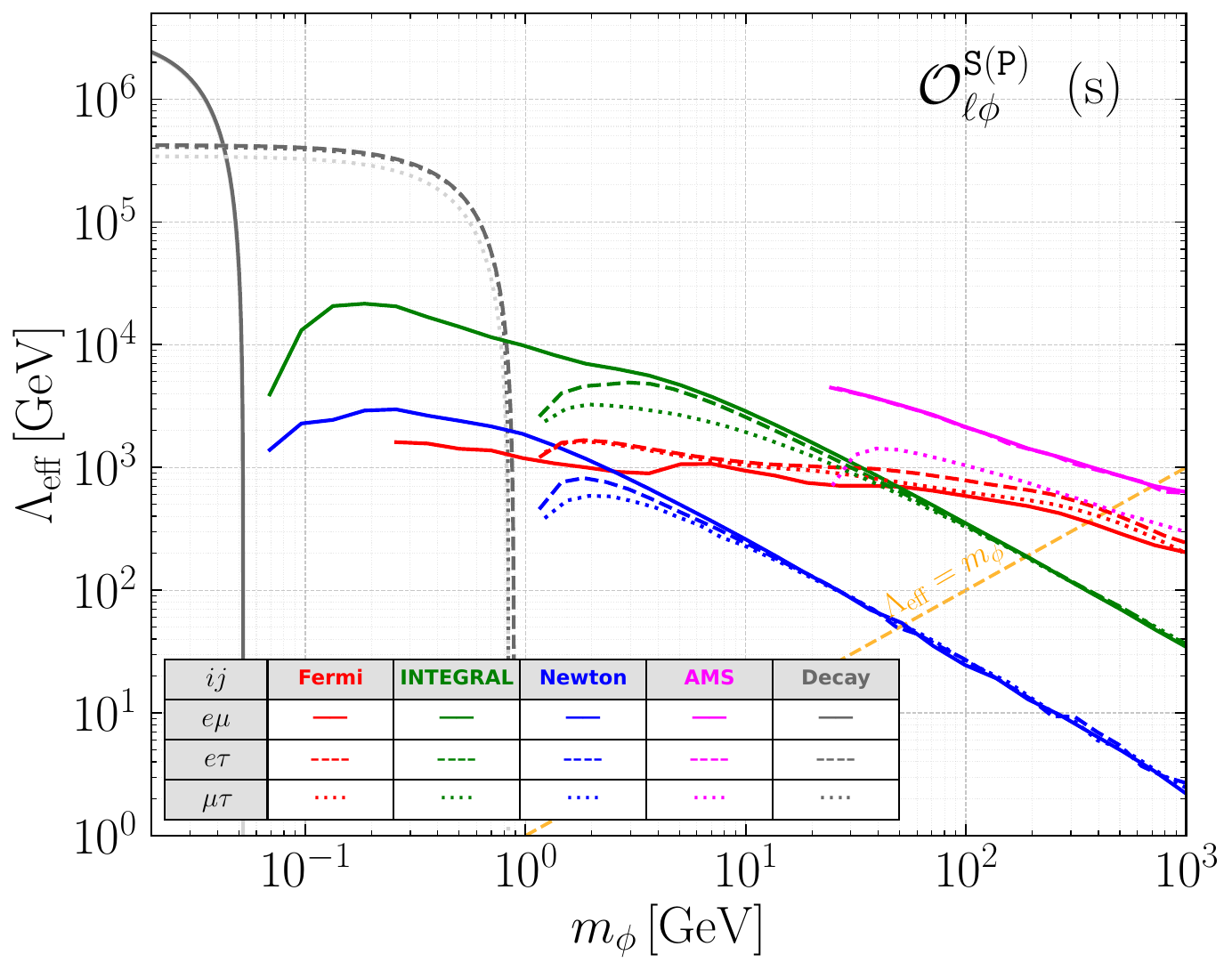} 
\includegraphics[width=0.329\textwidth]{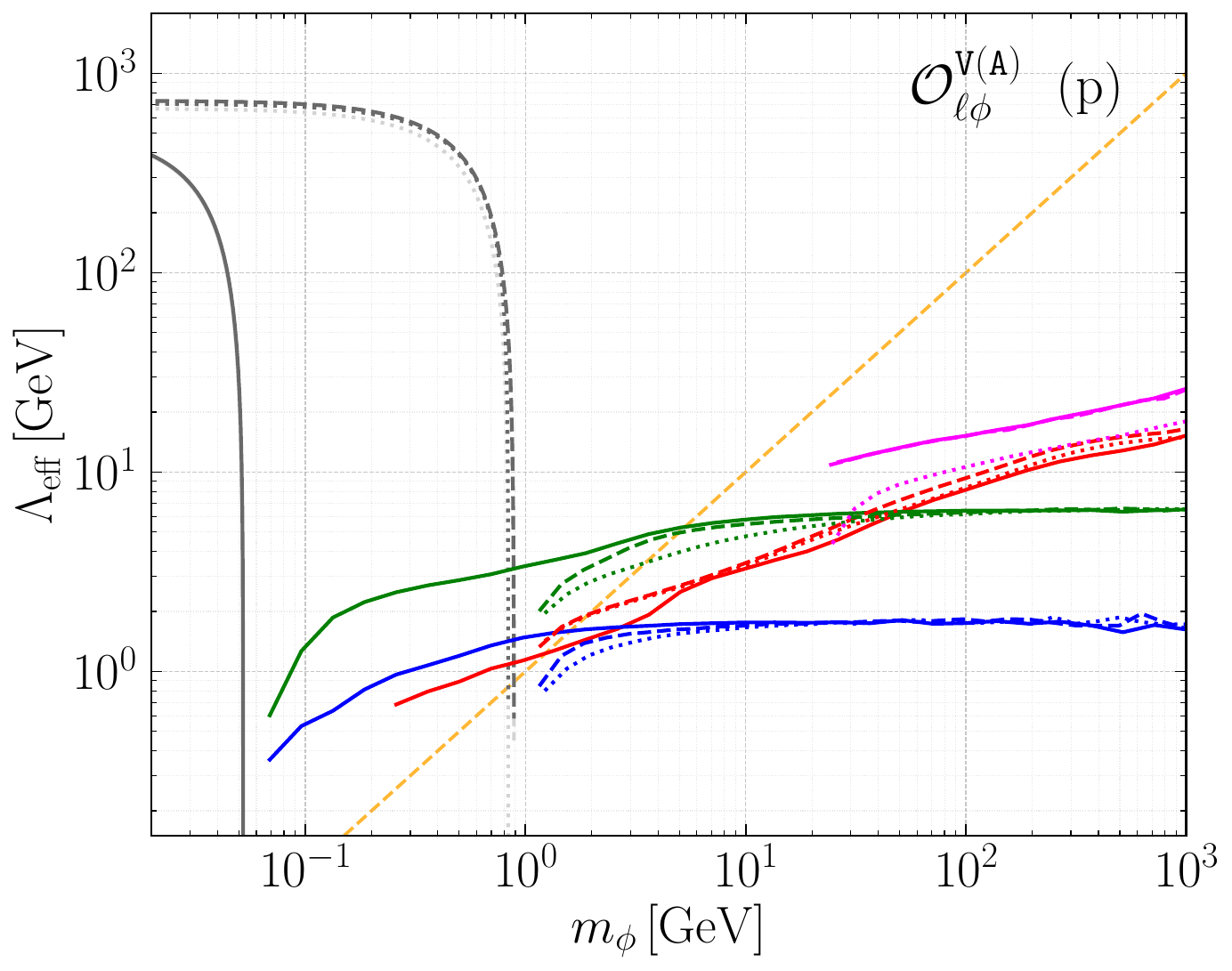} 
\caption{
Constraints on the effective scales associated with 
dim-5 (left panel) and dim-6 LFV DM-lepton operators in the complex scalar DM case. These constraints are derived from Fermi-LAT (red curves), INTEGRAL (green curves), XMM-Newton (blue curves), and AMS-02 (magenta curves). 
The solid, dashed, and dotted curves correspond to the $e\mu$, $e\tau$, and $\mu\tau$ flavor combinations, respectively. 
The gray curves denote constraints from $\mu$ and $\tau$ LFV decays into a pair of DM particles, reproduced from \cite{Jahedi:2025hnu}.The orange dashed straight line indicates the case where $\Lambda_{\rm eff}=m_\phi$.
}
\label{fig:scalar-DM-bounds}
\end{figure*}

So far, the derived constraints do not yet account for astrophysical uncertainties.
All photon and positron fluxes  are affected by uncertainties in the DM density profile.
Moreover, the photon flux induced by {\tt ICS} and the positron flux are
further influenced by uncertainties in the Galactic gas density, the interstellar radiation field, and the Galactic magnetic field~\cite{Cirelli:2020bpc,Cirelli:2023tnx}. Taken together, these effects would broaden the constraints on $\langle \sigma v_{\rm rel} \rangle$ into a band centered around the current values, with a varied width up to about one order of magnitude \cite{Cirelli:2020bpc,Cirelli:2023tnx}.

\subsection{Constraints on EFT operators}

In this subsection, we derive the indirect detection constraints on the LFV EFT operators listed in \cref{tab:operators}.
We discuss the constraints on an effective scale $\Lambda_{\rm eff}$ associated with each dim-$d$ operator $\calO_i^d$, defined through its corresponding WC $C_i^d$ via the relation $\Lambda_{\rm eff}\equiv |C_i|^{1/(d-4)}$. 
Using the leading-order expansions of $\sigma v_{\rm rel}$ induced by the EFT operators given in \cref{tab:cross-section}, 
the constraints shown in \cref{fig:sigma_spd} can be  
directly translated into bounds on the effective scales associated with the relevant EFT operators, with the exception of $\calO_{\ell\chi2}^{{\tt V (A)}}$ and $\calO_{\ell X3}^{{\tt V (A)}}$. 
For these two types of operators, the annihilation cross sections receive comparable contributions from two partial waves. Consequently, the constraints on their associated effective scales cannot be inferred from the results in \cref{fig:sigma_spd} and must be reevaluated separately using \cref{eq:sta} with $w = \Lambda_{\rm eff}$.
Our results are presented for operators related to the complex scalar or vector and Dirac fermion DM scenarios. 
In the cases of real scalar or vector and Majorana fermion DM, all operators marked with an \xmark in \cref{tab:operators} vanish, while for the remaining operators their effective scales are increased by a factor of ${(2\sqrt{2})}^{1/(d-4)}$  compared to those of their complex counterparts.
Below, we discuss the final constraints for each DM case individually.

\subsubsection{Scalar DM case}

\begin{figure*}[t]
\centering
\includegraphics[width=0.329\textwidth]{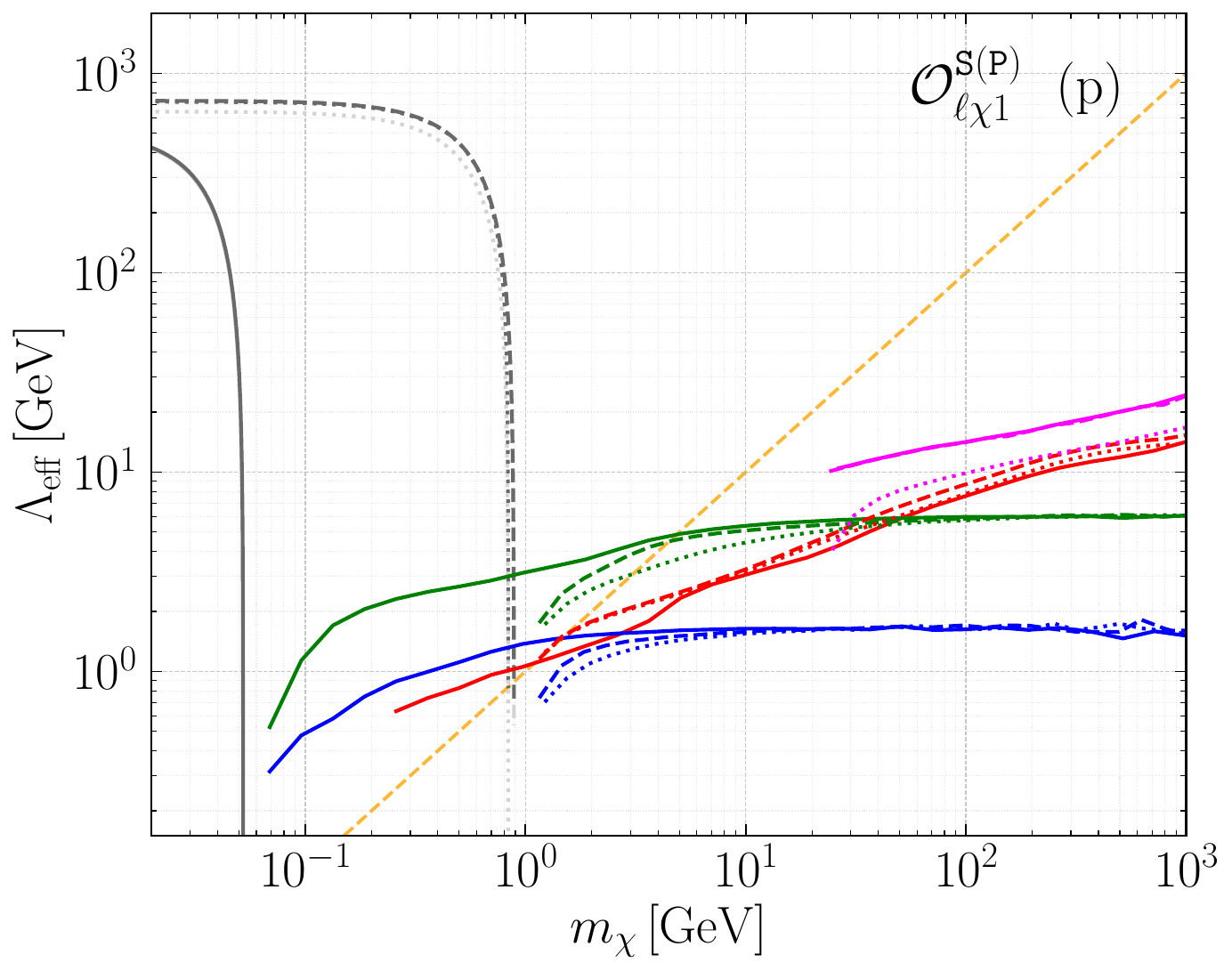} 
\includegraphics[width=0.329\textwidth]{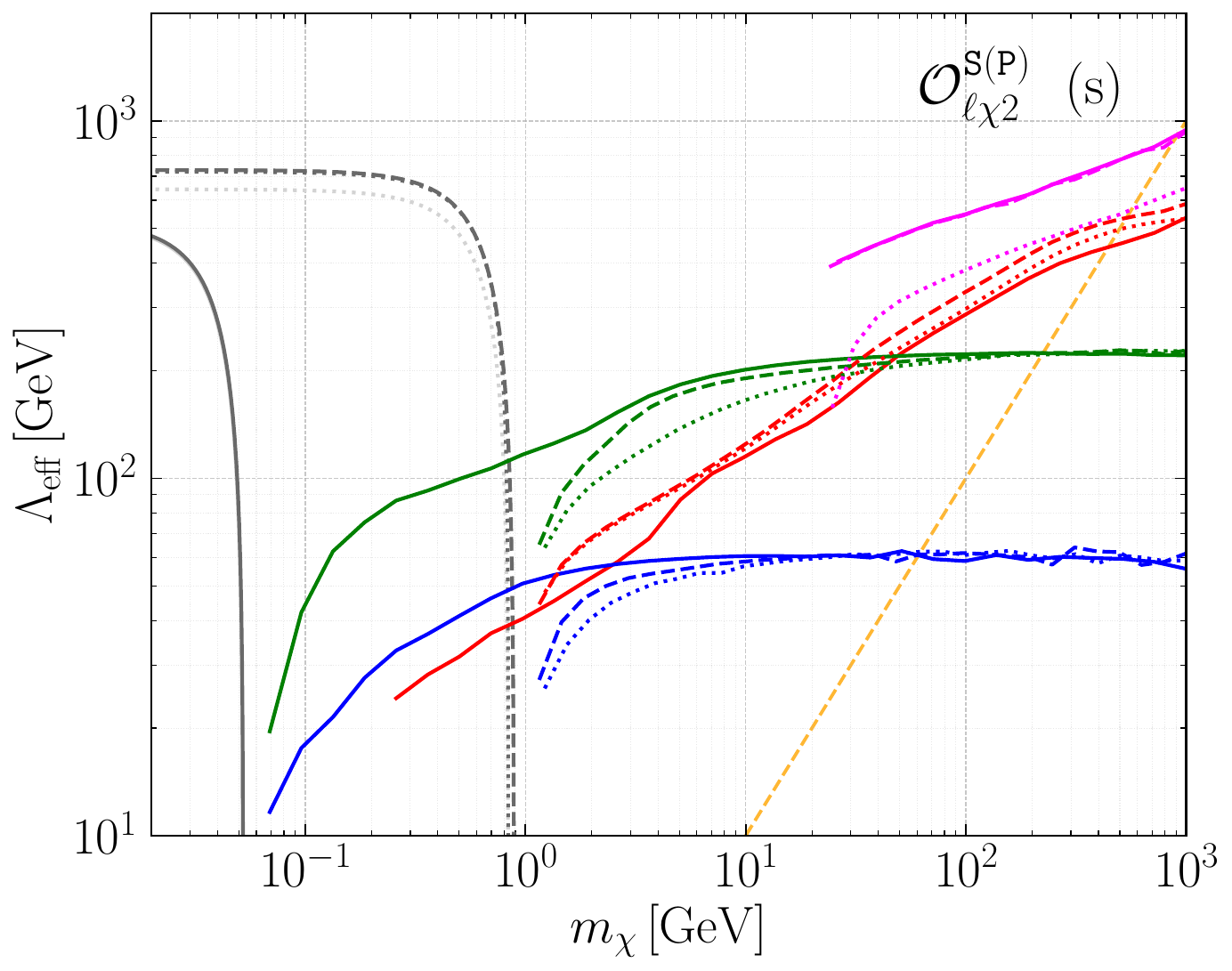}
\\
\includegraphics[width=0.329\textwidth]{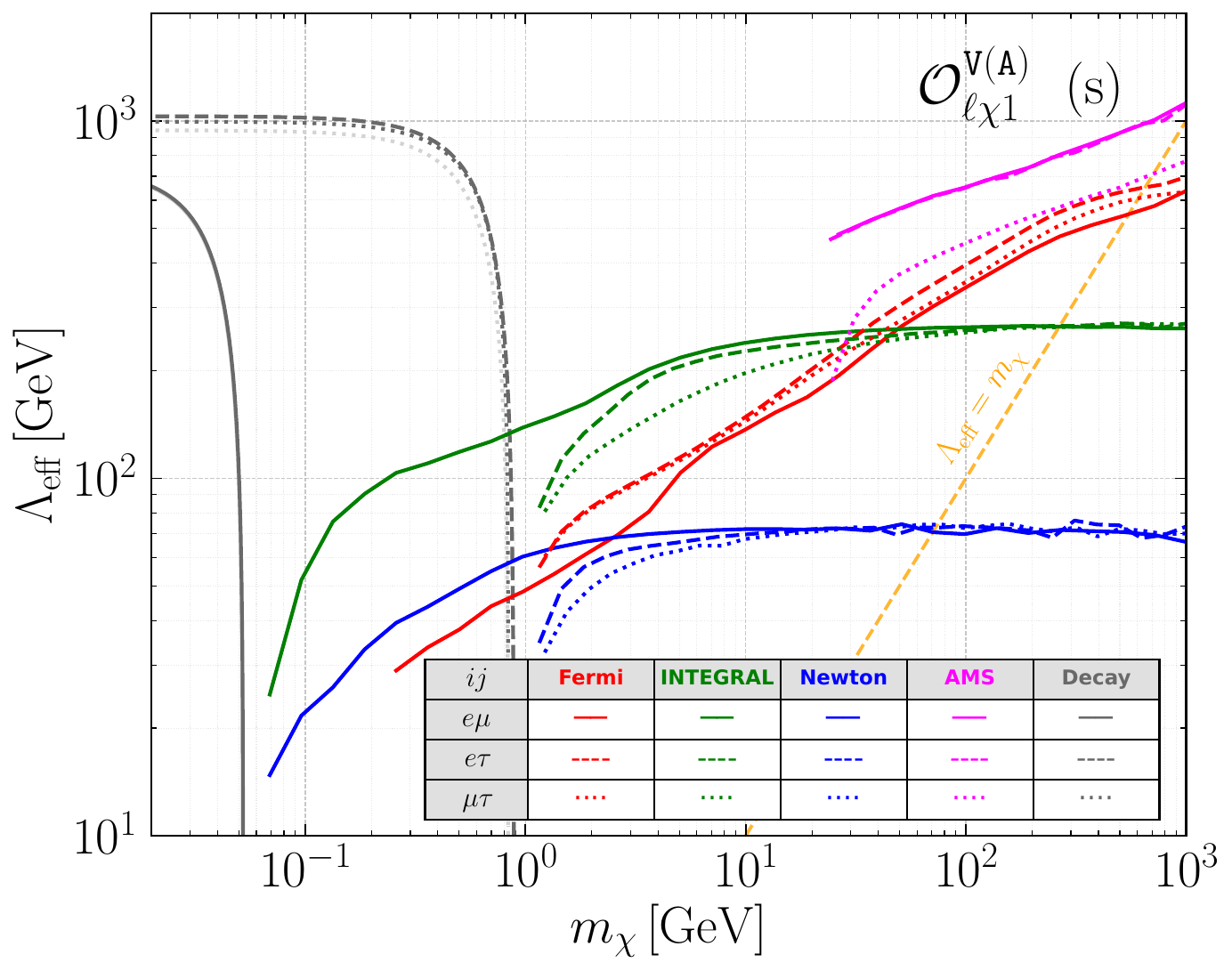}
\includegraphics[width=0.329\textwidth]{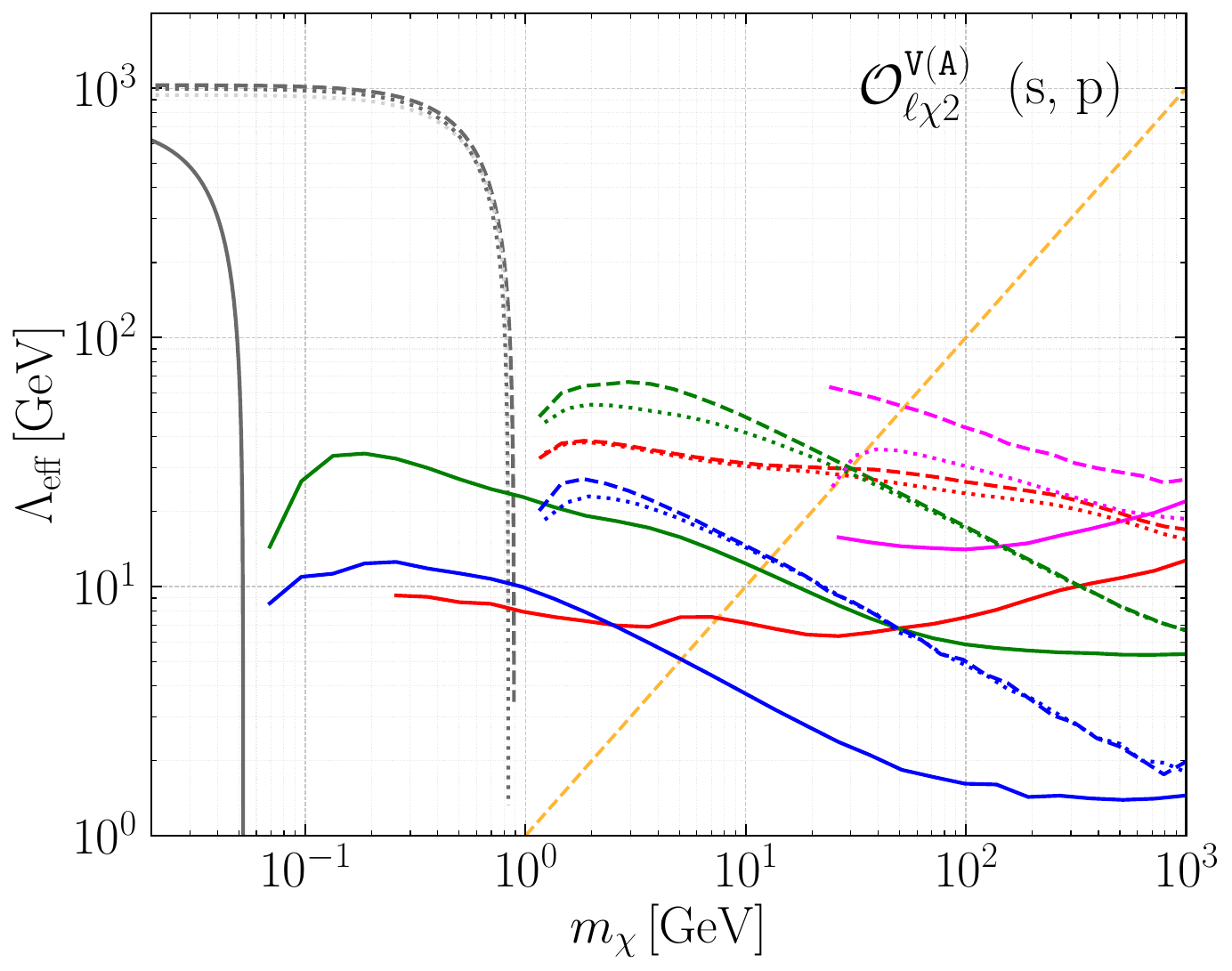} 
\includegraphics[width=0.329\textwidth]{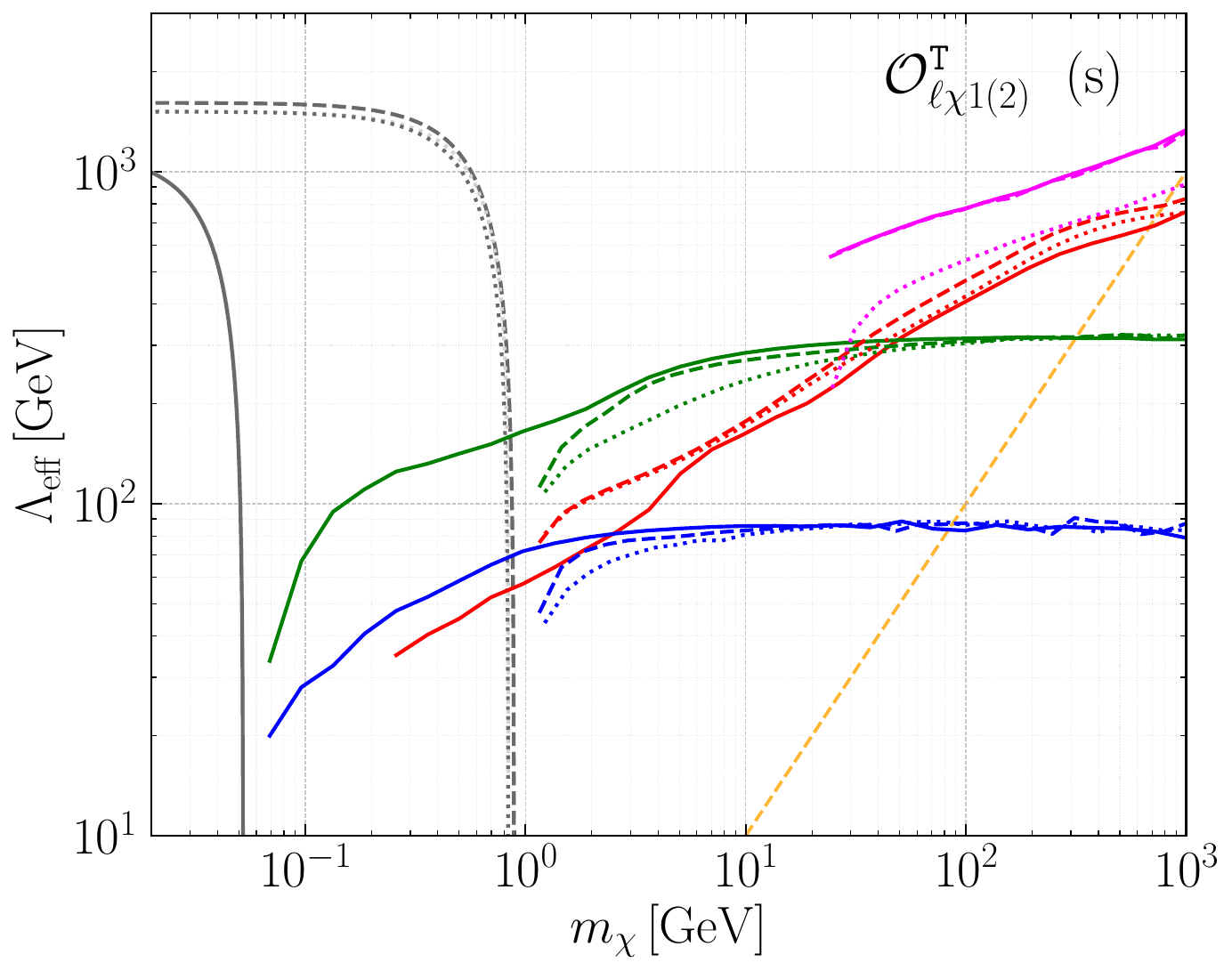} 
\caption{
The same as \cref{fig:scalar-DM-bounds} but for the Dirac fermion DM case.
}
\label{fig:fermion-DM-bounds}
\end{figure*}

In \cref{fig:scalar-DM-bounds}, we show the lower bounds on the effective scales for scalar DM operators that involve $e\mu$ (solid curves), $e\tau$ (dashed curves), and $\mu\tau$ (dotted curves) flavor combinations through the three LFV DM annihilation modes. 
Each pair of operators that differ by a $\gamma_5$ in the leptonic current is subject to nearly identical constraints because the mass of the lighter lepton is much smaller than the masses of the DM and heavier lepton partner in each flavor combination. Hence, they are grouped together in the plots throughout the paper. 
The x- and gamma-ray constraints obtained from Fermi-LAT, INTEGRAL, and XMM-Newton are represented by red, green, and blue curves, respectively, while the positron constraints from AMS-02 are depicted in magenta. 
In addition, we compile existing constraints from 
charged lepton-flavor-violating decays $\ell_j\to \ell_i+{\tt DM+DM}$~\cite{Jahedi:2025hnu} as gray curves for comparison. 
\footnote{
At the one-loop level, a double insertion of these operators may induce the LFV muonium-antimuonium oscillation through a DM loop.
A consistent treatment of such a process requires one to renormalize effective interactions that change the electron and muon flavor by two units and in opposite directions, and is beyond the scope of this work. 
Nevertheless, to give an estimate, we follow the standard practice by retaining only the logarithmic term in the ultraviolet divergent loop amplitude. Parametrizing the induced interaction by $\Lambda_{\mu e}^{-2}(\bar\mu \Gamma e)(\bar\mu\Gamma' e)$, we find that $\Lambda_{\mu e}^{-2} \sim (4\pi)^{-2}m_{\tt DM}^{2n-2}
\Lambda_{\rm eff}^{-2n}\ln(m_{\tt DM}^2/\Lambda_{\rm eff}^{2})$ with the double insertion of a dim-$(n+4)$ DSEFT operator. The current experimental bound $\Lambda_{\mu e}\gtrsim 3~\rm TeV$~\cite{Petrov:2022wau} yields a limit on $\Lambda_{\rm eff}\gtrsim 5\,\rm GeV(m_{\tt DM}/1~\rm GeV)^{1/2}$ for $n=2$, which is generally weaker than the INTEGRAL limit shown in \cref{fig:fermion-DM-bounds}. We, therefore, omit these constraints from the plots.}
Beyond these, more precise LFV bounds will also depend on the underlying models, since any underlying mediator generating LFV DSEFT interactions would likely induce other tree-level LFV processes as well.

As can be seen from the two panels, across all operator types, the INTEGRAL telescope provides the most stringent bound in the lower DM mass region ($m_\phi\lesssim 25$~GeV), whereas the AMS-02 experiment dominates in the higher mass region ($m_\phi\gtrsim 25$~GeV), 
consistent with the characteristics shown in \cref{fig:sigma_spd}.
In the plots, we also display the dominant partial-wave component in parentheses following the operator label to help understand the constraining features. 
In the left panel, the dim-5 operators $\calO_{\ell\phi}^{\tt S(P)}$ that induce s-wave contributions receive more stringent constraints, with $\Lambda_{\rm eff}$ reaching $\calO(20~\rm TeV)$ ($\calO(6~\rm TeV)$) for the $e\mu$ ($e\tau$ and $\mu\tau$) flavor case(s), when $m_\phi\approx 0.2~(3)~\rm GeV$. 
The constraints become weaker as the DM mass increases because their $\sigma v_{\rm rel}$ is essentially independent of the DM mass when $m_\phi\gg m_{i,j}$. 
On the other hand, the dim-6 operators $\calO_{\ell\phi}^{\tt V(A)}$ induce p-wave contributions at leading order, resulting in much weaker constraints on their associated $\Lambda_{\rm ref}$s,  reaching only up to tens of GeV when $m_\phi\gtrsim 200~\rm GeV$. 
The relatively flat behavior observed in the  constraints from INTEGRAL and Newton is a joint result of the DM mass enhancement in their $\sigma v_{\rm rel}$ expansion and the steeply decreasing sensitivity of both telescopes to $\hat b$, as demonstrated in \cref{fig:sigma_spd}. 
Although the constraints from charged LFV decays are stronger than the indirect detection by several orders of magnitude, 
the constraints from DM indirect detection experiments can probe the DM mass range beyond the decay threshold, making those results 
complementary to each other.

\subsubsection{Fermion DM case}

In \cref{fig:fermion-DM-bounds}, we present the lower bounds on the effective scale of each dim-6 operator in the fermion DM case. 
Similarly to the scalar DM case, the constraints from DM indirect detection experiments are complementary to those from charged LFV decays. 
From the third column in \cref{tab:cross-section}, it is clear that there are six operators that contribute dominantly through the s-wave annihilation for the entire kinematically allowed DM mass range: 
$\calO^{{\tt S(P)}}_{\ell \chi 2}$, $\calO^{{\tt V(A)}}_{\ell \chi 1}$, and $\calO^{{\tt T}}_{\ell \chi 1(2)}$. 
For these operators, the constraints on their effective scales are comparable and exhibit similar behavior as the DM mass increases, owing to their analogous quadratic dependence on $m_\chi$ when $m_\chi\gg m_{i,j}$. 
It should be noted that the constraints can reach up to hundreds of GeV when $m_\chi$ exceeds the GeV scale. 
In contrast, the operators $\calO^{{\tt S(P)}}_{\ell \chi 1}$ contribute to the p-wave annihilation in the velocity expansion, resulting in comparably weaker constraints similar to those of the dim-6 operators $\calO_{\ell \phi}^{\tt V(A)}$ in the scalar DM case, as shown in \cref{fig:scalar-DM-bounds}.

Finally, for the two operators $\calO^{{\tt V(A)}}_{\ell \chi 2}$, their contribution to $\sigma v_{\rm rel}$ is dominated by the s-wave annihilation up to DM masses around 90~GeV for the $e\mu$ flavor combination and 1.5~TeV for the $\tau$-related flavor combinations; see \cref{fig:sigmav}. Beyond these mass points, the p-wave contribution becomes dominant. 
Because of these nontrivial features, the corresponding constraints on their associated $\Lambda_{\rm eff}$'s exhibit significantly different behaviors compared to operators dominated solely by s- or p-wave contributions. 
The s-wave component of $\sigma v_{\rm rel}$ is proportional to the mass squared of the heavier final-state lepton through the ratio $r_{ij}$, as shown in the third column in \cref{tab:cross-section}. Consequently, the $e\tau$ and $\mu\tau$ flavor combinations are subject to more stringent bounds than the $e\mu$ case in all experiments. 
We find that the INTEGRAL telescope sets the strongest bounds, with values for $\Lambda_{\rm eff}$ of $e\tau$ flavor combination reaching up to 70\,GeV for DM masses around a few GeV.

\begin{figure*}[t]
\centering
\includegraphics[width=0.245\textwidth]{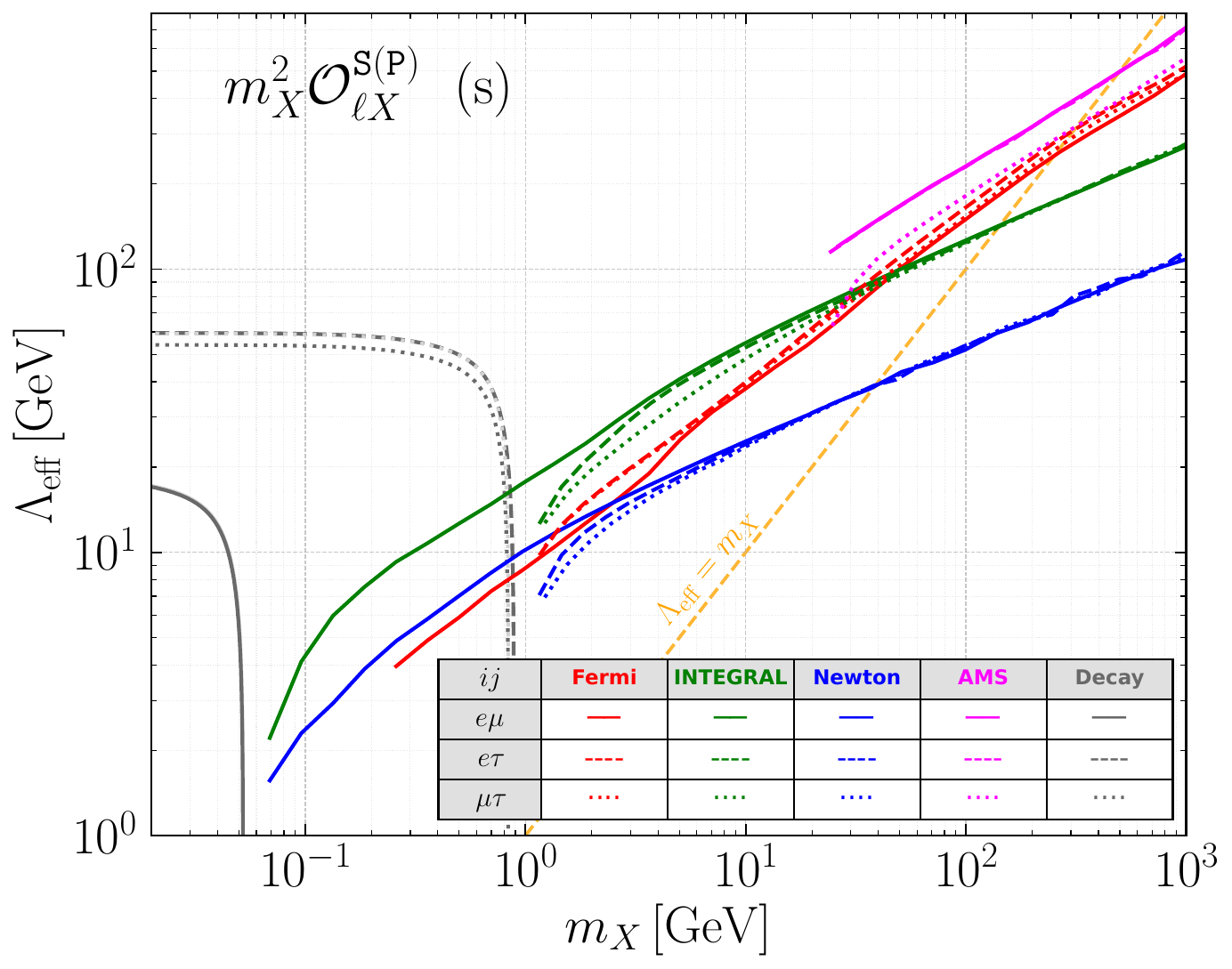} 
\includegraphics[width=0.245\textwidth]{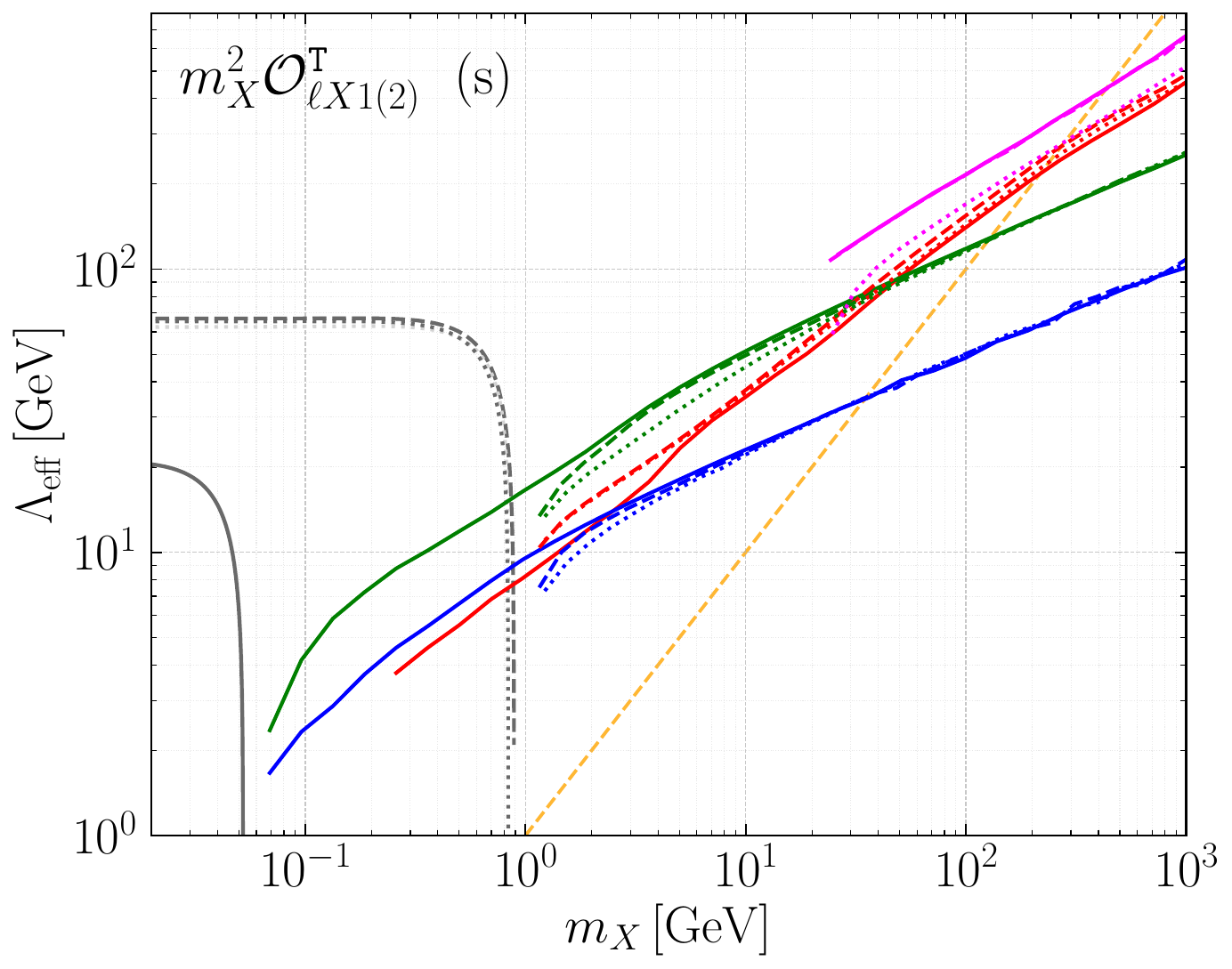} 
\includegraphics[width=0.245\textwidth]{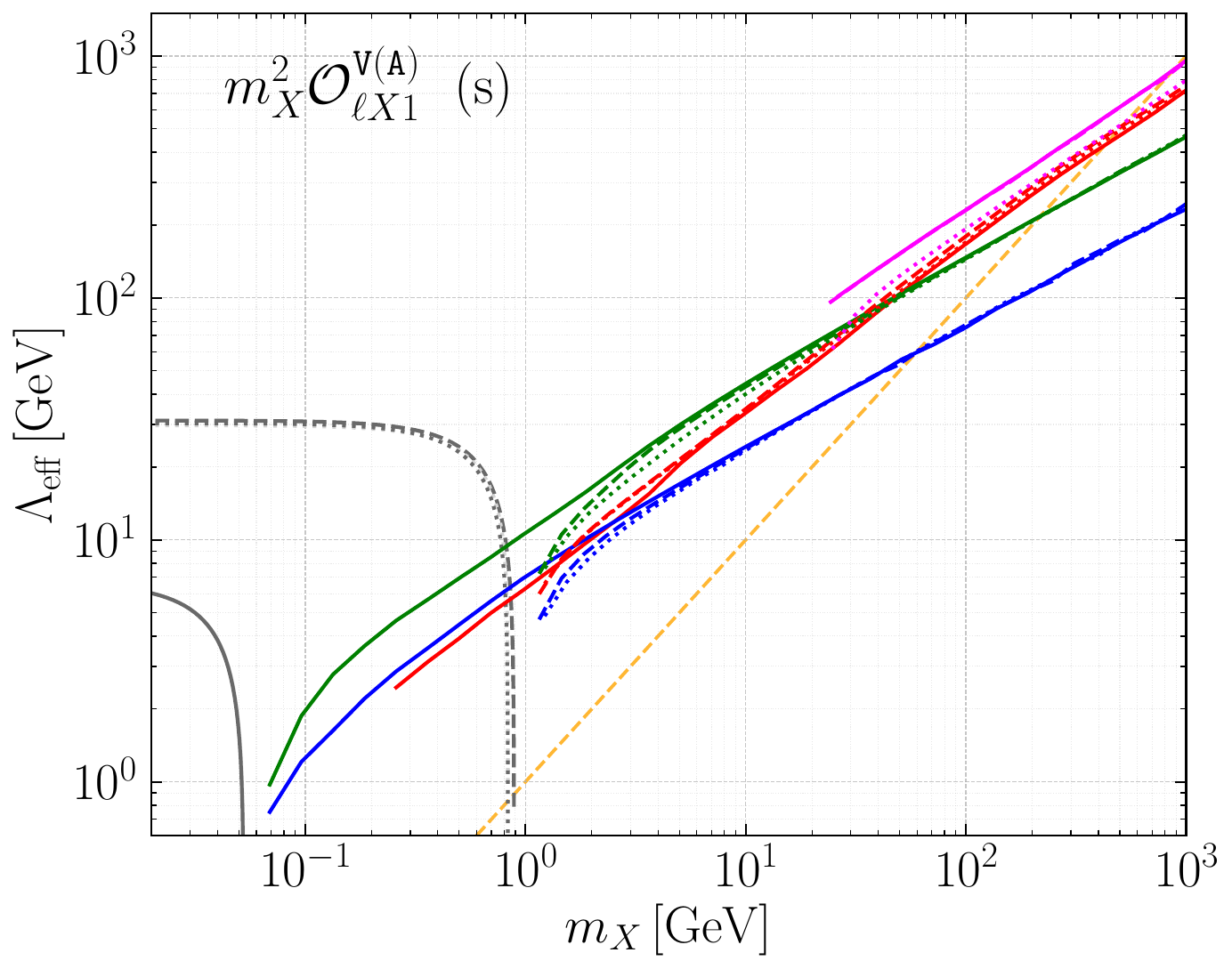} 
\includegraphics[width=0.245\textwidth]{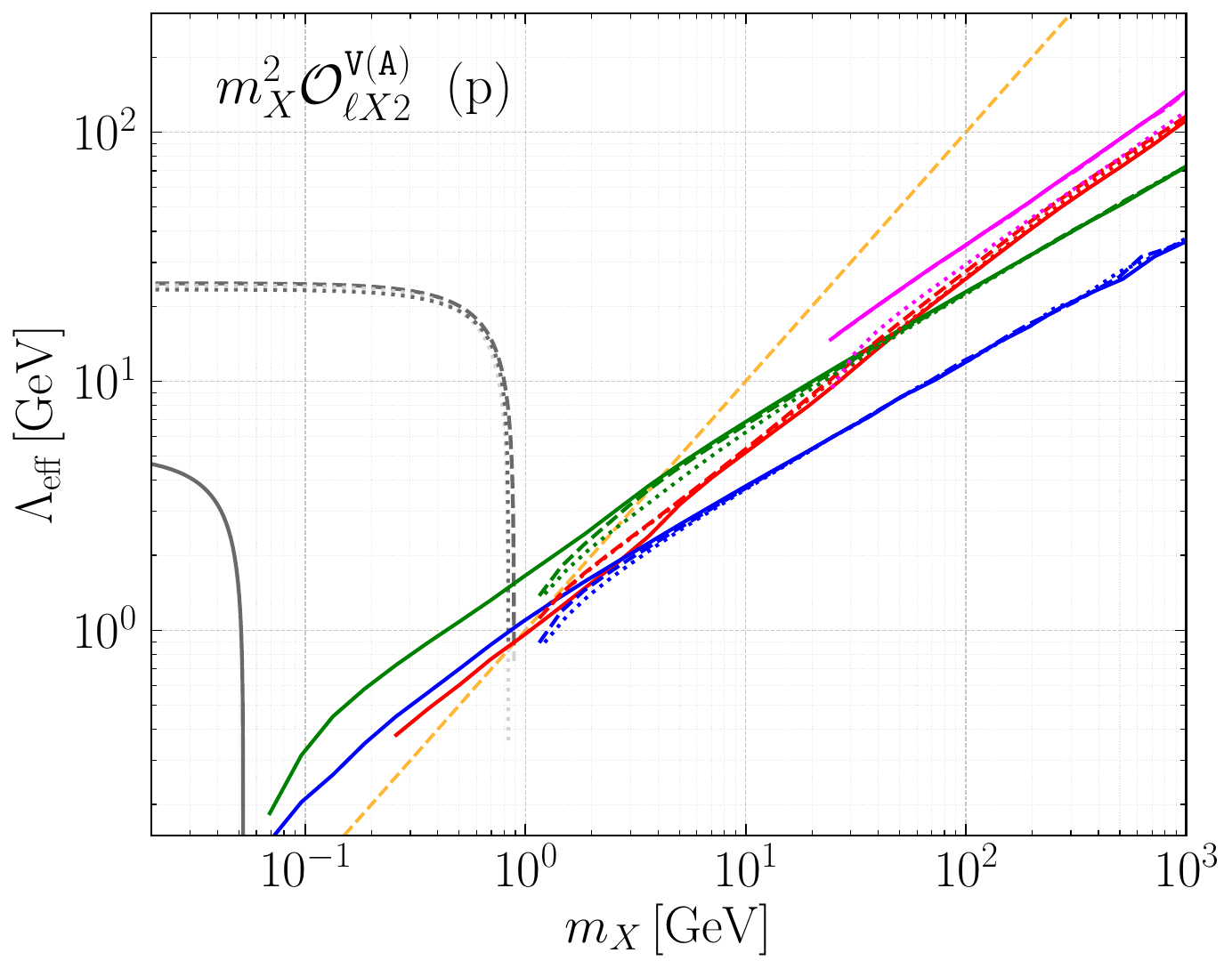}
\\
\includegraphics[width=0.245\textwidth]{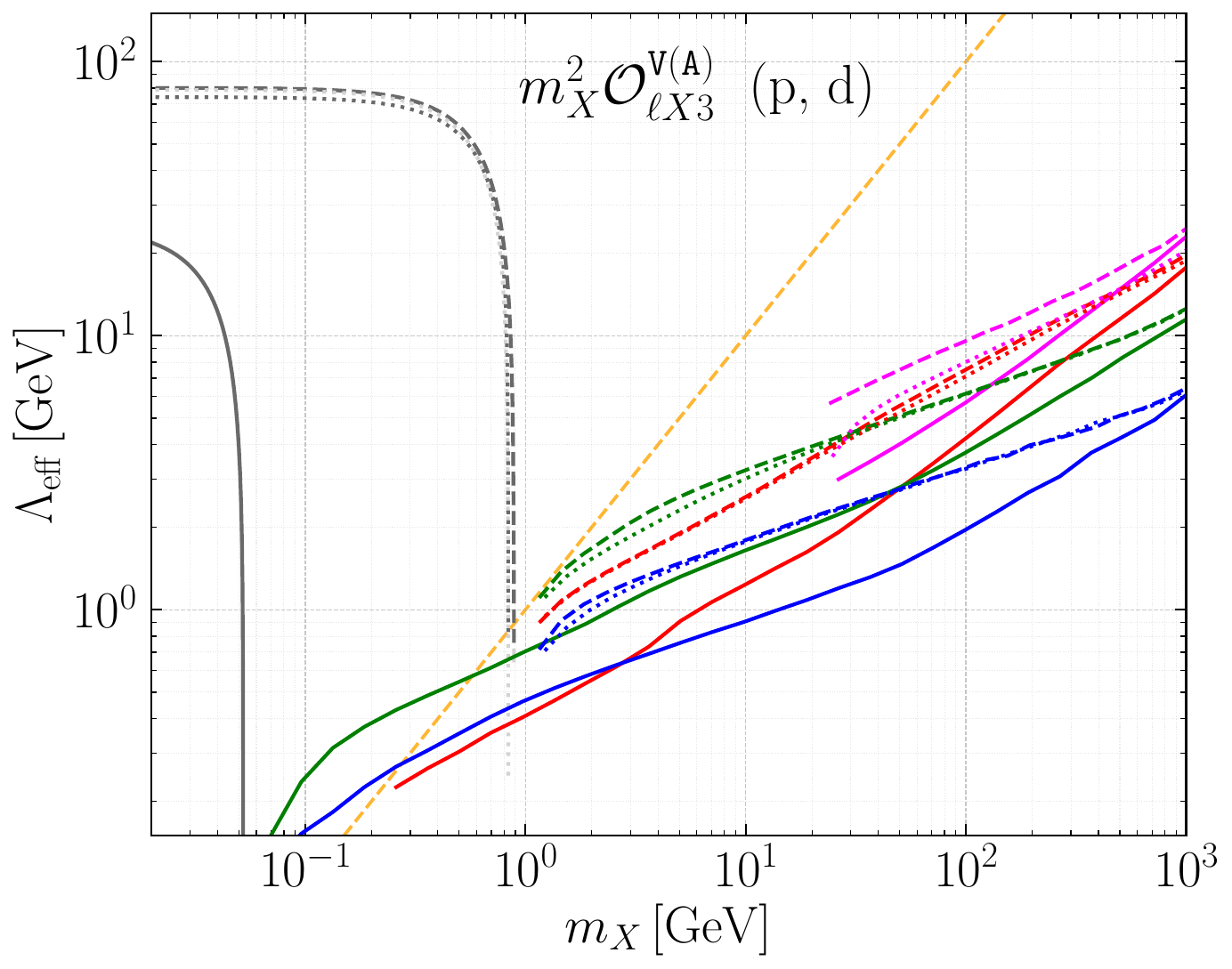}
\includegraphics[width=0.245\textwidth]{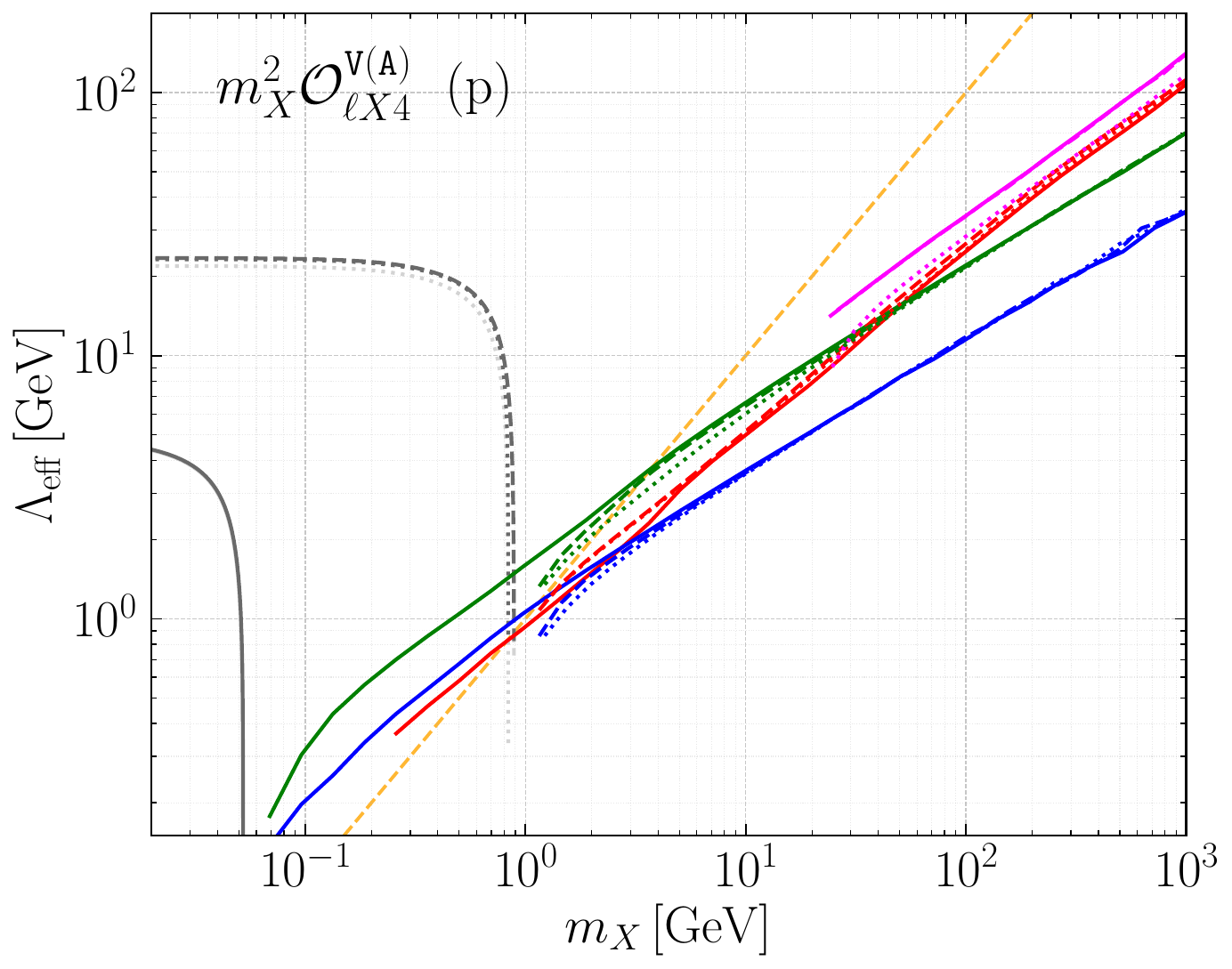} 
\includegraphics[width=0.245\textwidth]{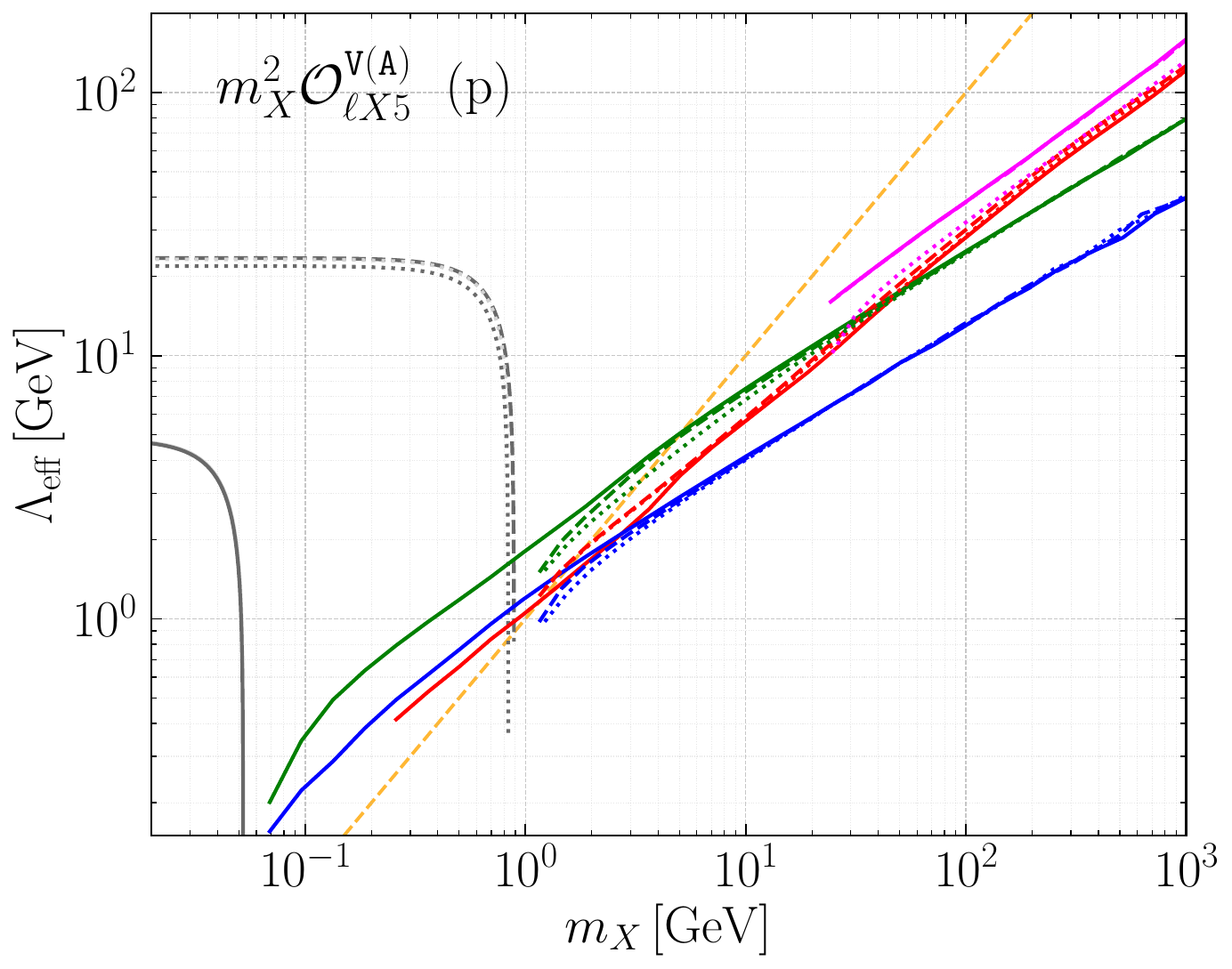}
\includegraphics[width=0.245\textwidth]{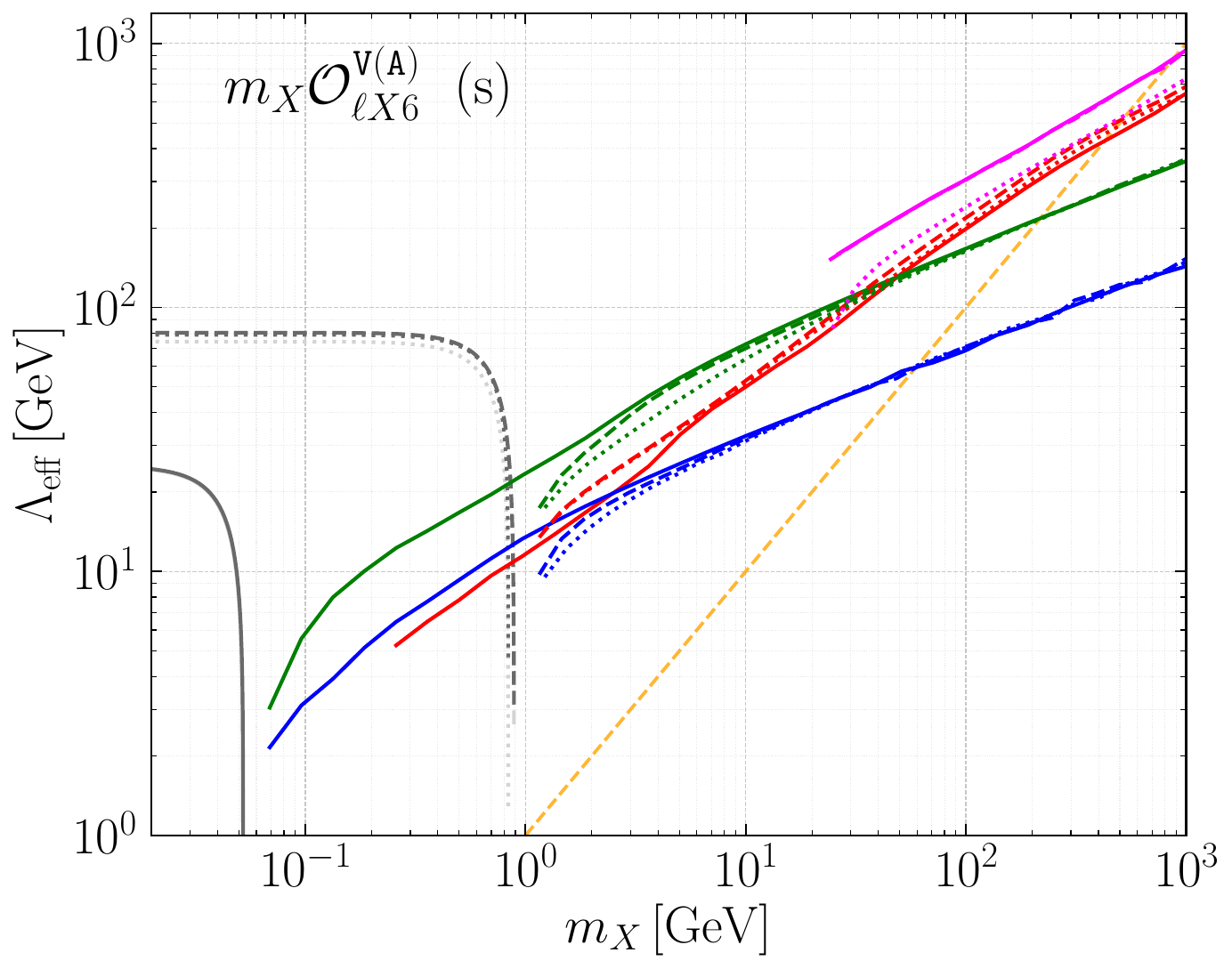}  
\caption{The same as \cref{fig:scalar-DM-bounds} but for the complex vector DM case A.
}
\label{fig:VDM-bounds-A}
\end{figure*}

\begin{figure*}[t]
\centering 
\includegraphics[width=0.329\textwidth]{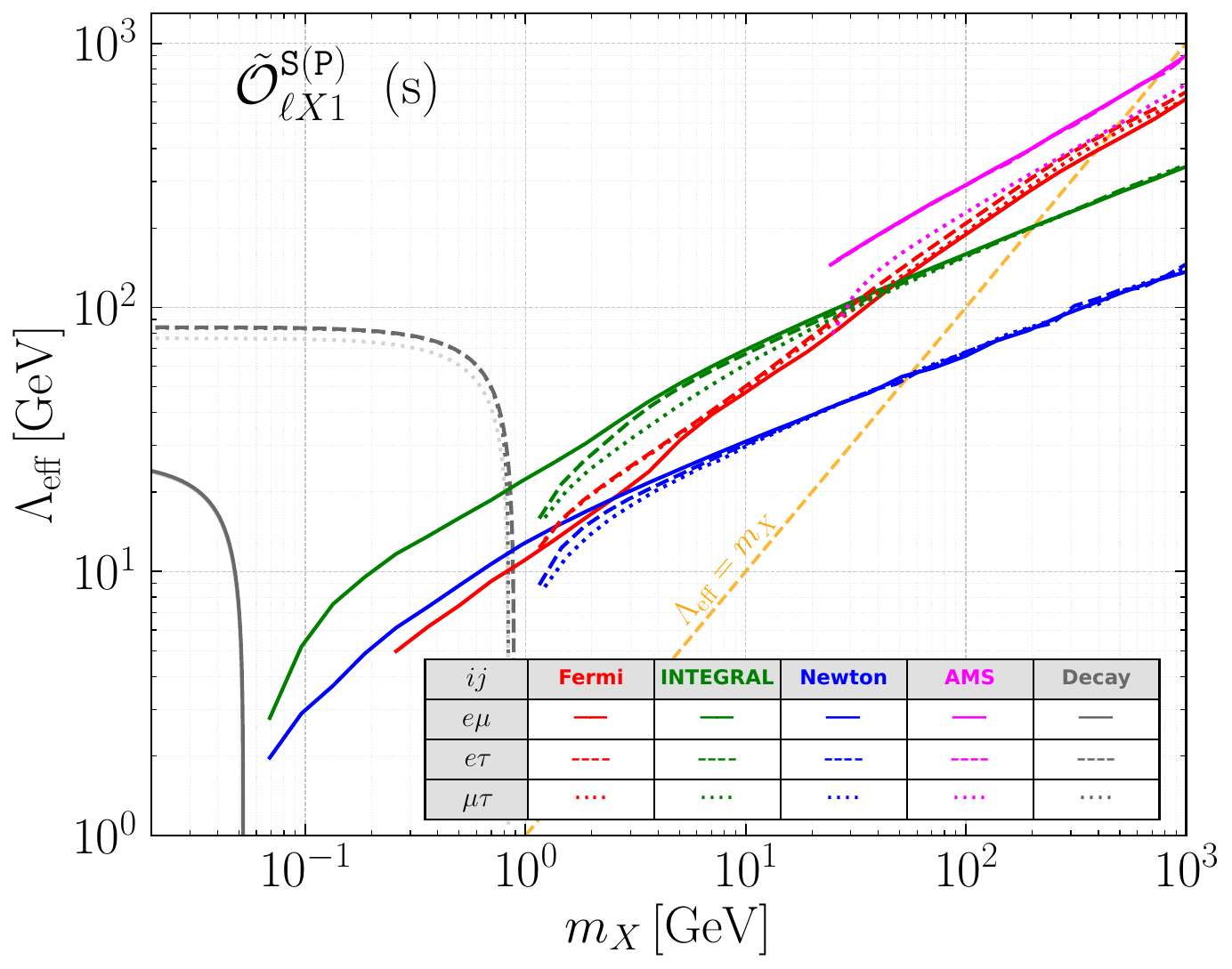}  
\includegraphics[width=0.329\textwidth]{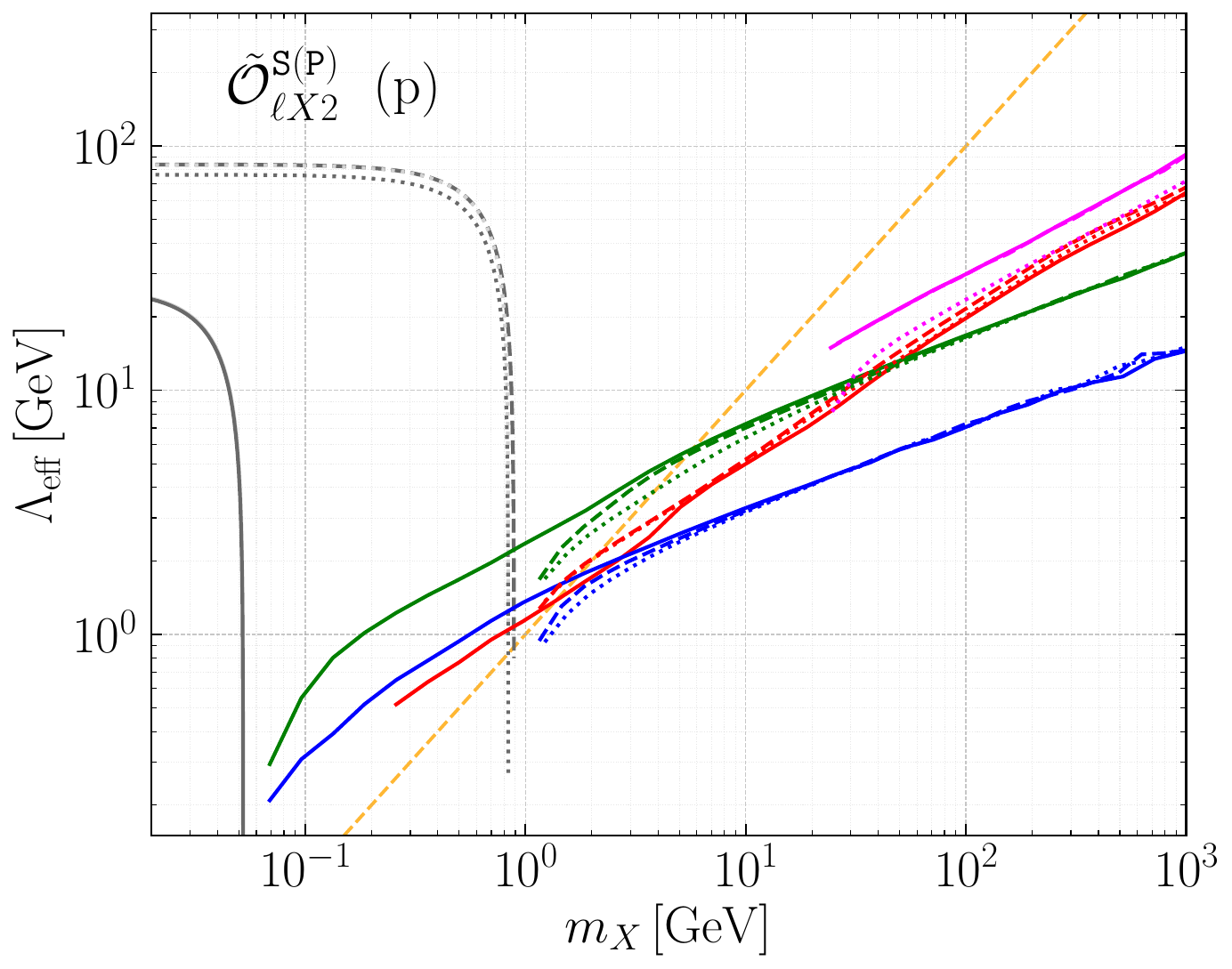} 
\includegraphics[width=0.329\textwidth]{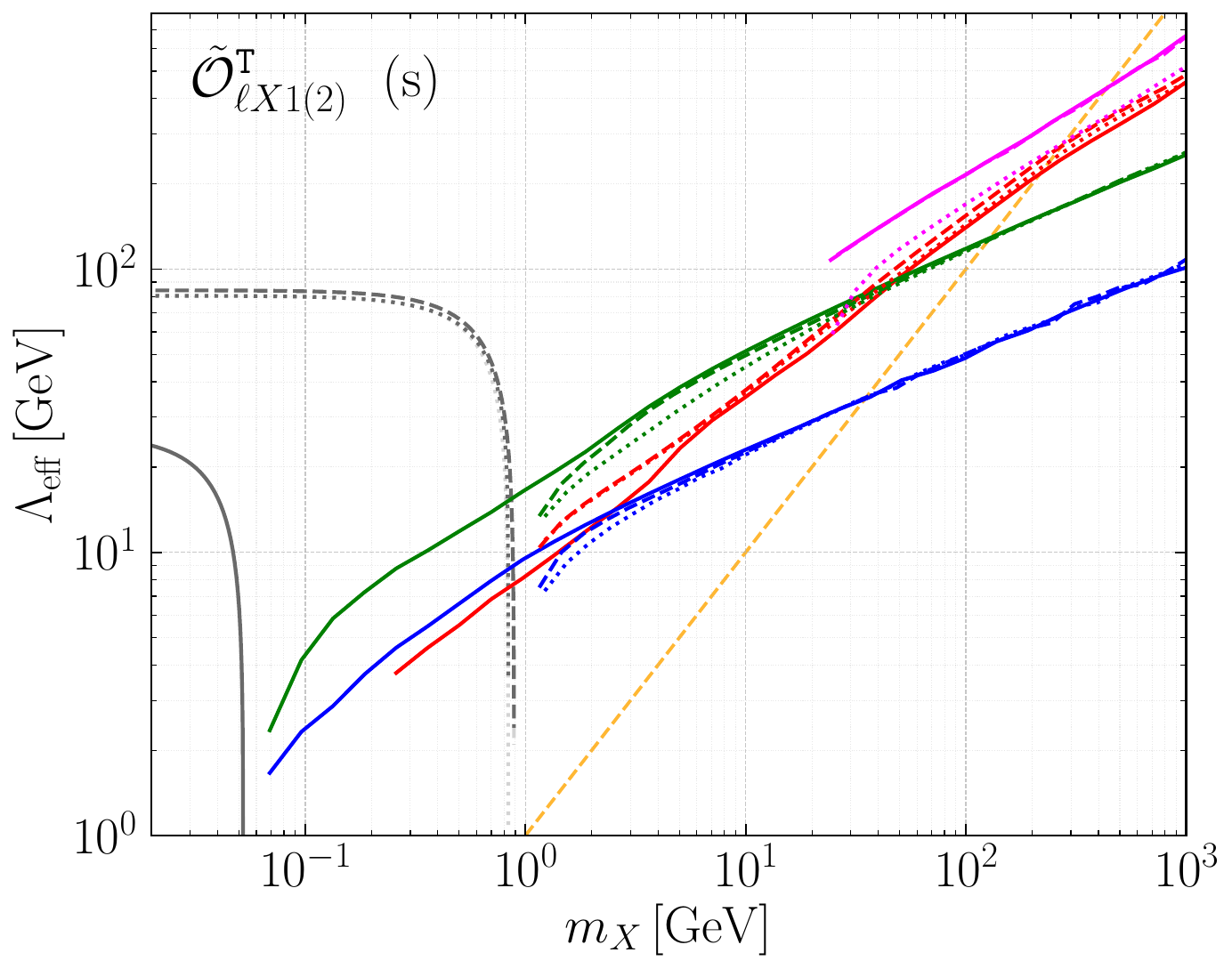}  
\caption{The same as \cref{fig:scalar-DM-bounds} but for the complex vector DM case B.
}
\label{fig:VDM-bounds-B}
\end{figure*}

\subsubsection{Vector DM case}

In the context of charged LFV decays involving light vector DM particles, $\ell_j\to \ell_i+X+X$, the total decay widths diverge as the DM mass approaches zero, $m_X \to 0$, in the vector case A, where the operators are formulated in terms of four-vector potentials. 
To circumvent this issue and set meaningful constraints at smaller values of $m_X$, 
Refs.\,\cite{He:2022ljo,Jahedi:2025hnu} have assumed that each WC in case A is proportional to a minimal power of $m_X$, which is determined by the number of independent four-vector potentials that cannot be expressed as field strength tensors. 
We adopt this convention and establish constraints from indirect detection for comparison.  
Following Ref.\,\cite{He:2022ljo}, we define the effective scale for each operator as follows:
\begin{align}
& C_{\ell X}^{\tt S(P)} \equiv \frac{m_X^2}{\Lambda_{\rm eff}^3},\quad
C_{\ell X1,2}^{\tt T} \equiv \frac{m_X^2}{\Lambda_{\rm eff}^3},\,
\nonumber\\
& C_{\ell X1,2,4,5}^{\tt V(A)} \equiv \frac{m_X^2}{\Lambda_{\rm eff}^4},\quad
C_{\ell X3,6}^{\tt V(A)} \equiv \frac{m_X}{\Lambda_{\rm eff}^3}.
\label{eq:VDMAWCs}
\end{align} 

In \cref{fig:VDM-bounds-A}, we present the constraints on $\Lambda_{\text{eff}}$s for vector DM case A based on the redefinitions in \cref{eq:VDMAWCs}. 
Because of the explicit mass factors in the parametrization, the indirect detection constraints on $\Lambda_{\rm eff}$s increase as the DM mass rises.  
Similarly to the scalar and fermion DM scenarios, we incorporate constraints from both astrophysical x- and gamma-ray and positron observations, as well as charged LFV decays. From the third column in \cref{tab:cross-section}, it is evident that there are eight operators in case A that contribute predominantly through the s-wave annihilation:
$\calO^{{\tt S(P)}}_{\ell X}$, $\calO^{{\tt T}}_{\ell X1,2}$, and $ \calO^{{\tt V(A)}}_{\ell X1,6}$. 
Consequently, the constraints on their associated  effective scales follow patterns similar to those of the s-wave operators in fermion DM case, with values ranging from a few GeV to TeV, depending on the DM masses.  
On the other hand, the six operators $\calO^{{\tt V(A)}}_{\ell X2,4,5}$ primarily contribute via p-wave annihilation, 
resulting in constraints that are nearly one order of magnitude weaker than those of s-wave operators at the same DM mass point. 
For the operators $\calO^{{\tt V(A)}}_{\ell X3}$, the contribution is dominated by p-wave annihilation up to DM masses of approximately 90\,GeV and 1.5\,TeV for the $e\mu$ and $\tau$-related flavor combinations, respectively. Beyond these mass thresholds, the d-wave contribution becomes dominant.
Similarly to the operators $\calO_{\ell\chi2}^{\tt V(A)}$ in the fermion DM case, the leading p-wave contribution is proportional to the mass squared of the heavier lepton, leading to more stringent bounds on operators involving the $e\tau$ and $\mu\tau$ flavor combinations.

Finally, \cref{fig:VDM-bounds-B} presents the results for the vector DM case B, where the operators are parametrized using field strength tensors.  
The four operators $\mathcal{\tilde{O}}^{{\tt S(P)}}_{\ell X1}$ and $\mathcal{\tilde{O}}^{{\tt T}}_{\ell X1,2}$ contribute through s-wave annihilation, while the two operators $\mathcal{\tilde{O}}^{{\tt S(P)}}_{\ell X2}$ are dominated by p-wave annihilation.
Since these operators are of dimension 7, the constraints on the effective scales of $\mathcal{\tilde{O}}^{{\tt S(P)}}_{\ell X1}$ and $\mathcal{\tilde{O}}^{{\tt T}}_{\ell X1,2}$ exhibit similar magnitudes and behavior to those of 
$m_X^2\calO_{\ell X}^{\tt S(P)}$ in \cref{fig:VDM-bounds-A} for the vector case A.

The EFT description is valid when the scale of underlying new physics $\Lambda_{\rm NP}$ is much larger than the masses of active degrees of freedom, i.e., $\Lambda_{\rm NP}\gg m_{e,\mu,\tau,{\tt DM}}$, which is related to the effective scale $\Lambda_{\rm eff}$ by $\Lambda_{\rm NP}=\hat c \Lambda_{\rm eff}$. Here, $\hat c$ is a product of the couplings of new heavy particles to both DM and SM particles, and may vary in a large range from small to large values close to perturbative limits. When the lower bound on $\Lambda_{\rm eff}$ is low, it does not necessarily mean that the effective field theory approach is not valid but means that the relevant data are not strong enough to set a useful bound. Our results indeed show that astrophysical photon and positron data can be used to set much better constraints on s-wave operators than p- or d-wave ones.
On the other hand, when the coupling $\hat c$ is $\calO(1)$, the parameter regions below the orange dashed straight lines (where $\Lambda_{\rm eff}=m_{\tt DM}$) in \cref{fig:scalar-DM-bounds,fig:fermion-DM-bounds,fig:VDM-bounds-A,fig:VDM-bounds-B} cannot be reliably described within the EFT framework. In these regions where the mediator mass approaches or even falls below $m_{\tt DM}$, a consistent treatment requires working directly with the underlying models. 
A more detailed discussion of this issue in the context of collider studies can be found in~\cite{Racco:2015dxa}.

\section{Summary}
\label{sec:summary}

In this study, we systemically explored the constraints on LFV effective interactions between a DM pair and a charged-lepton pair, utilizing the astrophysical photons and positrons observed by Fermi-LAT, INTEGRAL, XMM-Newton, and AMS-02. 
The DM candidates examined include a scalar, a fermion, and a vector. 
We collected the leading-order effective operators for each case and derived analytical expressions for the thermally averaged cross sections induced by these operators. 
For the photon flux, we considered three main contributions: final-state radiation, radiative decay, and inverse Compton scattering. For {\tt FSR}, we calculated the photon spectrum for each s-wave operator and found that the normalized spectrum exhibits a mild dependence on the operator structure. In particular, in the observationally relevant energy range, the spectra from all these s-wave operators tend to be degenerate, allowing the {\tt FSR} spectrum to be approximately described by lepton structure functions when DM mass is much larger than the heavier lepton mass. 
The comparison of photon flux contributions from these three mechanisms demonstrates that they complement each other in probing different DM mass regions. 

For the three LFV annihilation channels, we established meaningful constraints on DM-lepton interactions for different partial-wave cases (s, p, and d), and found that the resulting limits across different partial waves are related by almost constant velocity-suppression factors. 
For the three partial-wave cases, our results indicate that INTEGRAL provides the most stringent limits for DM masses below 25\,GeV or so, while AMS-02 yields the strongest limits for masses above. 
When we translated these general constraints into effective scales associated with the effective operators, the probed effective scale varies according to the DM candidates, their masses, and the operator structures involved. 
For dim-5 (dim-6) s-wave operators in the scalar DM scenario, the effective scale can be probed up to 10 TeV (20\,GeV). 
In contrast, for the operators related to fermion and vector DM cases, the probed effective scales can range from GeV to TeV, depending on the DM masses and operator structures.
Constraints on p-wave operators are typically weaker by 1--2 orders of magnitude compared to those on s-wave operators of the same dimension.
Interestingly, there are two types of operators, $\calO_{\ell\chi2}^{{\tt V(A)}}$ and $\calO_{\ell X3}^{{\tt V(A)}}$, which receive significant contributions from two partial waves rather than a single one, with the dominant contribution changing between different DM mass regions.
These results complement existing low-energy charged lepton decay constraints when DM mass is below the decay threshold.

In addition to the indirect detection searches examined in this study, these LFV DM scenarios can induce a variety of intriguing signals in different environments. 
Here, we discuss two possibilities. 
First, such DM scenarios can inject energy into cosmic fluids during the epoch between recombination and reionization in the early Universe, potentially affecting the anisotropy of the cosmic microwave background \cite{Slatyer:2015jla,Slatyer:2016qyl}
and leading to stringent constraints. 
The second avenue involves collider searches. 
At lepton colliders, such as Belle II, lepton-flavored DM interactions can be investigated through processes like $e^+ + e^-\to e^\pm \mu^\mp/e^\pm\tau^\mp/\mu^\pm\tau^\mp$, along with the production of DM particles. However, 
due to suppression by additional SM couplings and phase space factors, the resulting constraints are expected to be much weaker than those obtained via indirect detection. 
On the other hand, for the $\mu e$ flavor case, future facilities such as the Muon collider~\cite{InternationalMuonCollider:2025sys} and the $\mu$TRISTAN proposal~\cite{Hamada:2022mua} have the potential to probe the relevant effective scale up to the TeV level for dim-6 operators,
covering a wide range of DM masses and surpassing the  indirect bounds established in this work. 
We postpone these interesting topics in future research.

\acknowledgments
This work was supported in part by Grants No.\,NSFC-12305110, No.\,NSFC-12035008 and No.\,NSFC-12347112.

\appendix
\section{Calculation of the general velocity-dependent $\langle \sigma v_{\rm rel} \rangle$}
\label{app:thermal}
For DM annihilation into SM particles beyond the s-wave case, the velocity distribution of the initial-state DM particles must be taken into account. The thermally averaged annihilation cross section is then given by
\begin{align}
\langle \sigma v_{\rm rel} \rangle (\pmb{r}) & \equiv
\int d^3 \pmb{v}_1 ~f(\pmb{r}, \pmb{v}_1) \int d^3 \pmb{v}_2 ~f(\pmb{r}, \pmb{v}_2) \sigma v_{\rm rel}
\nonumber
\\
& \equiv \int d v_{\rm rel} ~F({\bm r}, v_{\rm rel}) \sigma v_{\rm rel}.
\label{eq:sigmavr}
\end{align}
Here, $\bm v_1$ and $\bm v_2$ are the lab-frame velocities of the two initial DM particles, 
whose velocity distribution is described by  $f(\bm r,\bm v)$ at position $\bm r$. 
Since $\sigma v_{\rm rel}$ depends only on the relative velocity $v_{\rm rel} = |\bm{v}_1 - \bm{v}_2|$ in the nonrelativistic limit, it is convenient to first integrate out all other velocity degrees of freedom in the distribution functions and then define a new distribution function $F({\bm r}, v_{\rm rel})$, as was done in the second step.

In this work, we follow Refs.\,\cite{Robertson:2009bh,Zhao:2016xie,Arguelles:2019ouk} and assume that DM particles obey a three-dimensional Maxwell-Boltzmann velocity distribution with an $r$-dependent velocity dispersion $v_0(r)$,
\begin{align}
f({\bm r}, {\bm v} ) = \frac{1}{(2\pi v_0^2)^{3/2}} \exp\left(-\frac{v^2}{2 v_0^2}\right).
\end{align}
Based on the definition in \cref{eq:sigmavr}, it is straightforward to see that
\begin{equation}
    F({\bm r}, v_{\rm rel}) = 4\pi v_{\rm rel}^2 \frac{1}{(2\pi(2v_0^2))^{3/2}}\exp\left(\frac{-v_{\rm rel}^2}{2(2v_0^2)}\right).
\end{equation}
This shows that the relative-velocity distribution retains the standard Maxwell-Boltzmann form but with an effective velocity dispersion $\sqrt{2} v_0 (r)$.
As a result, the moments of the relative velocity are
calculated to take the form
\begin{align}
\langle v_{\rm rel}^{2n} \rangle (\bm{r}) 
& \equiv  \int_{0}^{\infty} v_{\rm rel}^{2n}\, F(\bm{r}, v_{\rm rel})\, dv_{\rm rel}
\nonumber
\\
& =2^n (2n+1)!!\,v_0^{2n}(r).
\end{align}
Using the nonrelativistic expansion
\begin{equation}
    \sigma v_{\rm rel}  = \hat a + \hat b\, v_{\rm rel}^2 + \hat d\, v_{\rm rel}^4+\cdots,
\end{equation}
the thermally averaged annihilation cross section follows directly as
\begin{equation}
\langle \sigma v_{\rm rel} \rangle
= \hat{a} + 6\hat{b}\, v_0^2(r)  + 60\hat{d} \, v_0^4(r) + \cdots.
\end{equation}

The velocity dispersion $v_0 (r)$ can be obtained by solving the Jeans equation \cite{binney2011galactic,Ferrer:2013cla}
\begin{equation}
v_0^2(r) = \frac{1}{\rho(r)} \int_{\infty}^r dr ~\rho(r) \frac{d \phi}{d r},
\end{equation}
where $\rho(r)$ is the mass density of DM,
and $\phi(r)$ is the total gravitational potential at radius $r$, including contributions from both DM and ordinary matter.
Following \cite{Boddy:2018ike,Arguelles:2019ouk}, we model $\phi(r)$ as a sum of three components: the DM halo ($\phi_{\rm halo})$, the Milky Way (MW) bulge ($\phi_{\rm bulge})$, and the MW disk ($\phi_{\rm disk}$).
The DM halo potential with the NFW profile is given by
\begin{align}
\phi(r)_{\rm{halo}}&= -G_N \int_{0}^{r_{\rm vir}} \frac{\rho(r)}{|r-r'|}d^3r'
\notag\\
&=-4\pi G_N 
\Big[ \int_{0}^{r}\hspace{-1.5mm} \rho(r')\frac{r'^2}{r}dr'
+\int_{r}^{r_{\rm vir}}\hspace{-3mm} \rho(r') r' dr'\Big]
\notag\\
&=4\pi \rho_0 G_N r_s^3\left\{\frac{1}{r}\left[\frac{r}{r+r_s}-\log\left(\frac{r+r_s}{r_s}\right)\right]\right.
\notag\\
& \left.+\frac{r-r_{\rm vir}}{(r+r_s)(r_s+r_{\rm vir})}\right\},
\end{align}
where $G_N$ is Newton's gravitational constant.
The bulge and disk potentials are modeled as~\cite{Boddy:2018ike} 
\begin{align}
\phi(r)_{\rm{budge}}&=-\frac{G_N M_b}{r+c_b},\\
\phi(r)_{\rm{disc}}&=-\frac{G_N M_d}{r}\left(1-e^{-r/c_d}\right),
\end{align}
where $M_b = 1.5\times 10^{10}~M_\odot$ and $c_b = 0.6~{\rm kpc}$, while $M_d = 7\times 10^{10}~M_\odot$ and $c_d = 4~{\rm kpc}$, with $M_\odot$ denoting the solar mass.

\normalem
\bibliography{paper_refs}
\bibliographystyle{utphys28mod}

\end{document}